\documentclass[11pt]{article}

\usepackage[margin=2.75cm]{geometry}
\usepackage{authblk}
\usepackage{amsmath,amssymb,amsthm,mathtools}
\usepackage{bm}
\usepackage[expansion=false]{microtype}
\usepackage{multirow}
\usepackage{float}
\usepackage{tikz}
\usetikzlibrary{arrows.meta,positioning}
\usepackage{xcolor}
\definecolor{darkgreen}{RGB}{0,100,0}

\usepackage{chngcntr}

\usepackage{tocloft}
\usepackage[
    colorlinks=true,
    linkcolor=blue,
    citecolor=darkgreen,
    urlcolor=teal
]{hyperref}
\usepackage{thmtools}
\usepackage{tabularx}
\usepackage{array}
\usepackage{quantikz}

\newtheorem{theorem}{Theorem}
\newtheorem{lemma}{Lemma}

\theoremstyle{definition}
\newtheorem{definition}{Definition}

\newtheorem{problem}{Problem}
\usepackage{titlesec}

\allowdisplaybreaks
\newcommand{\N}{\mathcal N}
\newcommand{\M}{\mathcal M}
\newcommand{\E}{\mathbb E}
\newcommand{\C}{\mathbb C}

\newcommand{\Id}{\mathrm{Id}}
\newcommand{\tr}{\operatorname{tr}}
\newcommand{\tauD}{\tau_{d_{\mathrm{out}}}}
\newcommand{\diam}{\diamond}
\newcommand{\Hfin}{\mathcal H_{\mathsf{fin}}}
\newcommand{\Lop}{\mathcal L}

\usepackage[titletoc,title]{appendix}

\usepackage[normalem]{ulem}

\usepackage{xcolor}
\usepackage{tikz}
\usetikzlibrary{arrows.meta,decorations.pathreplacing}

\definecolor{stepgreen}{RGB}{0,110,60}

\newcommand{\supp}{\operatorname{supp}}

\newcommand{\acc}{\mathrm{acc}}
\newcommand{\din}{d_{\mathrm{in}}}
\newcommand{\dout}{d_{\mathrm{out}}}
\newcommand{\tauout}{\tau_{\dout}}

\title{Lower bounds for testing the completely depolarizing channel}

\author{Hirak Ghosh\thanks{\href{mailto:Hirak.Ghosh@warwick.ac.uk}{Hirak.Ghosh@warwick.ac.uk}}}
\author{Andrew Jackson}
\author{Animesh Datta}

\affil{\small Department of Physics, University of Warwick, Coventry, CV4 7AL, United Kingdom}

\date{}

\begin{document}
\maketitle

\begin{abstract}
We study lower bounds to the query complexity of testing whether an unknown quantum channel mapping a $\din$-dimensional input to a $\dout$-dimensional output is the completely depolarizing channel or is $\varepsilon$-far from it in diamond distance. We show that for this task 
(i) any ancilla-assisted, coherent, adaptive tester requires at least $\Omega(\din\dout/\varepsilon^2)$ queries;
(ii) any ancilla-assisted, coherent, non-adaptive tester requires at least $\max\{\Omega(\din^{1.5}\dout/\varepsilon^2), \Omega(\din^{2}/\varepsilon^2)\}$ queries. 
We also show that these lower bounds apply to the more general problem of determining whether an unknown channel belongs to the set of all state-replacement channels or is $\varepsilon$-far from the entire set in diamond distance.

\end{abstract}  

\setcounter{tocdepth}{2}

\newpage
\tableofcontents

\newpage
\section{Introduction}
We study the query complexity of testing for the completely depolarizing channel $\mathcal N:\mathcal L(\C^{\din})\to\mathcal L(\C^{\dout})$, defined by $\mathcal N(X):=\operatorname{tr}(X)I/\dout$. Given black-box access to an unknown quantum channel $\mathcal M:\mathcal L(\C^{\din})\to\mathcal L(\C^{\dout})$, the task is to distinguish $\mathcal M=\mathcal N$ from $\|\mathcal M-\mathcal N\|_{\diamond}\ge\varepsilon$ with constant success probability. This problem, commonly referred to as the certification of the completely depolarizing channel, is the quantum channel analogue of mixedness testing for quantum states \cite{10.1145/2746539.2746582, 9318006, o2025instance} and uniformity testing for classical distributions \cite{4626074, Goldreich2011, cj19-01}. When $\din =1,$ channel testing in the diamond distance reduces to mixedness testing of quantum states in the trace distance.

Unlike quantum state testing, where the tester receives a fixed product-state resource $\rho^{\otimes n}\in\mathcal L((\mathbb C^d)^{\otimes n})$, quantum channel testing affords the freedom of specifying the input state and the manner of channel access. In the coherent non-adaptive model, the $n$ input registers may be prepared jointly with an ancilla and the channel can be applied in parallel, followed by an arbitrary joint measurement. In the fully adaptive model, however, the channel uses are allowed to be queried sequentially, allowing later inputs and the retained quantum memory to depend coherently on earlier outputs. Incoherent strategies for testing the completely depolarizing channel have been studied recently~\cite{fawzi2023quantum}. 

In this paper, we consider the broader class of ancilla-assisted, coherent testers, including fully adaptive strategies. As such, our central question is:

\begin{center}
\textit{What is the minimum query complexity necessary to test whether a quantum channel is completely depolarizing under coherent, fully adaptive strategies?}
\end{center}

Formally, the testing task in the above question is defined as in Problem~\ref{prob:depolarizing-channel-testing} below: 
\begin{problem}
\label{prob:depolarizing-channel-testing}
Let $\din,\dout \in \mathbb{N}^{\ge2}$, $\tauout:=I_{\dout}/\dout$, and define $\N:\Lop(\C^{\din})\to\Lop(\C^{\dout})$ by $\N(X):=\tr(X)\tauout$.
Given black-box query access to an unknown quantum channel
$\M:\Lop(\C^{\din})\to\Lop(\C^{\dout})$, distinguish with success probability at least
$2/3$ between 
\begin{align}
    \text{[Null Hypothesis]}~~H_0&: \M=\N, ~~\text{or}\nonumber\\\text{[Alternate Hypothesis]}~~H_1&: \|\M-\N\|_\diam\ge\varepsilon.
    \label{Eqn:hypothesis}
\end{align}
\end{problem}
\subsection{Results}
\label{sec:results} 
We show that Problem~\ref{prob:depolarizing-channel-testing} requires $\max\{\Omega(\din^{1.5}\dout/\varepsilon^2), \Omega(\din^{2}/\varepsilon^2)\}$ queries in the coherent non-adaptive model (See Def.~\ref{def:nonadaptive-tester}) in Theorem~\ref{thm:max-non-adapt}. We also show that Problem~\ref{prob:depolarizing-channel-testing} requires $\Omega(\din\dout/\varepsilon^2)$ queries in the coherent adaptive model (See Def.~\ref{def:adaptive_tester}) in Theorem~\ref{thm:main_adapt}. 
Table~\ref{tab:channel-testing-results} compares our results with existing results.

In the ancilla-assisted coherent, non-adaptive model (See Def.~\ref{def:nonadaptive-tester}) we construct a channel family $\mathcal M_{v,U}$ that is hard to distinguish from the completely depolarizing channel $\mathcal N$. Let $\bm V$ be a Haar random unitary acting on the input Hilbert space and $\bm U$ be an independent Haar random unitary acting on the output Hilbert space. Let $\bm v:=\bm V|0\rangle$. For fixed $v,U$, define
\begin{equation}
    \rho_U
    := \tauout+\frac{\varepsilon}{\sqrt{\dout}}U\Delta_0U^\dagger,
    \quad
    \M_{v,U}(X)
    :=
    \tr(|v\rangle\langle v|X)\rho_U
    +
    \tr[(I-|v\rangle\langle v|)X]\tauout,
    \label{construction}
\end{equation}
where $\Delta_0$ is chosen so that our hard channel instance $\M_{v,U}$ satisfies 
$\|\M_{v,U}-\N\|_\diamond\ge\varepsilon$. Each application of $\mathcal{M}_{v,U}$ effectively implements a two-outcome projective measurement $\{|v\rangle\langle v|, I-|v\rangle\langle v|\}$. We refer to a projection onto the $\ket{v}\bra{v}$ as a \textit{hit}, a projection onto its orthogonal complement as a \textit{miss}. 

We then define $\sigma_0$ as the pre-measurement state under the null hypothesis (in Eqn.~\eqref{Eqn:hypothesis}) and $\bar{\sigma}$ as the averaged (over all $v,U$) pre-measurement state under the alternate hypothesis. By Lemma~\ref{lem:randomized-to-deterministic-tester} any tester with error probability at most $1/3$ must satisfy: $1/3 \le \frac{1}{2} \|\sigma_0 - \bar\sigma\|_1$. The proof strategy is then to find an upper bound on $\|\sigma_0 - \bar\sigma\|_1$ in terms of the number of queries $n$, the proximity parameter $\varepsilon$ and the input and output dimensions of the channel. We proceed by splitting the one norm into contributions from a low-hit record and a high-hit record, where a \textit{record} is a sequence of hit or miss outcomes across the $n$ channel calls and low-hit records contain at most $m$ hits and high-hit records contain more than $m$ hits. This yields
\[
\frac12\|\sigma_0-\bar{\sigma}\|_1
\leq
\frac12\|X_m\|_1
+\frac12\|Y_m\|_1,
\]
where $X_m$ denotes the low-hit contribution and $Y_m$ denotes the high-hit contribution. 

For the lower bound $\Omega(\din^{1.5}\dout/\varepsilon^2)$, the high-hit contribution is bounded by Markov's inequality, where we use the fact that the expected number of hits is $n/\din$ and the probability of observing at least $m$ hits is at most $n/m\din$. The low-hit contribution is then bounded by a quadratic expression in \(X_m\). Expanding it involves two independent draws of the random channel, allowing us to use Haar-moment identities to control their overlaps. The high-level idea is to bound the low-hit contribution over subsets of hit positions, indexed by their cardinality $j$. All odd-order terms in the low-hit moment expansion vanish under Haar averaging. This leaves the two-hit ($j=2$) term as the first nonzero contribution, followed by the $j=4,6,\ldots$ hit terms. These surviving terms are suppressed by factors depending
on $\dout$, and controlling their higher moments requires 
\[
    n = \Omega\left(
        \frac{\din^{3/2}\dout}
        {\varepsilon^2}
    \right).
\]

For the second bound $\Omega(\din^{2}/\varepsilon^2)$, we fix the output unitary $U=I$ in Eqn.~\eqref{construction}, retaining only the Haar randomness in $v$. The high-hit contribution can be still bounded by invoking Markov's inequality. For the low-hit contribution, we observe that single-hit term $j=1$ now survives, followed by contributions from every order $j=2,3,\ldots$. Unlike the previous case, these terms admit a direct geometric-series bound in the number of hits, with common ratio $m\varepsilon^2/\din$. Choosing $m=\Theta(\din/\varepsilon^2)$ and
combining the low-hit and high-hit estimates yields
\[
    n = \Omega\left(
        \frac{\din^2}{\varepsilon^2}
    \right).
\]

Derived from separate ``hard" cases, these two bounds exhibit a crossover at $\dout \sim \sqrt{\din}$ that may be either fundamental to the problem's complexity or simply a limitation of the current analysis. Since an alternative approach could potentially yield a bound surpassing both, a matching upper-bound algorithm is necessary to determine whether such a crossover indeed exists.

For the coherent adaptive lower bound (access model defined in Def.~\ref{def:adaptive_tester}), we work with the same input-output hard family as in Eqn.~\eqref{construction}. We again decompose the trace norm between the two pre-measurement states as in the non-adaptive case, with the only difference being on how we bound the low-hit contribution. Unlike in the non-adaptive setting, the tester can now use information from earlier channel outputs to adapt its subsequent inputs, potentially correlating them with the hidden direction $v$. Consequently, the fixed-input Haar-moment argument used previously does not directly apply. To overcome this difficulty, we introduce a \emph{truncated adaptive experiment} that follows the original experiment (described in Def.~\ref{def:adaptive_tester}) under the alternative hypothesis until $m$ hits have occurred, after which any subsequent hit produces the depolarizing output $\tau_{\dout}$. We then reduce the analysis of this truncated experiment to a non-adaptive, $m$-copy state-distinguishability problem by constructing a completely positive, trace-nonincreasing (CP-TNI) map. This is similar to the \textit{adaptive to non adaptive simulation} of Das and Wilde~\cite[Prop.~2]{DasWilde2019}, in which the parameter-dependent part of an adaptive experiment is isolated into auxiliary states and the remaining adaptive steps are absorbed into a parameter-independent map. 

We also provide lower bounds on the following membership problem:

\begin{restatable}[Membership in the set of state-replacement channels]{problem}{srproblem}

Define the set of all state-replacement channels
\[
    \mathsf{SR}(\C^{\din},\C^{\dout})
    :=
    \left\{
        \mathsf R_\sigma:
        \Lop(\C^{\din})\to\Lop(\C^{\dout})
        \;:\;
        \mathsf R_\sigma(X)=\tr(X)\sigma,\,
        \sigma\in\mathsf D(\C^{\dout})
    \right\}.
\]
where $\din,\dout \in \mathbb{N}^{\ge2}$. Given query access to an unknown channel
$ \mathcal M:
    \Lop(\C^{\din})\to\Lop(\C^{\dout}),
$ distinguish, with success probability at least $2/3$, between
\begin{align}
    \text{[Null Hypothesis]}~~ H_0^{\rm SR} &:
    \mathcal M\in\mathsf{SR}(\C^{\din},\C^{\dout}), ~~\text{or}\nonumber\\\text{[Alternate Hypothesis]}~~H_1^{\rm SR} &:
    \inf_{\mathsf R_\sigma\in
    \mathsf{SR}(\C^{\din},\C^{\dout})}
    \|\mathcal M-\mathsf R_\sigma\|_\diamond
    \ge\varepsilon.
    \label{Eqn:hypothesisSR}
\end{align}
\label{prob:SR}
\end{restatable}
Problem 2 reduces to Problem 1 when $\sigma = \tau_{\dout}$. To extend our bounds from Problem~\ref{prob:depolarizing-channel-testing}, the channel family in Eqn.~\eqref{construction} must remain sufficiently distant from all channels in the larger class for a general $\sigma$.  We show in Appendix~\ref{corollary_proof} that this holds for the family in Eqn.~\eqref{construction} with a slight modification in the proximity parameter. This leads to the following corollary.

\begin{restatable}[Lower bound for testing state-replacement channels]{corollary}{srcorollary}
\label{cor:state-replacement-lower-bound}

Any coherent (possibly adaptive) quantum tester
that solves Problem~\ref{prob:SR}
requires
$
    \Omega\left(
        {\din \dout}/{\varepsilon^2}
    \right)
$
queries. Moreover, any coherent non-adaptive tester solving the same problem requires
$
    \max \{\Omega(\din^{1.5} \dout/\varepsilon^2),\Omega(\din^{2} /\varepsilon^2)\}
$, for $0\le \varepsilon\le1/2$.
\end{restatable}
\begin{proof}
   Proof in Appendix~\ref{corollary_proof}. 
\end{proof}

This generalises the study of channel certification from a target pure-state replacement channel~\cite[Theorem~1.2]{rosenthal2024quantum} to a broader target class of all state-replacement channels in the coherent adaptive access model.

Our lower bounds are not known to be optimal as we do not establish a matching upper bound. 
As noted in \cite[Table 1]{rosenthal2024quantum}, the sufficient query complexity for certifying a general target channel (and the depolarizing channel) in the coherent non-adaptive model is $\mathcal{O}(\din^2 \dout/\varepsilon^2)$. This upper bound follows directly from \cite[Lemma~C.1]{fawzi2023quantum} and \cite[Theorem~1.4]{10.1145/3313276.3316344}.

\begin{table}[!htbp]
    \centering
    \renewcommand{\arraystretch}{0.55}
    \setlength{\tabcolsep}{0.1pt}
    \begin{tabularx}{\linewidth}{ >{\raggedright\arraybackslash}p{0.2\linewidth} >{\raggedright\arraybackslash}p{0.3\linewidth} >{\centering\arraybackslash}X }
        \hline
\rule{0pt}{3ex}
\textbf{Testing task} &
\textbf{Access model} &
\textbf{Query complexity}\\[1ex]
\hline
        General &
        {Anc. Free} + {Incoh.} + {Non-Adapt.} &
        $\widetilde{O}\left(
            \dfrac{\din^2\dout^{1.5}}{\varepsilon^2}
        \right)$ \cite{rosenthal2024quantum}
        \\

        General &
        {Anc. Assist.} + {Coh.} + {Non-Adapt.} &
        $O\left(
            \dfrac{\din^2\dout}{\varepsilon^2}
        \right)$
        \cite{fawzi2023quantum,10.1145/3313276.3316344}
        \\

        General &
        {Anc. Assist.} +  {Coh.} + {Adapt.} &
        $\Omega\left(
            \dfrac{\sqrt{\din}}{\varepsilon}
        \right)$ \cite{rosenthal2024quantum}
        \\
        \hline  
        Completely depolarizing &
        {Anc. Assist.} +  {Incoh.} +  {Non-Adapt.} &
        $\widetilde{\Theta}\left(
            \dfrac{\din^2\dout^{1.5}}{\varepsilon^2}
        \right)$ \cite{fawzi2023quantum, rosenthal2024quantum}
        \\
        Completely depolarizing &
        {Anc. Assist.} +  {Incoh.} +  {Adapt.} &
        $\Omega\left(
            \dfrac{\din^2\dout+\dout^{1.5}}{\varepsilon^2}
        \right)$ \cite{fawzi2023quantum}
        \\
        Completely depolarizing &
        {Anc. Assist.} + {Coh.} + {Non-Adapt.} &
        $O\left(
            \dfrac{\din^2\dout}{\varepsilon^2}
        \right)$
        \cite{fawzi2023quantum,10.1145/3313276.3316344}
        \\
        Completely depolarizing &
        {Anc. Assist.} +  {Coh.} +  {Non-Adapt.} &
        $\max \left\{\Omega\left(\dfrac{\din^{1.5}\dout}{\varepsilon^2}\right), \Omega\left(\dfrac{\din^{2}}{\varepsilon^2}\right) \right\}$  Theorem~\ref{thm:max-non-adapt}
        \\

        Completely depolarizing &
        {Anc. Assist.} +  {Coh.} +  {Adapt.} &
        $\Omega\left(
            \dfrac{\din\dout}{\varepsilon^2}
        \right)$  Theorem~\ref{thm:main_adapt}
        \\
        \hline
        Pure State Replacement Channel &
        {Anc. Assist.} +  {Coh.} + {Adapt.} &
        $\Omega\left(
            \dfrac{\din}{\varepsilon}
        \right)$ \cite{rosenthal2024quantum}
        \\
        Set of all state-replacement channels &
        {Anc. Assist.} +  {Coh.} + {Non-Adapt.} &
        $\max \left\{\Omega\left(\dfrac{\din^{1.5}\dout}{\varepsilon^2}\right), \Omega\left(\dfrac{\din^{2}}{\varepsilon^2}\right) \right\}$  Cor.~\ref{cor:state-replacement-lower-bound}
        \\
        Set of all state-replacement channels &
        {Anc. Assist.} +  {Coh.} + {Adapt.} &
        $\Omega\left(
        \dfrac{\din\dout}{\varepsilon^2}
        \right)$  Cor.~\ref{cor:state-replacement-lower-bound}
        \\

        \hline
    \end{tabularx}
\caption{Query complexity bounds for channel certification in diamond distance. The abbreviations {Anc. Free}, {Anc. Assist.}, {Incoh.}, {Coh.}, {Non-Adapt.}, and {Adapt.} denote ancilla-free,
ancilla-assisted, incoherent measurement, coherent measurement, non-adaptive, and adaptive access,
respectively. The notation $\widetilde{O}$ and $\widetilde{\Theta}$ suppresses
logarithmic factors.}
    \label{tab:channel-testing-results}
\end{table}

\subsection{Related Work}
\paragraph{State and Distribution Testing:}
In classical distribution identicality testing\footnote{This task is often referred to as \emph{identity testing} in the literature. Here, ``identity'' denotes the property of being identical to a target object.}, given sample access
to an unknown distribution $p$ over $[d]$ and a known reference distribution $q$, to distinguish $p=q$ from $\|p-q\|_1\ge\varepsilon$. \cite{Goldreich2011} introduced collision-based uniformity testing, and \cite{d7762e2d2cc8487da199713a486bdd30} subsequently studied certification for arbitrary reference distributions.
\cite{4626074} established the tight $\Theta(\sqrt d/\varepsilon^2)$ sample complexity for uniformity testing. Subsequent work obtained the same worst-case complexity for general identity testing \cite{7782983}, followed by instance-optimal characterizations
\cite{doi:10.1137/151002526}.

The quantum analogue is state certification: given copies of an unknown state $\rho\in\mathsf D(\mathbb C^d)$ and a known reference state $\sigma$, the task is to distinguish $\rho=\sigma$ from $\|\rho-\sigma\|_1\ge\varepsilon$. 
\cite{10.1145/2746539.2746582} established the optimal $\Theta(d/\varepsilon^2)$ copy complexity for the special case
$\sigma=I_d/d$, known as mixedness testing, under collective measurements. \cite{10.1145/3313276.3316344}
subsequently treated certification against arbitrary reference states. \cite{9318006} proved an $\Omega(d^{4/3}/\varepsilon^2)$ lower bound for mixedness testing, even with adaptive single-copy measurements. \cite{pmlr-v178-chen22b} developed essentially instance-optimal bounds for non-adaptive incoherent state certification. Nearly instance-optimal bounds for unrestricted entangled measurements were obtained more recently \cite{o2025instance}.

\paragraph{Channel Testing:}
\cite{fawzi2023quantum} studied quantum channel certification for unitary and completely depolarizing channels using measurements that are incoherent across channel uses. For the latter, they obtained a non-adaptive query complexity of $\widetilde{\Theta}(\din^2 \dout^{1.5}/\varepsilon^2)$ and derived an adaptive lower bound of $\Omega(\din^2 \dout/\varepsilon^2)$. \cite{rosenthal2024quantum} subsequently considered certifying an arbitrary reference channel in diamond distance, extending the depolarizing channel upper bound $\widetilde{\Theta}(\din ^2\dout^{1.5}/\varepsilon^2)$ to arbitrary reference channels. They also proved a $\Omega(\sqrt{\din}/\varepsilon)$ lower bound even for fully coherent adaptive testers and an $\Omega(d_{in}/\varepsilon)$ bound for target pure-state replacement channels, alongside several useful results in the average-case imitation diamond (ACID) distance. A non-adaptive upper bound for an arbitrary reference channel with coherent measurements follows as a corollary from the diamond distance to two-norm testing between their corresponding Choi states \cite[Lemma~C.1]{fawzi2023quantum}, combined with the two-norm state testing from \cite[Theorem~1.4]{10.1145/3313276.3316344}, yielding $\mathcal{O}(\din^2 \dout / \varepsilon^2)$. These results are also detailed against our results in Table~\ref{tab:channel-testing-results}.

Recent work has studied unitary channel certification under structural promises, including promised coherent noise \cite{jeon2025query}, unitary targets under different access models \cite{chen2026strict}, and depolarizing deviations, for which optimal tolerant-testing bounds are known \cite{ghosh2026unitarychanneltestingdepolarizing}.

\section{Notation and Preliminaries}

\subsection{Operator and channel notation}

\paragraph{Basic notation:}

For a positive integer $d$, write $[d]:=\{1,\ldots,d\}$. All Hilbert spaces are finite-dimensional over $\mathbb C$. Throughout the paper, we use $A_x$, $B_x$, and $R_x$ to denote the input, output, and tester memory registers associated with the $x$-th channel use, respectively. Formally, let $\mathcal H_{A_x}\cong\mathbb C^{\din}$, $\mathcal H_{B_x}\cong\mathbb C^{\dout}$, and let $\mathcal H_{R_x}$ be an arbitrary finite-dimensional Hilbert space. 
To reduce burden on the notation, we use the same symbols $A_x$, $B_x$, and $R_x$ for the corresponding Hilbert spaces $\mathcal H_{A_x}$, $\mathcal H_{B_x}$, and $\mathcal H_{R_x}$, respectively. Dimensions are denoted as $\dim.$

We write $\Lop(A)$ for the space of linear operators on $A$, $I_A$ for the identity operator on $A$, and
\[\mathsf D(A) := \{\rho\in\Lop(A):\rho\succeq0,\ \tr\rho=1\}\]
for the set of density operators on $A$. 

Boldface symbols denote random variables, while plain symbols denote fixed realizations. We write $\tau_d:=I_d/d$ for the maximally mixed state on $\mathbb C^d$, and $\E$ for expectation, with a subscript indicating the random variable being averaged, as in $\E_{U}$.

\paragraph{Sub-normalized states:}

A \emph{sub-normalized state} on $A$ is an operator $X\in\Lop(A)$ satisfying $X\succeq0$ and $\tr X\le1$. Since $X\succeq0$, $\|X\|_1=\tr X$.

\emph{Example.}
Consider the qubit measure-and-prepare channel $\Phi(X)=\langle0|X|0\rangle\,\rho_0+\langle1|X|1\rangle\,\rho_1,$ where $\rho_0,\rho_1\in\mathsf D(B)$. Call the outcome associated with $|1\rangle\langle1|$ a \emph{hit} and that associated with $|0\rangle\langle0|$ a \emph{miss}. For an input state $\omega$, the
one-step hit-miss record is therefore $s\in\{0,1\}$. Define
$X_0:=\langle0|\omega|0\rangle\,\rho_0,~X_1:=\langle1|\omega|1\rangle\,\rho_1.
$Then $X_0$ and $X_1$ are the sub-normalized contributions associated
with the miss and hit records, respectively, with
$\tr X_0=\langle0|\omega|0\rangle,~\tr X_1=\langle1|\omega|1\rangle,
$
and
$X_0+X_1=\Phi(\omega).$ 

We use \emph{hit-miss sequence} and \emph{hit-miss record} interchangeably for a binary string indexing such hit-miss decompositions.
\paragraph{Norms:}
\label{sec:norms}

For $X\in\Lop(A)$, we use the Schatten norms
\[
    \|X\|_1:=\tr\sqrt{X^\dagger X},
    \quad\|X\|_2:=\sqrt{\tr(X^\dagger X)},
    \quad
    \|X\|_\infty:=\sup_{\|v\|=1}\|Xv\|.\]
The Hilbert--Schmidt inner product is
$\langle X,Y\rangle_{\mathrm{HS}}:=\tr(X^\dagger Y),$
with
\begin{equation}
    \bigl|\tr(X^\dagger Y)\bigr|
    \le
    \|X\|_2\|Y\|_2.
    \label{HS-norm}
\end{equation}
We will also use
\begin{equation}
    \|XY\|_2
    \le
    \|X\|_\infty\|Y\|_2,
    \label{eq:2 to operator norm}
\end{equation}
which follows from
$
    X^\dagger X\preceq\|X\|_\infty^2I_A.
$

For $\rho,\sigma\in\mathsf D(A)$, the trace distance is
\[
    D(\rho,\sigma)
    :=
    \frac12\|\rho-\sigma\|_1.
\]
Its variational characterization is
\[
    D(\rho,\sigma)
    =
    \max_{0\preceq M\preceq I_A}
    \tr[M(\rho-\sigma)],
\]
with the maximum attained by the projector onto the positive eigenspace of
$\rho-\sigma$ \cite[Theorem ~6.1]{khatri2020principles}. A POVM on outcome set $\mathcal Z$ is a family $\{M_z\}_{z\in\mathcal Z}$ satisfying $M_z\succeq0$ and $\sum_z M_z=I_A$.

\begin{lemma}[Quantum chi-squared divergence~\cite{Temme_2010}]
\label{lem:chi}
Let $\rho,\sigma$ be Hermitian trace-class operators, with $\rho\succeq0$ a state and
$\supp(\sigma)\subseteq\supp(\rho)$. Then
\[
    \|\sigma-\rho\|_1^2
    \le
    \tr\left[
        (\sigma-\rho)\rho^{-1}(\sigma-\rho)
    \right]
    =:
    \chi^2(\sigma\|\rho),
\]
where $\rho^{-1}$ denotes the inverse on $\supp(\rho)$.
\end{lemma}
\paragraph{Linear maps and quantum channels:}

For a linear map $\Phi:\Lop(A)\to\Lop(B)$, define its \emph{unnormalized} Choi matrix, relative to a fixed orthonormal basis $\{|i\rangle\}_i$ of $A$, by
\[
    J_\Phi:=\sum_{i,j}|i\rangle\langle j|\otimes\Phi(|i\rangle\langle j|).
\]
Then $\Phi$ is completely positive (CP) iff $J_\Phi\succeq0$, and it is trace-preserving (TP) iff $\tr_B J_\Phi=I_A$. If $\Phi$ is CP, it is trace-nonincreasing (TNI) iff $\tr_B J_\Phi\preceq I_A$. Hence $\Phi$ is a quantum channel iff
\[
    J_\Phi\succeq0
    \quad\text{and}\quad
    \tr_B J_\Phi=I_A.
\]

The diamond norm of a linear map $\Phi:\Lop(A)\to\Lop(B)$ is
\[
    \|\Phi\|_\diamond
    :=
    \sup_{\substack{X\in\Lop(A\otimes R)\\ \|X\|_1\leq 1}}
    \bigl\|(\Phi\otimes\Id_R)(X)\bigr\|_1,
\]
where it suffices to take $\dim R=\dim A$ (See \cite[Theorem 3.46]{watrous2018theory}). If $\Phi$ is Hermiticity-preserving, this simplifies to
\[
    \|\Phi\|_\diamond
    =
    \max_{\rho\in\mathsf D(A\otimes R)}
    \bigl\|(\Phi\otimes\Id_R)(\rho)\bigr\|_1. ~~~\cite[\text{Theorem 3.51}]{watrous2018theory}
\]
In particular, since the difference of two channels is Hermiticity-preserving, we define their (unhalved) diamond distance as $\|\Phi-\Psi\|_\diamond$.
\subsection{Haar-random unitaries}
\label{sec:Haar-prelim}

We write $\bm U\sim\operatorname{Haar}(U(d))$ for a Haar-random unitary, where
$U(d):=\{U\in\mathbb C^{d\times d}:U^\dagger U=UU^\dagger=I_d\}$
is the unitary group on $\mathbb C^d$. The Haar measure is the unique probability measure on $U(d)$ invariant under left and right multiplication. Consequently, for fixed $X\in\Lop(\mathbb C^d)$,
\begin{align}
    \E_{\bm U}[\bm U X\bm U^\dagger]=
    \frac{\tr X}{d}I_d.
    \label{eqn:Haarfirstmoment}
\end{align}
(See \cite[Cor. 13]{Mele2024introductiontoHaar}) We will also use the standard moment identity for Haar-random pure states.
If $\bm v=\bm U|0\rangle$ and
$\bm P:=|\bm v\rangle\langle\bm v|$, then for every $j\ge1$,
\begin{equation}
    \E_{\bm v}\bm P^{\otimes j}
    =\frac{\Pi_{\mathrm{sym}}^{(j)}}
         {\tr(\Pi_{\mathrm{sym}}^{(j)})}=
    \binom{d+j-1}{j}^{-1}
    \Pi_{\mathrm{sym}}^{(j)},
    \label{eq:Haar-pure-moment}
\end{equation}
where $\Pi_{\mathrm{sym}}^{(j)}$ is the projector onto the symmetric subspace of $(\mathbb C^d)^{\otimes j}$~\cite[Theorems~17,~22]{Mele2024introductiontoHaar}.

\subsection{Lower bound nomenclature}
\label{nomenclature}
A \emph{hard family} is a collection of channels that are sufficiently far from the target in the diamond distance, yet a random channel sampled according to a suitable distribution is hard to distinguish from the target. This distribution is referred to as the hard distribution, and a \emph{hard instance} is a fixed channel sampled from this family. A \emph{realized channel} is the deterministic channel obtained after fixing all random parameters.

A bound $n=\Omega(g)$ means that every tester satisfying the required success guarantee must make at least $cg$ channel queries, for some universal constant $c>0$. Equivalently, we may write $n\gtrsim g$, where $\gtrsim$ denotes an inequality up to a universal positive multiplicative constant. A lower bound is a \textit{necessary} condition on the number of queries required to solve a problem.

We next specify the channel access models for which we establish our lower bounds. 

\subsection{Access models}

In Section~\ref{sec:NACLB}, we analyze the diamond distance lower bound within a non-adaptive, ancilla-assisted coherent query model. Section~\ref{sec:ACLB} extends this analysis to the fully adaptive coherent model.  We now formally define the deterministic testers in both models with the standard observation that randomized testers are convex combinations of deterministic ones. Ref.~\cite{rosenthal2024quantum} defines these models. We restate them for completeness.

\paragraph{Coherent non-adaptive tester:}
\label{sec:nonadaptive-tester}

At query $i$, the tester sends in the state on the register $A_i\cong\C^{\din}$ through the unknown channel and receives a state on the register $B_i\cong\C^{\dout}$. In the non-adaptive model, all input registers are prepared jointly before any channel use, and the unknown channel is applied in parallel to $A_1,\ldots,A_n$. The register $R$ is an arbitrary finite-dimensional ancilla retained throughout the interaction, and the tester's pre-measurement register is $\Hfin:=B_1\otimes\cdots\otimes B_n\otimes R$.

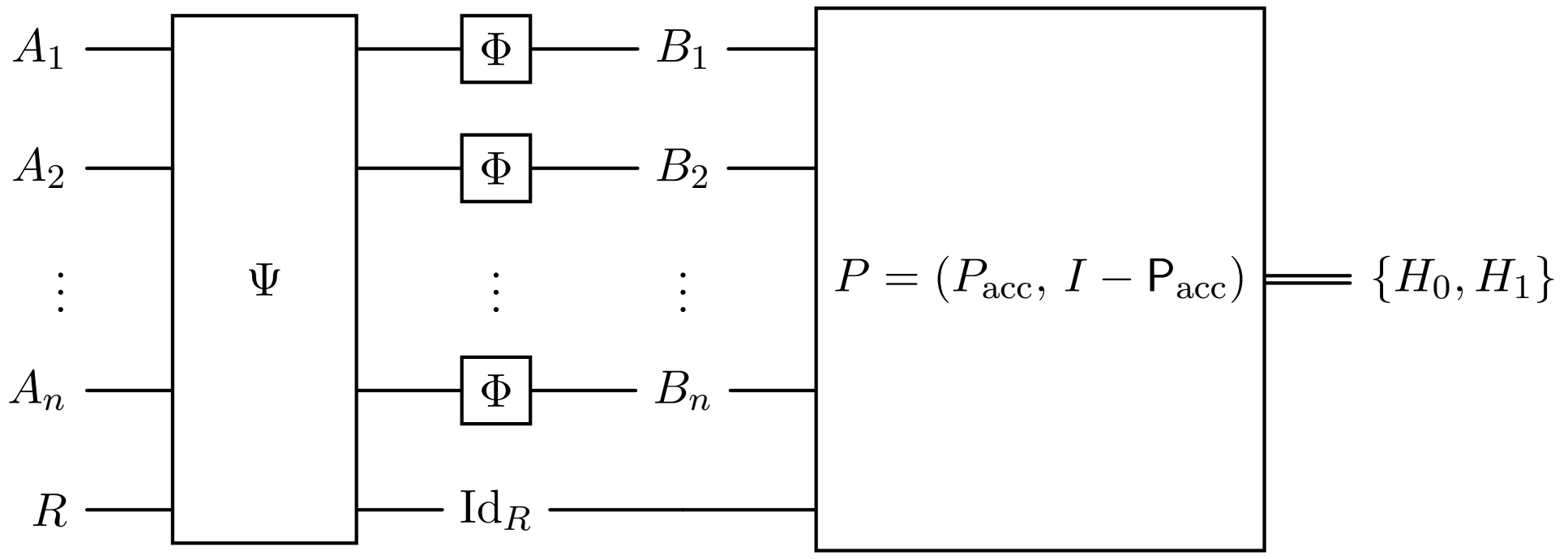
\begin{figure}[ht]
\vspace{+10pt}
\centering
\begin{quantikz}[row sep=0.34cm, column sep=0.7cm, wire types={q,q,n,q,q}]
\lstick{$A_1$}
  & \gate[5][1.5cm]{\Psi}
  & \gate{\Phi}
  & \gate[style={draw=none,fill=none}]{B_1}
  & \gate[5][2.4cm]{P=(P_{\acc},\,I-\mathsf P_{\acc})}
  & \setwiretype{n} \\
\lstick{$A_2$} & & \gate{\Phi} & \gate[style={draw=none,fill=none}]{B_2} & & \setwiretype{n} \\
\lstick{$\vdots$} & & \vdots & \vdots & & \setwiretype{c}\rstick{$\{{H_0},{H_1}\}$} \\
\lstick{$A_n$} & & \gate{\Phi} & \gate[style={draw=none,fill=none}]{B_n} & & \setwiretype{n} \\
\lstick{$R$} & & \gate[style={draw=none,fill=none}]{\Id_R} & & & \setwiretype{n}
\end{quantikz}
\caption{Coherent non-adaptive ancilla-assisted tester. The unknown channel
$\Phi:\Lop(A_i)\to\Lop(B_i)$ is applied in parallel to the $n$ input
registers, followed by a joint two-outcome measurement.}
\label{fig:tester}
\end{figure}

\begin{definition}[Coherent non-adaptive ancilla-assisted tester]
\label{def:nonadaptive-tester}
A deterministic $n$-query coherent non-adaptive tester consists of an initial state $\Psi\in\mathsf D(A_1\otimes\cdots\otimes A_n\otimes R)$ and a two-outcome POVM $\{P_{\mathrm{acc}},I_{\Hfin}-P_{\mathrm{acc}}\}$ on $\Hfin$, where $0\preceq P_{\mathrm{acc}}\preceq I_{\Hfin}$ and the outcome $ P_{\mathrm{acc}}$ denotes acceptance of the null hypothesis. The tester is shown in Figure~\ref{fig:tester}.
\end{definition}

Allowing $\Psi$ to be pure is without loss of generality, because any mixed input state can be purified by enlarging the unrestricted ancilla $R$~\cite[Theorem 2.10]{watrous2018theory}.

In Section~\ref{sec:tester}, we consider the more general fully adaptive, ancilla-assisted query model.
\paragraph{Coherent adaptive tester:}
\label{sec:tester}
 At query $i$, the tester sends a state on the register $A_i\cong\C^{\din}$ through the unknown channel and receives an output state on register $B_i\cong\C^{\dout}$. The register $R_{i-1}$ contains all quantum memory retained from the preceding queries, where $R_0,\ldots,R_{n-1}$ are arbitrary finite-dimensional Hilbert spaces. The tester's pre-measurement register is denoted by $\Hfin$.

\begin{definition}[Coherent adaptive ancilla-assisted tester]
\label{def:adaptive_tester}
A deterministic $n$-query tester consists of an initial state $\Psi\in\mathsf D(A_1\otimes R_0)$, CPTP maps $\mathsf V_i:\Lop(B_i\otimes R_{i-1})\to\Lop(A_{i+1}\otimes R_i)$ for $1\le i<n$, a final CPTP map $\mathsf V_n:\Lop(B_n\otimes R_{n-1})\to\Lop(\Hfin)$, and a two-outcome POVM $\{ P_{\mathrm{acc}},I_{\Hfin}- P_{\mathrm{acc}}\}$, where $0\preceq P_{\mathrm{acc}}\preceq I_{\Hfin}$ and the outcome $ P_{\mathrm{acc}}$ denotes acceptance of the null hypothesis. The resulting adaptive interaction is shown in Figure~\ref{fig:adaptive-tester}.
\end{definition}
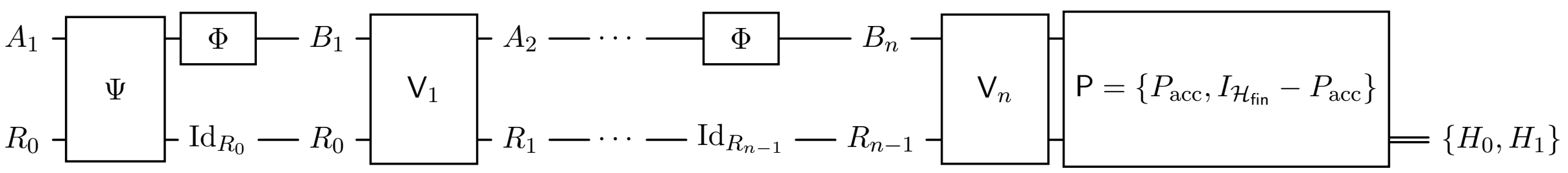
\begin{figure}[htbp]
\centering
\resizebox{\textwidth}{!}{%
\begin{quantikz}[
    row sep=0.6cm,
    column sep=0.5cm,
    wire types={q,q}
]
\lstick{$A_1$}
  & \gate[2,style={minimum width=1.25cm,minimum height=0.75cm}]{\Psi}
  & \gate[style={minimum width=0.95cm,minimum height=0.65cm}]{\Phi}
  & \gate[style={draw=none,fill=none,inner sep=1pt}]{B_1}
  & \gate[2,style={minimum width=1.35cm,minimum height=0.75cm}]{\mathsf V_1}
  & \gate[style={draw=none,fill=none,inner sep=1pt}]{A_2}
  & \gate[style={draw=none,fill=none}]{\cdots}
  & \gate[style={minimum width=0.95cm,minimum height=0.65cm}]{\Phi}
  & \gate[style={draw=none,fill=none,inner sep=1pt}]{B_n}
  & \gate[2,style={minimum width=1.35cm,minimum height=0.75cm}]{\mathsf V_n}
  & \gate[2,style={minimum width=2.3cm,minimum height=0.75cm}]
      {\mathsf P=
      \{ P_{\mathrm{acc}},
      I_{\Hfin}- P_{\mathrm{acc}}\}}
  & \setwiretype{n}
\\
\lstick{$R_0$}
  &
  & \gate[style={draw=none,fill=none,inner sep=1pt}]
      {\operatorname{Id}_{R_0}}
  & \gate[style={draw=none,fill=none,inner sep=1pt}]{R_0}
  &
  & \gate[style={draw=none,fill=none,inner sep=1pt}]{R_1}
  & \gate[style={draw=none,fill=none}]{\cdots}
  & \gate[style={draw=none,fill=none,inner sep=1pt}]
      {\operatorname{Id}_{R_{n-1}}}
  & \gate[style={draw=none,fill=none,inner sep=1pt}]{R_{n-1}}
  &
  &
  & \setwiretype{c}\rstick{$\{H_0,H_1\}$}
\end{quantikz}%
}

\caption{Coherent adaptive ancilla-assisted tester. The unknown channel $\Phi$ is queried sequentially, with CPTP maps $\mathsf V_i$ acting between successive uses and $\mathsf V_n$ producing the final state on $\Hfin$.}
\label{fig:adaptive-tester}
\end{figure}
\paragraph{Randomized Channel Tester:} A randomized channel tester can be viewed as a probability distribution over deterministic testers. Here a deterministic tester means that all internal classical randomness has been fixed. i.e., its input preparation, adaptive
operations, and final POVM are fixed, although the final quantum measurement remains probabilistic.\footnote{For example, let $T_0$ be a tester with one fixed input preparation and final POVM, and let $T_1$ be another tester with a different fixed preparation and POVM. A randomized tester $\mathcal T$ may sample a seed $r\in\{0,1\}$ with $\Pr(r=0)=\Pr(r=1)=1/2$, and then run $T_r$. Thus the randomness of $\mathcal T$ is exactly the random choice between the two deterministic testers $T_0$ and $T_1$ i.e. once $r$ is fixed, the chosen tester is fixed, although its final quantum measurement remains probabilistic. }

Thus, if a randomized tester succeeds on every channel in the null and alternate hypotheses, one may average over its internal randomness and conclude that some deterministic tester in its support has the corresponding acceptance-probability gap against random channels drawn from the two hypotheses. The following lemma formalizes this reduction.

\begin{lemma}[{\cite[Lemma 2.6]{rosenthal2024quantum}}]
\label{lem:randomized-to-deterministic-tester}
Let $\mathcal T$ be a randomized channel tester, let $A$ be a set of channels such that
$\mathcal T(\mathcal A)$ accepts with probability at least $p$ for all $\mathcal A\in A$,
and let $B$ be a set of channels such that $\mathcal T(\mathcal B)$ accepts with probability
at most $q$ for all $\mathcal B\in B$, where the probabilities are over both the
choice of $\mathcal T$ and the randomness of the output measurement. Let $\boldsymbol{\mathcal A}$
and $\boldsymbol{\mathcal B}$ be random channels with support in $A$ and $B$,
respectively. Then there exists a deterministic channel tester $T$ in the support of
the randomized tester such that
\[
    \Pr \bigl(T(\boldsymbol{\mathcal A})\text{ accepts}\bigr)
    -
    \Pr \bigl(T(\boldsymbol{\mathcal B})\text{ accepts}\bigr)
    \ge p-q,
\]
where the probability is over both the choice of
$\boldsymbol{\mathcal A}$ and $\boldsymbol{\mathcal B}$ and the randomness of the
output measurement.
\end{lemma}

\section{Non-adaptive Coherent Lower Bound}
\label{sec:NACLB}

\begin{theorem}[Non-Adaptive Lower bound]
\label{thm:max-non-adapt}
For $\dout\ge2$ and $0<\varepsilon\le1$, every ancilla-assisted, coherent, non-adaptive algorithm (from Def.~\ref{def:nonadaptive-tester}) requires $\max\{\Omega(\din^{1.5}\dout/\varepsilon^2), \Omega(\din^{2}/\varepsilon^2)\}$ queries to solve Problem~\ref{prob:depolarizing-channel-testing}.
\end{theorem}

The following sections develop the technical ingredients needed for the argument, culminating in the proof of $\Omega(\din^{1.5}\dout/\varepsilon^2)$ in Section~\ref{proof:thm_non-adapt} and the second lower bound $\Omega(\din^{2}/\varepsilon^2)$ in Section~\ref{proof:thm_non-adapt2}.

We first prove the lower bound for even $\dout$. The odd-dimensional proof is given in Section~\ref{app:odd-analysis}.

\subsection{Hard family and validity checks}\label{sec:hard_non-adapt}

\paragraph{Hard channel family:}

We define the random channel
\begin{align}\label{eq:hardchannel}
    \bm\M(X)
    &=\tr(\bm P X)\bm\rho+\tr((I_A-\bm P)X)\tauout \\
    &=\N(X)+\frac{\varepsilon}{\sqrt{\dout}}\,\tr(\bm P X)\bm\Delta,
    \label{eqn:hard-case-non-adapt-v1}
\end{align}
which is constructed in the following manner:

For even $\dout$, define the traceless Hermitian operator\footnote{For odd $\dout$ we defer the traceless Hermitian operator to section~\ref{app:odd-analysis}.}
\begin{equation}\label{eq:Delta0}
    \Delta_0
    :=
    \frac{1}{\sqrt{\dout}}
    \sum_{j=1}^{\dout/2}
    \left(
        |2j-1\rangle\langle 2j-1|
        -
        |2j\rangle\langle 2j|
    \right),
\end{equation}
where $\{|k\rangle\}_{k=1}^{d_{\text{out}}}$ forms an orthonormal basis for $\mathbb{C}^{d_{\text{out}}}$.
 The construction proceeds by sampling $\bm v = \bm W\ket{0}$ with $ \bm W\sim\operatorname{Haar}(U(\din))$ and $
    \bm U\sim\operatorname{Haar}(U(\dout))$ independently
and defining
\[
    \bm P:=|\bm v\rangle\langle\bm v|,
    \quad
    \bm\Delta:=\bm U\Delta_0\bm U^\dagger,
    \quad
    \bm\rho:=\tauout+\frac{\varepsilon}{\sqrt{\dout}}\bm\Delta.
\]
Here $\Delta_0$ (and by unitary invariance, $\bm \Delta$) satisfies the following properties:
\[
    \tr(\Delta_0)=0,\quad
    \|\Delta_0\|_2=1,\quad
    \|\Delta_0\|_\infty=\frac{1}{\sqrt{\dout}},\quad
    \|\Delta_0\|_1=\sqrt{\dout}.
\]

\paragraph{CPTP check:}

Let $J_{\bm\M}=\bm P^{\mathsf T}\otimes\bm\rho+(I_A-\bm P)^{\mathsf T}\otimes\tauout$ be the (unnormalized) Choi matrix of $\bm\M$. Since the eigenvalues of $\bm\Delta$ are $\pm1/\sqrt{\dout},$ the eigenvalues of $\bm\rho$ are $(1\pm\varepsilon)/\dout$. Therefore, for $\varepsilon\ge0$,
\[
    J_{\bm\M}\succeq0
    \quad\Longleftrightarrow\quad
    \bm\rho\succeq0
    \quad\Longleftrightarrow\quad
    (1-\varepsilon)/\dout\ge0
    \quad\Longleftrightarrow\quad
    0\le\varepsilon\le1.
\]
Therefore, every realization of the hard family, i.e., every $\M_{v,U}$ with $v$ and $U$ fixed, is CP for $0<\varepsilon\le1$. Moreover, $\tr(\bm\Delta)=\tr(\Delta_0)=0$, so $\tr(\bm\rho)=1$, and hence $\tr_B J_{\bm\M}=\bm P^{\mathsf T}+(I_A-\bm P)^{\mathsf T}=I_A$. Thus every realization of $\bm\M$ is TP as well.
\paragraph{Diamond-distance check:}

Since
$(\bm\M-\N)(X)={\varepsilon}/{\sqrt{\dout}}\,\tr(\bm P X)\bm\Delta,$ evaluating the channel difference on the state $\bm P$ yields 
   $ \|\bm\M-\N\|_\diam
    \ge
    \|(\bm\M-\N)(\bm P)\|_1
    =
    \frac{\varepsilon}{\sqrt{\dout}}\|\bm\Delta\|_1
    =
    \varepsilon.$

\paragraph{Connection to mixedness testing in \cite{o2025instance}:} The output hardness of our hard family is closely related to the
\textit{quantum Paninski construction} used by~\cite{o2025instance}
for mixedness testing. Writing
$
    \Sigma=\operatorname{diag}(+1,-1,\ldots,+1,-1),$ and representing our traceless Hermitian as $
    \Delta_0=\frac{\Sigma}{\sqrt{\dout}},
$
the output state corresponding to the projector $\bm P$, can be written as
\[
    \bm\rho
    =
    \frac{I}{\dout}
    +
    \frac{\varepsilon}{\dout}
    \bm U\Sigma\bm U^\dagger,
\]
thereby recovering the Haar-random perturbation in their paper. 

We now apply Lemma~\ref{lem:randomized-to-deterministic-tester} to convert the tester's success guarantee into a constant separation between the averaged pre-measurement states under the null and alternate hypotheses.

\subsection{Reduction to an averaged output trace distance}
\label{sec:non-adapt_testing}
Let $A:=\{\N\}$ be the null set, and let $B$ be the set of channels in the alternate hypothesis. Define the random channels $\boldsymbol{\mathcal A}:=\N$ deterministically and $\boldsymbol{\mathcal B}:=\bm\M_{\bm v,\bm U}, $ where $(\bm v,\bm U)$ is sampled according to the hard distribution above.

For every fixed realization $v,U$, we have proved that $\|\M_{v,U}-\N\|_\diam\ge\varepsilon,$ and $\M_{v,U}\in B$. Hence $\boldsymbol{\mathcal B}$ is supported entirely on the alternative hypothesis.

Now suppose that a possibly randomized $n$-query non-adaptive tester $\mathcal T$ satisfies the required \textit{completeness} and \textit{soundness} guarantees, that is, it accepts $\N$ with probability at least $2/3$, and accepts every channel in $B$ with probability at most $1/3$. The probabilities here include both the internal randomness of $\mathcal T$ and the randomness of the final measurement.

Applying Lemma~\ref{lem:randomized-to-deterministic-tester} with $p=2/3$ and $q=1/3$, there exists a deterministic tester $T$ in the support of $\mathcal T$ such that
\begin{equation}
    \Pr\bigl(T(\boldsymbol{\mathcal A})\text{ accepts}\bigr)
    -
    \Pr\bigl(T(\boldsymbol{\mathcal B})\text{ accepts}\bigr)
    \ge
    \frac13.
    \label{eq:deterministic-average-gap}
\end{equation}
From this point onward, we fix this deterministic tester $T$. In
particular, there is no longer any randomness coming from the choice of the
tester.

Since $T$ is non-adaptive and deterministic, it is specified by an input
state
$\Psi\in\mathsf D(A_1\otimes\cdots\otimes A_n\otimes R)$ and a final binary POVM whose acceptance element satisfies
$
    0\preceq\mathsf P_{\acc}\preceq I.
$
Define pre-measurement states under  the null and alternative, and averaged alternative by
\[
    \sigma_0
    :=
    (\N^{\otimes n}\otimes\Id_R)(\Psi),
    \quad
    \sigma_{v,U}
    :=
    (\M_{v,U}^{\otimes n}\otimes\Id_R)(\Psi),
    \quad
    \bar\sigma
    :=
    \E_{\bm v,\bm U}
    [\bm\sigma_{\bm v,\bm U}],
\]
where
$    \bm\sigma_{\bm v,\bm U}:=
    (\bm\M_{\bm v,\bm U}^{\otimes n}\otimes\Id_R)(\Psi).
$ For each fixed $v,U$, Born rule gives
\[
    \Pr\bigl(T(\N)\text{ accepts}\bigr)
    =
    \tr(\mathsf P_{\acc}\sigma_0),
    \quad
    \Pr\bigl(T(\M_{v,U})\text{ accepts}\bigr)
    =
    \tr(\mathsf P_{\acc}\sigma_{v,U}).
\]

Since $\boldsymbol{\mathcal A}=\N$ deterministically, we have 
$
\Pr\bigl(T(\boldsymbol{\mathcal A})\text{ accepts}\bigr)=\tr(\mathsf P_{\acc}\sigma_0).
$
For the random alternative, the channel realization is first drawn according
to $(\bm v,\bm U)$, and the final measurement is then performed. Hence, by the law of total probability,
\begin{align}
    \Pr\bigl(T(\boldsymbol{\mathcal B})\text{ accepts}\bigr)
    &=
    \E_{\bm v,\bm U}\left[
        \Pr\bigl(
            T(\boldsymbol{\mathcal B})\text{ accepts}
            \mid \bm v,\bm U
        \bigr)
    \right]
    =
    \E_{\bm v,\bm U}\left[
        \tr\left(
            \mathsf P_{\acc}\bm\sigma_{\bm v,\bm U}
        \right)
    \right].
\end{align}
Substituting these expressions into Eqn.~\eqref{eq:deterministic-average-gap} gives
\begin{align}
    \frac13
    &\le
    \Pr\bigl(T(\boldsymbol{\mathcal A})\text{ accepts}\bigr)
    -
    \Pr\bigl(T(\boldsymbol{\mathcal B})\text{ accepts}\bigr)
    \nonumber\\
    &=
    \tr(\mathsf P_{\acc}\sigma_0)
    -
    \E_{\bm v,\bm U}
    \tr\left(
        \mathsf P_{\acc}\bm\sigma_{\bm v,\bm U}
    \right)
    =
    \tr\left[
        \mathsf P_{\acc}
        \left(
            \sigma_0-
            \E_{\bm v,\bm U}[\bm\sigma_{\bm v,\bm U}]
        \right)
    \right]
    =
    \tr\left[
        \mathsf P_{\acc}(\sigma_0-\bar\sigma)
    \right].
    \label{eq:yao-gap}
\end{align}

Finally, $\sigma_0$ and $\bar\sigma$ are both density operators, so $\sigma_0-\bar\sigma$ is Hermitian and traceless. By the variational characterization of trace distance,
$
    \tr\left[
        \mathsf P_{\acc}(\sigma_0-\bar\sigma)
    \right]
    \le
    \frac12\|\sigma_0-\bar\sigma\|_1.
$

Combining this with Eqn.~\eqref{eq:yao-gap} yields
\begin{equation}
    \frac13
    \le
    \frac12\|\bar\sigma-\sigma_0\|_1.
    \label{eq:trace-need}
\end{equation}

Thus the deterministic tester $T$ obtained from Lemma~\ref{lem:randomized-to-deterministic-tester} satisfies Eqn.~\eqref{eq:trace-need}. It remains to prove an upper bound on this trace distance, in terms of $n$, $\varepsilon$, $\din$, and $\dout$, that holds for every deterministic $n$-query non-adaptive tester, and hence for
$T$ in particular.

We next decompose the tester's output according to a hit-miss record induced by the hidden input direction $\bm v$.

\subsection{Hit and miss record}
\label{sec:hit,miss}

Recall that
\[
    \bm\M(X)
    =
    \tr(\bm P X)\bm\rho
    +
    \tr((I_A-\bm P)X)\tauout,
    \quad
    \bm P=|\bm v\rangle\langle\bm v|.
\]
Therefore, each query undergoes the projective measurement
$\{\bm P,I_A-\bm P\}$ before the input register is discarded. The outcomes
$s\in\{0,1\}$ are labeled as a hit ($s=1$), associated with $\bm P$ and
output $\bm\rho$, or a miss ($s=0$), associated with $I_A-\bm P$ and
output $\tauout$. Define
\[
    \bm P_1:=\bm P,
    \quad
    \bm P_0:=I_A-\bm P.
\]

For $s=(s_1,\ldots,s_n)\in\{0,1\}^n$, define
\begin{equation}\label{eq:Ps}
    \bm P_s
    :=
    \bigotimes_{i=1}^n \bm P_{s_i},
    \quad
    k(s):=\sum_{i=1}^n s_i,
\end{equation}
where $s_i=1$ denotes a hit, $s_i=0$ denotes a miss, and $k(s)$ is the total
number of hits. These labels are introduced only for proof
bookkeeping and are not revealed to the tester. Define
\begin{equation}\label{eq:ps-def}
    \bm p(s)
    :=
    \tr \left[
        (\bm P_s\otimes I_R)
        \Psi
    \right].
\end{equation}
Equivalently, since $\bm P_s$ is a projector,
$
    \bm p(s)
    =
    \tr \left[
        (\bm P_s\otimes I_R)
        \Psi
        (\bm P_s\otimes I_R)
    \right].
$
For $\bm p(s)>0$, define the normalized conditional ancilla state
\begin{equation}\label{eq:omega-sv}
    \bm\omega_R^s
    :=
    \frac{1}{\bm p(s)}
    \tr_{A_1\cdots A_n}
    \left[
        (\bm P_s\otimes I_R)
        \Psi
        (\bm P_s\otimes I_R)
    \right]
\end{equation}
where
\[
    \bm\omega_R^s\succeq0,
    \quad
    \tr\bm\omega_R^s=1,
    \quad
    \sum_s \bm p(s)=1,
\]
and
\begin{equation}\label{eq:omega-average}
    \sum_s \bm p(s)\bm\omega_R^s
    =
    \Psi_R,
    \quad
    \Psi_R:=\tr_{A_1\cdots A_n}\Psi.
\end{equation}

If $\bm p(s)=0$, choose $\bm\omega_R^s$ arbitrarily. No independence
among the bits $s_1,\ldots,s_n$ is assumed.
\medskip

\paragraph{Pre-measurement states:} Recall that the state
$
    \bm\sigma_{\bm v,\bm U}
    :=
    (\bm\M^{\otimes n}\otimes\Id_R)(\Psi)
$
is the pre-measurement state on
$B_1\otimes\cdots\otimes B_n\otimes R$. For a fixed hit-miss sequence $s$, the
$k(s)$ hit outputs $\bm\rho$ may occupy arbitrary positions among the
$n$ output registers. We therefore introduce a permutation unitary
$\Pi_s$ on $B_1\otimes\cdots\otimes B_n$ that reorders these tensor
factors so that the hit positions appear first, followed by the miss
positions: 
\begin{equation}\label{eq:hard-output-decomp}
    \bm\sigma_{\bm v,\bm U}
    =
    \sum_s
    \bm p(s)\,
    \Pi_s^\dagger
    \left[
        \bm\rho^{\otimes k(s)}
        \otimes
        \tauout^{\otimes(n-k(s))}
    \right]
    \Pi_s
    \otimes
    \bm\omega_R^s.
\end{equation}

Since $\bm p(s)$ and $\bm\omega_R^s$ depend only on $\bm v$, conditioning on $\bm v$, and averaging over the independent Haar unitary $\bm U$ yields 
\begin{equation}\label{eq:sigmav-decomp}
    \bar{\bm\sigma}
    :=
    \E_{\bm U} \left[
        \bm\sigma_{\bm v,\bm U}
        \,\middle|\,\bm v
    \right]=
    \sum_s
    \bm p(s)\,
    \Pi_s^\dagger
    \left[
        \bar\rho_{k(s)}
        \otimes
        \tauout^{\otimes(n-k(s))}
    \right]
    \Pi_s
    \otimes
    \bm\omega_R^s.
\end{equation}
Where $\bar\rho_k := \E_{\bm U}[\bm\rho^{\otimes k}], \text{ for }k:=k(s)\geq 0$.
Under the null hypothesis $H_0$ (see Eqn.~\eqref{Eqn:hypothesis}), the pre-measurement state is
$
    \sigma_0
    =
    \tauout^{\otimes n}\otimes\Psi_R.
$
Substituting Eqn.~\eqref{eq:omega-average}, we obtain \[ \sigma_0
    =
    \sum_s
    \bm p(s)\,
    \tauout^{\otimes n}
    \otimes
    \bm\omega_R^s = \sum_s
    \bm p(s)\,
    \Pi_s^\dagger
    \left[
        \tauout^{\otimes k(s)}
        \otimes
        \tauout^{\otimes(n-k(s))}
    \right]
    \otimes
    \bm\omega_R^s\]  where the last equality follows since $\tauout^{\otimes n}$ is invariant under permutations. 

Therefore taking the difference of the pre-measurement states we get
\begin{equation}\label{eq:sigmav-minus-sigma0}
    \bar{\bm\sigma}-\sigma_0
    =
    \sum_s
    \bm p(s)\,
    \Pi_s^\dagger
    \left[
        \left(
            \bar\rho_{k(s)}
            -
            \tauout^{\otimes k(s)}
        \right)
        \otimes
        \tauout^{\otimes(n-k(s))}
    \right]
    \Pi_s
    \otimes
    \bm\omega_R^s.
\end{equation}

With $\bar\sigma - \sigma_0 = \E_{\bm v}\left[\bar{\bm\sigma} - \sigma_0\right],$ we now split the contributions of sub-normalized states by the number of hits.

\paragraph{Splitting into two sums:} Fix $m\ge1$. Partitioning the sum into $k(s) < m$ and $k(s) \ge m$ 
\begin{align}
    \bar\sigma-\sigma_0
    &=
    \underbrace{
    \E_{\bm v}
    \sum_{k(s)<m}
    \bm p(s)
    \Bigg(
        \Pi_s^\dagger
        \left[
            \bar\rho_{k(s)}
            \otimes
            \tauout^{\otimes(n-k(s))}
        \right]
        \Pi_s
        -
        \tauout^{\otimes n}
    \Bigg)
    \otimes
    \bm\omega_R^s
    }_{:=\,X_m}
    \nonumber\\
    &\quad+
    \underbrace{
    \E_{\bm v}
    \sum_{k(s)\ge m}
    \bm p(s)
    \Bigg(
        \Pi_s^\dagger
        \left[
            \bar\rho_{k(s)}
            \otimes
            \tauout^{\otimes(n-k(s))}
        \right]
        \Pi_s
        -
        \tauout^{\otimes n}
    \Bigg)
    \otimes
    \bm\omega_R^s
    }_{:=\,Y_m}.
\end{align}
Taking the trace norm and applying the triangle inequality on the second summand gives
\begin{align}
    \|\bar\sigma-\sigma_0\|_1
    &\le
    \|X_m\|_1
    +
    \E_{\bm v}
    \sum_{k(s)\ge m}
    \bm p(s)
    \left\|
        \Bigg(
            \Pi_s^\dagger
            \left[
                \bar\rho_{k(s)}
                \otimes
                \tauout^{\otimes(n-k(s))}
            \right]
            \Pi_s
            -
            \tauout^{\otimes n}
        \Bigg)
        \otimes
        \bm\omega_R^s
    \right\|_1
    \nonumber\\
    &=
    \|X_m\|_1
    +
    \E_{\bm v}
    \sum_{k(s)\ge m}
    \bm p(s)
    \left\|
        \Pi_s^\dagger
        \left[
            \bar\rho_{k(s)}
            \otimes
            \tauout^{\otimes(n-k(s))}
        \right]
        \Pi_s
        -
        \tauout^{\otimes n}
    \right\|_1,
    \label{eqn:trace_split}
\end{align}
where we have used $\|\bm\omega_R^s\|_1=\tr(\bm\omega_R^s)=1$, since $\bm\omega_R^s$ is a normalized state. For $k=k(s)$, permutation invariance of the trace norm gives
\begin{align}
\left\|
    \Pi_s^\dagger
    \left[
        \bar\rho_k
        \otimes
        \tauout^{\otimes(n-k)}
    \right]
    \Pi_s
    -
    \tauout^{\otimes n}
\right\|_1
=
\left\|
    \left(
        \bar\rho_k-\tauout^{\otimes k}
    \right)
    \otimes
    \tauout^{\otimes(n-k)}
\right\|_1
=
\left\|
    \bar\rho_k-\tauout^{\otimes k}
\right\|_1,
\end{align}
Define\footnote{We could alternatively combine the trivial bound
$\delta_k\le1$ with Lemma~\ref{lem:output-mixing}, which gives
$
    \delta_k
    \le
    \min\left\{
        1,\,
        \sqrt K\,\frac{k\varepsilon^2}{\dout}
    \right\}.
$
Therefore the bound on the high-hit contribution becomes
\[
    \E_{\bm v}\sum_{k(s)\ge m}p_{\bm v}(s)\delta_{k(s)}
    \le
    \min\left\{
        \frac1m,\,
        \sqrt K\,\frac{\varepsilon^2}{\dout}
    \right\}
    \E_{\bm v}\sum_{k(s)\ge m}p_{\bm v}(s)k(s)\\
    \le
    \frac{n}{\din}
    \min\left\{
        \frac1m,\,
        \sqrt K\,\frac{\varepsilon^2}{\dout}
    \right\}.
\]
Later in the proof we will go on to show that the low-hit term 
$\|X_m\|_1\lesssim{m\varepsilon^2}/{\sqrt{\din}\,\dout}.$
For the choice $m=\sqrt{\din/(108K)}\,\dout/\varepsilon^2$, the minimum between the two high-hit bounds is $\sqrt K\,\varepsilon^2/\dout$ for $2\le\din<108$, and $1/m$ for $\din\ge108$. In the former regime,
$n \ge {\din\dout}/{\varepsilon^2}\ge{\din^{3/2}\dout}/{\sqrt{108}~\varepsilon^2},
$
so both regimes yield the claimed scaling up to a universal constant.}
\begin{align}
    \delta_k
    :=
    \frac12
    \left\|
        \bar\rho_k-\tauout^{\otimes k}
    \right\|_1 \le 1.
\end{align}

Therefore Eqn.~\eqref{eqn:trace_split} can now be expressed as 
\begin{align}
    \frac12\|\bar\sigma-\sigma_0\|_1
    &\le
    \frac12\|X_m\|_1
    +
    \frac12 \cdot 2 \cdot \E_{\bm v}
    \sum_{k(s)\ge m}
    \bm p(s)\delta_{k(s)}
   \nonumber\\
    &\le
    \frac12\|X_m\|_1
    +
    \E_{\bm v}
    \sum_{k(s)\ge m}
    \bm p(s)
    \nonumber\\
    &\le
    \frac12 \|X_m\|_1
    +
    \frac{1}{m}
    \E_{\bm v}
    \sum_s
    \bm p(s)k(s) \nonumber\\
    &= \frac12\|X_m\|_1 +
    \frac{1}{m}\sum_{i=1}^n
    \E_{\bm v}
    \tr\left[
        (|\bm v\rangle\langle\bm v|_{A_i}\otimes I)
        \Psi
    \right]\
    \nonumber
    \\&= \frac12\|X_m\|_1 +
    \frac{1}{m}\sum_{i=1}^n
    \tr\left[
        \left(
            \frac{I_{A_i}}{\din}\otimes I
        \right)
        \Psi
    \right]
    \nonumber
    \\&= 
    \frac12 \|X_m\|_1
    +
    \frac{n}{m\din}.
    \label{eq:high-hit-bound-coherent}
\end{align}

\begin{restatable}[Low Hit Contribution Bound]{lemma}{LowHit}
\label{lem:low-hit-bound}
Let $\mu_j$ be the $j$-th output-overlap moment,
defined by
\begin{equation}
    \mu_j :=
    \E_{\bm U,\bm U'}
    \left[
        \dout
        \tr\left(
            (\rho_{\bm U}-\tau_{\mathrm{out}})
            (\rho_{\bm U'}-\tau_{\mathrm{out}})
        \right)
    \right]^j,
\end{equation}
where $\bm U,\bm U'$ are independent unitaries sampled from the haar measure and
$\tau_{\mathrm{out}}=I/\dout$.
For each $j\geq1$, let $c_j\geq0$ that satisfies
\begin{equation}
    \E_{\bm v}
    \sum_{k(s)<m}
    \left(\prod_{i\in S}s_i\right)
    \bm p(s)\bm\omega_R^s
    \preceq c_j\Psi_R,
    \quad \text{where}~~|S|=j,
\end{equation}
where $\bm\omega_R^s$ is the conditional reference state
on the ancilla $R$ and $\Psi_R$ its unconditional state. Then the low-hit contribution $X_m$ satisfies
\begin{equation}
    \|X_m\|_1^2
    \leq
    \E_{\bm v}
    \sum_{k(s)<m}
    \bm p(s)
    \sum_{j=1}^{k(s)}
    \binom{k(s)}{j}\mu_jc_j.
    \label{eqn:Xm-before-h}
\end{equation}
\end{restatable}
\begin{proof}
    Proof in Appendix~\ref{app:Low-Hit-Contri}.
\end{proof}

Thus, the low-hit term is controlled by the scalar sum
$
    \sum_{j=1}^{k(s)}
    \binom{k(s)}{j}\mu_jc_j,
$
which depends on a branch $s$ only through its number of hits $k(s)$. Since every low-hit branch satisfies $k(s)<m$, it therefore suffices to bound this quantity uniformly for $k\le m-1$. Accordingly, for $k\ge0$, define
\[
    h(k)
    :=
    \sum_{j=1}^k
    \binom{k}{j}\mu_jc_j,
    \quad
    h(0):=0.
\]

\begin{restatable}[Bounding the low-hit series]{lemma}{HkBound}
\label{lem:hk-bound}
Let
\[
    h(k):=\sum_{j=1}^{k}\binom{k}{j}\mu_jc_j,
    \qquad~~~
    \lambda_k
    :=
    \frac{Kk^2\varepsilon^4}{2\dout^2},
    \quad K=96.
\]
For $k\ge 2$ and $\din\ge 2$, we have
\begin{equation}
\label{Eqn:h(k)cases}
    h(k)
    <
    \begin{cases}
        \displaystyle
        \frac{2\lambda_k e^{\lambda_k}}
             {\din(\din+1)},
        & \din\in\{2,3\},\\[8pt]
        \displaystyle
        \frac{2\lambda_k(\din-1)}
             {(\din-2-\lambda_k)^3},
        & \din\ge4,
    \end{cases}
\end{equation}
where the second bound holds provided
$\lambda_k<\din-2$.
\end{restatable}
\begin{proof}
    Proof in Appendix~\ref{app:h(k)_bound}
\end{proof}

Since $\mu_jc_j\ge0$, $h(k)$ is nondecreasing in $k$, and hence
\begin{align}
    \|X_m\|_1^2
    \le
    \E_{\bm v}
    \sum_{k(s)<m}\bm p(s)h(k(s))
    \le
    h(m-1).
    \label{eq:Xm-h-bound}
\end{align}
Substituting $k=m-1$ in Eqn.~\eqref{Eqn:h(k)cases} and using Eqn.~\eqref{eq:Xm-h-bound} gives
\begin{align}
    \|X_m\|_1^2
    <
    \begin{cases}
        \displaystyle
        \frac{
            2\lambda_{m-1}e^{\lambda_{m-1}}
        }{
            \din(\din+1)
        },
        & \din\in\{2,3\},\\
        \displaystyle
        \frac{
            2\lambda_{m-1}(\din-1)
        }{
            (\din-2-\lambda_{m-1})^3
        },
        & \din\ge4,
    \end{cases}
    \label{eq:Xm-explicit-bound}
\end{align}
where the second bound holds whenever
$
    \lambda_{m-1}<\din-2
$.

We choose $m$ such that
\begin{equation}
    \lambda_{m-1}
    =
    \frac{K(m-1)^2\varepsilon^4}{2\dout^2}
    \le
    \frac{\din}{216}.
    \label{eq:lambda-m-condition}
\end{equation}
For $\din\in\{2,3\}$, since
$
    \lambda_{m-1}\le\din/216\le1/72,
$
Eqn.~\eqref{eq:Xm-explicit-bound} gives
\[
    \|X_m\|_1^2
    <
    \frac{
        2\lambda_{m-1}e^{\lambda_{m-1}}
    }{
        \din(\din+1)
    }
    \le
    \frac{
        e^{1/72}
    }{
        108(\din+1)
    }
    <
    \frac19.
\]
For $\din\ge4$, since
$
    \lambda_{m-1}\le\din/216\le(\din-2)/2,
$
Eqn.~\eqref{eq:Xm-explicit-bound} gives
\begin{align}
    \|X_m\|_1^2
    &<
    \frac{2\lambda_{m-1}(\din-1)}
         {(\din-2-\lambda_{m-1})^3}
    \le
    \frac{16\lambda_{m-1}(\din-1)}
         {(\din-2)^3}
    \nonumber\\
    &\le
    \frac{2\din(\din-1)}
         {27(\din-2)^3}
    \le
    \frac19.
\end{align}
Thus Eqn.~\eqref{eq:lambda-m-condition} guarantees
$
    \|X_m\|_1<1/3
$
for every $\din\ge2$.

\subsection[Proof of the first bound]{Proof of the $\Omega(\din^{1.5} \dout/\varepsilon^2)$ bound }

\label{proof:thm_non-adapt}
\begin{theorem}[Non-Adaptive Coherent Lower Bound]
\label{thm:main_non-adapt}
For $\dout\ge2$ and $0<\varepsilon\le1$, every ancilla-assisted, coherent, non-adaptive algorithm (from Def.~\ref{def:nonadaptive-tester}) requires $\Omega(\din^{1.5} \dout/\varepsilon^2)$ queries to solve Problem~\ref{prob:depolarizing-channel-testing}.
\end{theorem}
\begin{proof}
Choose $m-1
    :=
    \left\lfloor
        {\dout\sqrt{\din}}/
        {\sqrt{108K}\,\varepsilon^2}
    \right\rfloor,
    \text{ where } K=96.$
Then
\begin{align}
    \lambda_{m-1}
    &=
    \frac{K(m-1)^2\varepsilon^4}{2\dout^2}
    \le
    \frac{\din}{216},
    \quad
    \|X_m\|_1^2
    \le
    h(m-1)
    <
    \frac19 \implies
    \|X_m\|_1<\frac13.
    \label{eq:Xm-final}
\end{align}
Combining Eqs.~\eqref{eq:trace-need},
\eqref{eq:high-hit-bound-coherent}, and \eqref{eq:Xm-final} we get
\begin{align}
    \frac13
    &\le
    \frac12\|\bar\sigma-\sigma_0\|_1
    <
    \frac16+\frac{n}{\din m},
\end{align}
Substituting the choice of $m$ above
proves the theorem for even $\dout$.  See Appendix~\ref{app:odd-analysis} for odd $\dout$.
\end{proof}

\subsection[Proof of the second bound]{Proof of the $\Omega(\din^{2}/\varepsilon^2)$ bound}
\label{proof:thm_non-adapt2}

We now prove the lower bound $\Omega(\din^2/\varepsilon^2)$ in Theorem~\ref{thm:main_non-adapt}. For this we choose the worst-case construction in Eqn.~\eqref{eqn:hard-case-non-adapt-v1} and fix the output unitary $U=I$ thereby retaining only the Haar randomness in the input direction $v$. Physically, this leaves the input direction as the sole source of uncertainty where every hit produces a known output perturbation $\Delta$. Consequently, the low-hit contribution admits a geometric-series bound,
\[
\|X_m\|_1^2
\leq
\sum_{j=1}^{m-1}
\left(\frac{m\varepsilon^2}{\din}\right)^j.
\]
This is in contrast with the first lower bound, where $U$ was Haar-random and the distinguishing information was instead carried by correlations between successive hits, requiring higher-order Haar-moment bounds. These correlations then introduced the dependence on $\dout$ that the crude geometric-series estimate would lose in $U=I$ case. The second lower bound for the $U=I$ case is stated below.

\begin{restatable}[Independent $\din^2/\varepsilon^2$ lower bound]{theorem}{dinbound}
    \label{thm:din2-lower-bound}
Let $\din,\dout\ge2$ and $0<\varepsilon\le1$. Every
ancilla-assisted, coherent, non-adaptive algorithm from
Def.~\ref{def:nonadaptive-tester} requires
$\Omega(\din^2/\varepsilon^2)$ queries to solve
Problem~\ref{prob:depolarizing-channel-testing}.
\end{restatable}

\begin{proof}
    Proof in Appendix~\ref{app:din2bound}
\end{proof}

The two constructions reveal complementary sources of hardness, although the distinction between their bounds is partly methodological. The fixed-output construction admits a sharper analysis of input uncertainty, whereas the Haar-random output construction exploits cancellations that introduce an additional dependence on $\dout$. Together, they yield
\[
    n=\Omega\left(
    \frac{\max\{\din^{3/2}\dout,\din^2\}}
    {\varepsilon^2}
    \right).
\]

A crossover happens around $\dout\sim \sqrt{\din}$. Fixing the output perturbation isolates the uncertainty in the input direction and yields the stronger bound when \(\dout\leq\sqrt{\din}\). Randomizing the output introduces an additional uncertainty whose effect becomes more pronounced as the output dimension grows, yielding the stronger bound when \(\dout\geq\sqrt{\din}\). This crossover reflects the relative strengths of our two lower-bound arguments. Whether it also marks a genuine change in the difficulty of the underlying instances remains unclear at present. A matching upper bound algorithm will establish if such a crossover indeed exists.

\section{Adaptive Coherent Lower Bound}
\label{sec:ACLB}
\begin{theorem}[Adaptive Coherent Lower Bound]
\label{thm:main_adapt}
For $\dout\ge2$ and all $0<\varepsilon \le 1$, every ancilla-assisted, coherent, adaptive algorithm (from Def.~\ref{def:adaptive_tester}) requires $\Omega(\din \dout/\varepsilon^2)$ queries to solve Problem~\ref{prob:depolarizing-channel-testing}.
\end{theorem}

This proof follows the same construction and high-hit bound as the non-adaptive coherent bound in Section~\ref{sec:hard_non-adapt}. The low-hit contribution requires a separate analysis, however, because adaptive steps correlate hits and misses within a record, and the identity in Eqn.~\eqref{eqn:pairbound} in the proof of Lemma~\ref{lem:low-hit-bound} may no longer hold. Here, instead we construct a \emph{truncation} map in Lemma~\ref{lem:truncation} that upper bounds the low-hit contribution by an $m$-copy state-distinguishability problem. The following sections develop the technical ingredients needed for the argument, culminating in the proof of the main lower bound in Section~\ref{sec:lower}.

\subsection{Hard family}
\label{sec:hard_adapt}

We use the same hard family as in Section~\ref{sec:hard_non-adapt}, defined as
\[
    \bm\M_{\bm v,\bm U}(X)
    =
    \N(X)
    +
    \frac{\varepsilon}{\sqrt{\dout}}\,
    \tr\left(|\bm v\rangle\langle\bm v|X\right)\bm\Delta,
\]
where $\bm\Delta=\bm U\Delta_0\bm U^\dagger$,
$
    \bm v=\bm W\ket{0},
    \ \bm W\sim\operatorname{Haar}(U(\din))
$
and
$
    \bm U\sim\operatorname{Haar}(U(\dout))
$
are sampled independently. For fixed $v,U$, the corresponding realization
is $\M_{v,U}$ in Eqn~\eqref{construction}.

\subsection{Reduction to an averaged output trace distance}
\label{sec:adapt_testing}

The reduction to bounding a constant trace-distance separation is identical to the one we used in Section~\ref{sec:non-adapt_testing} for the non-adaptive lower bound. Under the null hypothesis (Defined in Problem~\ref{prob:depolarizing-channel-testing}, Eqn.~\eqref{Eqn:hypothesis}), let $\boldsymbol{\mathcal A}=\mathcal N$ deterministically, and under the alternate let $\boldsymbol{\mathcal B}=\bm\M_{\bm v,\bm U}$.

By Lemma~\ref{lem:randomized-to-deterministic-tester}, any possibly randomized coherent adaptive tester succeeding with probability at least $2/3$ yields a deterministic adaptive tester (from Def.~\ref{def:adaptive_tester}) whose average acceptance-probability gap is at least $1/3$. Fix such a deterministic adaptive tester. The pre-measurement states under the null and the alternate hypothesis are respectively
\begin{align}
    \sigma_0
    &:=
    \mathsf V_n
    \circ(\N\otimes\Id_{R_{n-1}})
    \circ\mathsf V_{n-1}
    \circ\cdots\circ
    \mathsf V_1
    \circ(\N\otimes\Id_{R_0})(\Psi),\nonumber\\
    \bm\sigma_{\bm v,\bm U}
    &:=
    \mathsf V_n
    \circ(\bm\M_{\bm v,\bm U}\otimes\Id_{R_{n-1}})
    \circ\mathsf V_{n-1}
    \circ\cdots\circ
    \mathsf V_1
    \circ(\bm\M_{\bm v,\bm U}\otimes\Id_{R_0})(\Psi),
    \label{eqn-pre-meas-states-adapt}
\end{align}
Let
\[
    \bar\sigma
    :=
    \E_{\bm v,\bm U}
    [\bm\sigma_{\bm v,\bm U}]
\]
denote the averaged pre-measurement state under the alternate hypothesis. Using Lemma~\ref{lem:randomized-to-deterministic-tester}, the Born rule and the variational characterization of trace distance we get
\begin{equation}
    \frac13
    \le
    \frac12\|\bar\sigma-\sigma_0\|_1.
    \label{eq:adapt-trace-need}
\end{equation}
Thus it remains to upper bound the RHS of Eqn.~\eqref{eq:adapt-trace-need} for an arbitrary deterministic coherent adaptive tester.
\subsection{Splitting the trace distance into low-hit and high-hit events}
\label{sec:split}

Recall that, for fixed $v,U$, the hard channel has the measure-and-prepare form \[\mathcal M_{v,U}(X)=\langle v|X|v\rangle\,\rho_U+\tr[(I_A-|v\rangle\langle v|)X]\tauD,\] where $\tauD:=I_{\dout}/\dout$ and $\rho_U:=\tauD +{\varepsilon}/{\sqrt{\dout}}U\Delta_0U^\dagger$. Thus, each channel use can be naturally decomposed into the two possible outcomes $\{|v\rangle\langle v|,\,I_A-|v\rangle\langle v|\}$, which correspond to whether the output state is $\rho_U$ or $\tauD$, respectively. We call the $|v\rangle\langle v|$ outcome a \emph{hit}. We then split the pre-measurement states in Eqn.~\eqref{eqn-pre-meas-states-adapt} according to whether the outcome record contains at most $m$ hits or more than $m$ hits.

\paragraph{Outcome-record expansion:}

Suppose the state entering query $i$ is a positive operator
$X$ on $A_i\otimes R_{i-1}$. For the fixed unit vector
$v\in\mathbb C^{\din}$, let
$s_i\in\{0,1\}$ denote the hit-miss outcome, with
$s_i=1$ for a hit and $s_i=0$ for a miss. Define the corresponding residual memory states on $R_{i-1}$ by
\begin{align}
    \mathsf{Res}_i^{\,1,v}[X]
    &:=
    (\langle v|\otimes I_{R_{i-1}})
    X
    (|v\rangle\otimes I_{R_{i-1}}),
    \label{eq:rstate-hit}\\
    \mathsf{Res}_i^{\,0,v}[X]
    &:=
    \tr_{A_i}\left[
        \bigl((I_{A_i}-|v\rangle\langle v|)\otimes I_{R_{i-1}}\bigr)X
    \right]
    \label{eq:rstate-miss}
\end{align}
respectively. Then
$
    \mathsf{Res}_i^{\,1,v}[X],
    \mathsf{Res}_i^{\,0,v}[X]\succeq0
$,
and $\mathsf{Res}_i^{\,1,v}[X]
    +
    \mathsf{Res}_i^{\,0,v}[X]
    =
    \tr_{A_i}X.$ If $X$ is normalized, their traces are the probabilities of a hit and a
miss, respectively.

Using these memory states, one use of $\M_{v,U}$ and $\N$ can be written
as
\begin{align}
    (\M_{v,U}\otimes\Id_{R_{i-1}})(X)
    &=
    \rho_U\otimes\mathsf{Res}_i^{\,1,v}[X]
    +
    \tauD\otimes\mathsf{Res}_i^{\,0,v}[X],
    \label{eq:channels-alt}\\
    (\N\otimes\Id_{R_{i-1}})(X)
    &=
    \tauD\otimes
    \left(
        \mathsf{Res}_i^{\,1,v}[X]
        +
        \mathsf{Res}_i^{\,0,v}[X]
    \right).
    \label{eq:channels-null}
\end{align}
The complete adaptive experiment can therefore be expanded over hit-miss
records $s_{1:n}=(s_1,\ldots,s_n)\in\{0,1\}^n$. Let
$\psi_i^{v,U}(s_{1:i})$ and $\psi_i^{v,0}(s_{1:i})$ denote the
corresponding sub-normalized tester states after the first $i$ uses under
$\M_{v,U}$ and $\N$, respectively, where
$
    \psi_0^{v,U}(\varnothing)
    =
    \psi_0^{v,0}(\varnothing)
    =
    \Psi,
$
and $\varnothing$ denotes the empty outcome record before any channel use. Define recursively
\begin{align}
    \psi_i^{v,U}(s_{1:i})
    &:=
    \begin{cases}
    \mathsf V_i\left(
        \rho_U\otimes
        \mathsf{Res}_i^{\,1,v}
        [\psi_{i-1}^{v,U}(s_{1:i-1})]
    \right),
    & s_i=1,\\[4pt]
    \mathsf V_i\left(
        \tauD\otimes
        \mathsf{Res}_i^{\,0,v}
        [\psi_{i-1}^{v,U}(s_{1:i-1})]
    \right),
    & s_i=0,
    \end{cases}
    \label{eq:branch-rec-alt}\\
    \psi_i^{v,0}(s_{1:i})
    &:=
    \mathsf V_i\left(
        \tauD\otimes
        \mathsf{Res}_i^{\,s_i,v}
        [\psi_{i-1}^{v,0}(s_{1:i-1})]
    \right).
    \label{eq:branch-rec-null}
\end{align}
Summing over all hit-miss records recovers the pre-measurement states:
\begin{equation}
    \sigma_{v,U}
    =
    \sum_{s_{1:n}\in\{0,1\}^n}
    \psi_n^{v,U}(s_{1:n}),
    \quad
    \sigma_0
    =
    \sum_{s_{1:n}\in\{0,1\}^n}
    \psi_n^{v,0}(s_{1:n}).
    \label{eq:branch-sum}
\end{equation}

\paragraph{Low- and High-Hit Contributions:}
By Eqns.~\eqref{eq:channels-alt} and \eqref{eq:channels-null}, the two channel evolutions may differ at query $i$ solely in the outcome corresponding to $|v\rangle\langle v|$, where $\M_{v,U}$ subsequently prepares $\rho_U$, whereas $\N$ prepares $\tauD$. For a fixed record  $s_{1:i}\in\{0,1\}^i$, define
\begin{equation}
    \mathsf K_v(s_{1:i})
    :=
    \sum_{t=1}^i s_t,
    \label{eq:Kdef}
\end{equation}
the number of hits encountered up to query $i$. Fix $m\ge1$. Recall that, for a fixed realization $v,U$ the channel outcomes under the alternate and the null hypothesis were respectively:
\begin{align}
    \M_{v,U}(X)
    &=
    \underbrace{\langle v|X|v\rangle\,\rho_U}_{\text{hit}}
    +
    \underbrace{\tr\left[(I_A-|v\rangle\langle v|)X\right]\tauD}_{\text{miss}},
    \\
    \N(X)
    &=
    \underbrace{\langle v|X|v\rangle\,\tauD}_{\text{hit}}
    +
    \underbrace{\tr\left[(I_A-|v\rangle\langle v|)X\right]\tauD}_{\text{miss}}.
\end{align}

We separate records with at most $m$ hits from those with more than $m$:
\begin{align}
    \sigma_{\le m}^{v,U}
    &:=
    \sum_{\mathsf K_v(s_{1:n})\le m}
    \psi_n^{v,U}(s_{1:n}),
    &
    \sigma_{>m}^{v,U}
    &:=
    \sum_{\mathsf K_v(s_{1:n})>m}
    \psi_n^{v,U}(s_{1:n}),
    \\
    \sigma_{\le m}^{v,0}
    &:=
    \sum_{\mathsf K_v(s_{1:n})\le m}
    \psi_n^{v,0}(s_{1:n}),
    &
    \sigma_{>m}^{v,0}
    &:=
    \sum_{\mathsf K_v(s_{1:n})>m}
    \psi_n^{v,0}(s_{1:n}).
\end{align}
Accordingly,
\begin{equation}
    \sigma_{v,U}
    =
    \sigma_{\le m}^{v,U}
    +
    \sigma_{>m}^{v,U},
    \quad
    \sigma_0
    =
    \sigma_{\le m}^{v,0}
    +
    \sigma_{>m}^{v,0}.
    \label{eqn:states_sigma}
\end{equation}
The superscripts denote the underlying channel evolution. Specifically, $\sigma_{\le m}^{v,U}$ and $\sigma_{>m}^{v,U}$ correspond to the fixed hard instance $\M_{v,U}$, while the superscript $0$ indicates evolution under $\N$. As in the non-adaptive proof, denote the low-hit contribution by
\begin{equation}
    X_m
:=
\E_{\bm v}
\left(
    \bm\sigma_{\le m}^{\bm v,0}
    -
    \E_{\bm U}\bm\sigma_{\le m}^{\bm v,\bm U}
\right).
    \label{eq:adaptive-Xm}
\end{equation}
Therefore,
\begin{equation}
    \sigma_0-\bar\sigma
    =
    X_m
    +
    \E_{\bm v}
\left(
    \bm\sigma_{>m}^{\bm v,0}
    -
    \E_{\bm U}\bm\sigma_{>m}^{\bm v,\bm U}
\right).
    \label{eq:adaptive-low-high}
\end{equation}

Because $\sigma_0$ and $\bar\sigma$ are normalized states,
\begin{equation}
    \frac12\|\bar\sigma-\sigma_0\|_1
    =
    \frac12\|\sigma_0-\bar\sigma\|_1
    =
    \sup_{0\preceq\mathsf P\preceq I}
    \tr\left[\mathsf P(\sigma_0-\bar\sigma)\right].
    \label{eq:variational-split}
\end{equation}
For every POVM element $0\preceq\mathsf P\preceq I$, using
Eqn.~\eqref{eq:adaptive-low-high} gives
\begin{align}
    \tr\left[\mathsf P(\sigma_0-\bar\sigma)\right]
    &=
    \tr[\mathsf P X_m]
    +
    \E_{\bm v}
    \Big(
        \tr[\mathsf P\bm\sigma_{>m}^{\bm v,0}]
        -
        \E_{\bm U}\tr[\mathsf P\bm\sigma_{>m}^{\bm v,\bm U}]
    \Big).
    \label{eq:low-hit-high-hit-split}
\end{align}

\textit{Low Hit Contribution: }
The first term in Eqn.~\eqref{eq:low-hit-high-hit-split} is the contribution to the acceptance probability gap from outcome records $s_{1:n}$ satisfying $\mathsf K_v(s_{1:n})\le m$. For this low-hit contribution, we construct a \emph{truncation} map that reduces the comparison to an $m$-copy state-distinguishability problem. This allows us to use the $\chi^2$-divergence and Haar-moment bounds from Section~\ref{sec:output}.

\textit{High-Hit Contribution: }
The second term in Eqn.~\eqref{eq:low-hit-high-hit-split} is the contribution from outcome records satisfying $\mathsf K_v(s_{1:n})>m$. We bound this contribution by the total probability of the high-hit records, which is controlled using Markov's inequality.

\subsection{Upper bound on the low-hit contribution}
\label{sec:low-hit}

Recall that the low-hit contribution is the sum over outcome records $s_{1:n}$ satisfying $\mathsf K_v(s_{1:n})\le m$, namely $\sigma_{\le m}^{v,U}$ under $\M_{v,U}$ and $\sigma_{\le m}^{v,0}$ under $\N$. Along each such record, the two evolutions can differ at most $m$ hit locations. Following the adaptive-to-non-adaptive simulation principle of Das and Wilde~\cite[Prop.~2]{DasWilde2019}, the following lemma groups the truncated adaptive interaction and memory-state bookkeeping into a single common CP trace-nonincreasing map $\Gamma_{v,m}$. Its construction is described in Appendix~\ref{app:gamma}.

\begin{lemma}[Truncation map]
\label{lem:truncation}
For every unit vector $v\in\C^{\din}$ and $m\ge1$, there exists a completely positive, trace non-increasing (CP-TNI) map
\begin{equation}
    \Gamma_{v,m}:\Lop\big((\C^{\dout})^{\otimes m}\big)\longrightarrow\Lop(\Hfin),
    \label{eq:Gamma-type}
\end{equation} which is independent of $U$ and satisfies
\begin{equation}
    \Gamma_{v,m}(\rho_U^{\otimes m})=\sigma_{\le m}^{v,U},
    \quad
    \Gamma_{v,m}(\tauD^{\otimes m})=\sigma_{\le m}^{v,0}.
    \label{eq:Gamma-endpoints}
\end{equation}
\end{lemma}
\medskip
Define $\bar\rho_m:=\E_{\bm U}[\bm\rho_{\bm U}^{\otimes m}]$ and $\delta_m:=\frac12\|\bar\rho_m-\tauD^{\otimes m}\|_1.$ Since $\Gamma_{v,m}$ is independent of $U$, we show
\begin{align}
    \E_{\bm U}[\bm\sigma_{\le m}^{v,\bm U}]
    &=
    \E_{\bm U}\left[
        \Gamma_{v,m}\left(\bm \rho_{\bm U}^{\otimes m}\right)
    \right]
    && \text{by Eqn.~\eqref{eq:Gamma-endpoints}}
    \nonumber\\
    &=
    \Gamma_{v,m}\left(
        \E_{\bm U}\left[\bm\rho_{\bm U}^{\otimes m}\right]
    \right)
    && \text{by linearity of $\Gamma_{v,m}$}
    \nonumber\\
    &=
    \Gamma_{v,m}(\bar\rho_m),
    && \text{by the definition of $\bar\rho_m$.}
    \label{eq:Gamma-average}
\end{align}

\begin{proof}
    The lemma follows from the definition of $\Gamma_{v,m}$ in Section~\ref{formal_const}, the end-point identities established in Eqns.~\eqref{eq:Gamma-rho-endpoint} and \eqref{eq:Gamma-tau-endpoint}, and the CP-TNI property established in Eqn.~\eqref{eq:TNI}.
\end{proof}

\paragraph{A mathematical tool:}
The map $\Gamma_{v,m}$ is an auxiliary CP trace non-increasing map used only in the proof and is not visible to the tester. The map simulates the true adaptive experiment up to $m$ hits. Its input consists of $m$ registers $C_1,\ldots,C_m$, one for each possible hit. Along a record with at most $m$ hits, the map follows the same adaptive
evolution as the original tester. On the $r$-th hit, the state in $C_r$ is used as the channel output. On a miss, the output is $\tauD$ and the same adaptive maps $\mathsf V_i$ are applied. 

Evaluating the same map on $\rho_U^{\otimes m}$ reproduces the low-hit contribution under $\M_{v,U}$, while evaluating it on $\tauD^{\otimes m}$ reproduces the low-hit contribution under $\N$ as shown in Eqn.~\eqref{eq:Gamma-endpoints}. For an intuitive understanding of the map please see Section~\ref{app:gamma}.

\begin{lemma}[low-hit bound]
\label{lem:low-hit}
For every unit vector $v\in\C^{\din}$, every $m\ge1$, and every POVM element
$0\preceq \mathsf P\preceq I$,
\begin{equation}
    \tr[\mathsf P\sigma_{\le m}^{v,0}]
-
\E_{\bm U}\tr[\mathsf P\bm\sigma_{\le m}^{v,\bm U}]
\le\delta_m.
    \label{eq:low-hit-bound}
\end{equation}
\end{lemma}

\begin{proof}
Let
    $\mathsf D_m:=\tauD^{\otimes m}-\bar\rho_m.$ Since $\mathsf D_m$ is Hermitian and traceless, let
$\mathsf D_m= \mathsf D_m^+- \mathsf D_m^-$ be its Jordan decomposition.  Then
\[
    \mathsf D_m^\pm\succeq0,
    \quad
    \tr \mathsf D_m^+=\tr \mathsf D_m^-=\frac12\|\mathsf D_m\|_1=\delta_m.
\]
Therefore,
\begin{align}
    &\tr[\mathsf P\sigma_{\le m}^{v,0}]
-
\E_{\bm U}\tr[\mathsf P\bm\sigma_{\le m}^{v,\bm U}]
    \nonumber\\
    &\quad=
    \tr \left[
        \mathsf P\,\Gamma_{v,m}(\mathsf D_m)
    \right]
    && \text{by the endpoint identities}
    \nonumber\\
    &\quad=
    \tr \left[
        \mathsf P\,\Gamma_{v,m}(\mathsf D_m^+)
    \right]
    -
    \tr \left[
        \mathsf P\,\Gamma_{v,m}(\mathsf D_m^-)
    \right]
    && \text{by linearity of $\Gamma_{v,m}$}
    \nonumber\\
    &\quad\le
    \tr \left[
        \mathsf P\,\Gamma_{v,m}(\mathsf D_m^+)
    \right]
    && \text{since $\Gamma_{v,m}(\mathsf D_m^-)\succeq0$}
    \nonumber\\
    &\quad\le
    \tr\Gamma_{v,m}(\mathsf D_m^+)
    && \text{since $0\preceq\mathsf P\preceq I$}
    \nonumber\\
    &\quad\le
    \tr \mathsf D_m^+ =
    \frac12\|\mathsf D_m\|_1
    =
    \delta_m.
    && \text{since $\Gamma_{v,m}$ is TNI (See Eqn.~\eqref{eq:TNI})}
\end{align}
\end{proof}

Note that $\delta_m$ is independent of $v$, hence the low-hit contribution 
\[\E_{\bm v}
    \Big(
        \tr[\mathsf P\bm\sigma_{>m}^{\bm v,0}]
        -
        \E_{\bm U}\tr[\mathsf P\bm\sigma_{>m}^{\bm v,\bm U}]
    \Big) \le \E_v \delta_m \le \delta_m\].
    It now remains to bound $\delta_m$. Since
$\bar\rho_m=\E_{\bm U}[\bm\rho_{\bm U}^{\otimes m}]$, if
$\bm U,\bm U'$ are independent Haar unitaries then
$\tr(\bar\rho_m^2)=\E_{\bm U,\bm U'}[\tr(\bm\rho_{\bm U}\bm\rho_{\bm U'})^m]$.
Using
$\rho_U=\tauD+(\varepsilon/\sqrt{\dout})\Delta_U$ and
$\tr\Delta_U=0$, we have
$\tr(\rho_{\bm U}\rho_{\bm U'})
=\dout^{-1}\bigl(1+\varepsilon^2
\tr(\bm\Delta_{\bm U}\bm\Delta_{\bm U'})\bigr)$.
Moreover,
$\tr(\bm\Delta_{\bm U}\bm\Delta_{\bm U'})
=
\tr(\Delta_0\bm U^\dagger\bm U'\Delta_0\bm U'^\dagger\bm U)$,
and $\bm V:=\bm U^\dagger\bm U'$ is Haar distributed. We therefore set
\[
    \bm \beta:=\tr(\Delta_0\bm V\Delta_0\bm V^\dagger),
    \quad
    \bm V\sim\operatorname{Haar}(U(\dout)).
\]

\begin{lemma}[the $m$-copy estimate]
\label{thm:delta}
Let
$\lambda_m:=96\cdot m^2\varepsilon^4/\dout^2$.  Then
\begin{align}
    4\delta_m^2 
    &\le
    \chi^2(\bar\rho_m\|\tauD^{\otimes m})
    :=\dout^m\tr \left[(\bar\rho_m-\tauD^{\otimes m})^2\right] =\E[(1+\varepsilon^2\bm \beta)^m]-1 \nonumber
    \le e^{\lambda_m}-1.
    \label{eq:delta-chi}
\end{align}
Consequently,
\begin{equation}
    \delta_m
    \le
    \sqrt{\frac{{96}}{2}}\,
    \frac{m\varepsilon^2}{\dout}
    \quad\text{whenever }\lambda_m\le1
    \label{eq:delta-facts}
\end{equation}
And \[\delta_1=0
 \emph{\text{ since }}
\E_{\bm U}[\bm\Delta_{\bm U}]
=\frac{\tr\Delta_0}{\dout}I=0.\]
\end{lemma}

\begin{proof}
Proof is established in Appendix~\ref{sec:output}. 
\end{proof}

\subsection{Upper bound on the high-hit contribution}
\label{sec:high-hit}

The following analysis doesn't depend on the truncation map. Recall that for an outcome record $s_{1:n}\in\{0,1\}^n$, we defined the total number of hits by $\mathsf K_v(s_{1:n}):=\sum_{i=1}^n s_i.$

For fixed $v$, let $\bm S_{1:n}=(\bm S_1,\ldots,\bm S_n)$ denote the random hit-miss outcome record under the null hypothesis, with realizations $s_{1:n}=(s_1,\ldots,s_n)\in\{0,1\}^n$, where $s_i=1$ denotes a hit and $s_i=0$ a miss. Define
\begin{equation}
    \bm{\mathsf K}_v
    :=
    \mathsf K_v(\bm S_{1:n})
    =
    \sum_{i=1}^n \bm S_i,
    \quad
    \tr\sigma_{>m}^{v,0}
    =
    \Pr\nolimits_{0}(\bm{\mathsf K}_v>m).
    \label{eq:pa-def}
\end{equation} 
\begin{lemma}[high-hit-event bound]
\label{lem:high-hit}
For every POVM element $0\preceq \mathsf P\preceq I$,
\begin{equation}
    \tr[\mathsf P\sigma_{>m}^{v,0}]
    -\E_{\bm U}\tr[\mathsf P\bm\sigma_{>m}^{v,\bm U}]
    \le
    \Pr\nolimits_{0}(\bm{\mathsf K}_v>m).
    \label{eq:high-hit-bound}
\end{equation}
\end{lemma}

\begin{proof}
Since
$\E_{\bm U}\tr[\mathsf P\bm\sigma_{>m}^{v,\bm U}]\ge 0,$
we have
\[
    \tr[\mathsf P\sigma_{>m}^{v,0}]
    -
    \E_{\bm U}\tr[\mathsf P\bm\sigma_{>m}^{v,\bm U}]
    \le
    \tr[\mathsf P\sigma_{>m}^{v,0}]
    \le
    \tr\sigma_{>m}^{v,0}
    =
    \Pr\nolimits_0(\bm{\mathsf K}_v>m).
\]
\end{proof}

\begin{lemma}
\label{lem:occupation}
For every $m\ge1$,
\begin{equation}
    \E_{\bm v}
    \Pr\nolimits_{0}(\bm{\mathsf K}_{\bm v}>m)
    \le\frac{n}{\din(m+1)}.
    \label{eq:occupation-bound}
\end{equation}
\end{lemma}
\begin{proof}
We use $\Pr_0$ and $\E_0$ for probability and expectation under the null hypothesis. For each $i\in[n]$, let $\varphi_i\in\mathsf D(A_i)$ denote the unconditional input state at query $i$ under the null hypothesis. Since neither the tester nor $\N$ depends on $v$, the state $\varphi_i$ is independent of $v$. Then
\begin{align}
    \E_{\bm v}
    \Pr\nolimits_{0}(\bm{\mathsf K}_{\bm v}>m)
    &=
    \E_{\bm v}\Pr\nolimits_{0}\left(\bm{\mathsf K}_{\bm v}\ge m+1\right)
    && \text{since $\bm{\mathsf K}_{\bm v}$ is integer valued}
    \nonumber\\
    &\le
    \frac{1}{m+1}\,
    \E_{\bm v}\E_0[\bm{\mathsf K}_{\bm v}]
    && \text{by Markov's inequality}
    \nonumber\\
    &=
    \frac{1}{m+1}
    \sum_{i=1}^n
    \E_{\bm v}\E_0[\bm S_i]
    && \text{since $\bm{\mathsf K}_{\bm v}=\sum_{i=1}^n\bm S_i$}
    \nonumber\\
    &=
    \frac{1}{m+1}
    \sum_{i=1}^n
    \E_{\bm v}
    \left[
        0\cdot\Pr_0(\bm S_i=0)
        +
        1\cdot\Pr_0(\bm S_i=1)
    \right]
    && \text{since $\bm S_i\in\{0,1\}$}
    \nonumber\\
    &=
    \frac{1}{m+1}
    \sum_{i=1}^n
    \E_{\bm v}
    \tr\left(
        |\bm v\rangle\langle\bm v|\varphi_i
    \right)
    && \text{by the Born rule}
    \nonumber\\
    &=
    \frac{1}{m+1}
    \sum_{i=1}^n
    \tr\left(
        \E_{\bm v}[|\bm v\rangle\langle\bm v|]\varphi_i
    \right)
    && \text{since $\varphi_i$ is independent of $\bm v$}
    \nonumber\\
    &=
    \frac{1}{m+1}
    \sum_{i=1}^n
    \frac{\tr(\varphi_i)}{\din}
    && \text{$\E_{\bm v}[|\bm v\rangle\langle\bm v|]=I_A/\din$}
    \nonumber\\
    &=
    \frac{n}{\din(m+1)}
    && \text{since $\tr(\varphi_i)=1$.}
\end{align}
\end{proof}

By Eqns.~\eqref{eq:adapt-trace-need}, \eqref{eq:variational-split},
\eqref{eq:low-hit-high-hit-split} and Lemmas~ \ref{lem:low-hit},
\ref{lem:high-hit} and \ref{lem:occupation}, every $m\ge1$
satisfies
\begin{equation}
     \frac13
    \le\frac12\|\bar\sigma-\sigma_0\|_1
    \le \delta_m+\frac{n}{\din(m+1)}.
    \label{eq:master-necessary}
\end{equation}

With this, the proof of Theorem~\ref{thm:main_adapt} follows almost immediately.
\subsection{Proof of Theorem~\ref{thm:main_adapt}}
\label{sec:lower}

\begin{proof}
Set
$
    \alpha
    :=
   \dfrac{1}{3\sqrt{2 \cdot 96}}.
$ 

\medskip\noindent
\textbf{Case 1: $\alpha \dout/\varepsilon^2\ge1$.}
Choose
$
    m
    :=
    \left\lfloor
        \alpha\frac{\dout}{\varepsilon^2}
    \right\rfloor.
$ We first verify the hypothesis required for the linear estimate on
$\delta_m$ in Theorem~\ref{thm:delta}
\begin{align}
    \lambda_m
    &=
    \frac{96 \cdot m^2\varepsilon^4}
         {\dout^2}
    && \text{definition of $\lambda_m$ in Theorem~\ref{thm:delta}}
    \nonumber\\
    &\le
    \frac{96 \cdot \alpha^2
          \dout^2\varepsilon^4}
         {\varepsilon^4\dout^2}
    && m\le
       \alpha \dout/\varepsilon^2
    \nonumber\\
    &=
    96 \cdot \alpha^2
    =
    \frac{1}{18}
    <1,
    && \alpha=
       \frac{1}{3\sqrt{2 \cdot 96}}.
    \label{eq:case1-lambda}
\end{align}
Thus the condition $\lambda_m\le1$ in
Theorem~\ref{thm:delta} is satisfied, so
\begin{align}
    \delta_m
    &\le
    \sqrt{\frac{96}{2}}\,
    \frac{m\varepsilon^2}{\dout}
    \le
    \sqrt{\frac{96}{2}}\,
    \alpha =
    \frac16.
    && \text{Since}~~ m\le
       \alpha \dout/\varepsilon^2 ~~\text{and}~~\alpha=
       \frac{1}{3\sqrt{2 \cdot 96}}
    \label{eq:case1-delta}
\end{align}
Substituting this into \eqref{eq:master-necessary} yields
\begin{align}
    n
    &\ge
    \frac{\din(m+1)}6
    >
    \frac{\alpha}{6}
    \frac{\din\dout}
         {\varepsilon^2} =
    \frac{1}{18\sqrt{2 \cdot 96}}
    \frac{\din\dout}
         {\varepsilon^2}.
    && \because ~~m :=\left\lfloor
        \alpha\frac{\dout}{\varepsilon^2}
    \right\rfloor\le\alpha\frac{\dout}{\varepsilon^2}
    <
    m+1.
    \label{eq:case1-final}
\end{align}

\medskip\noindent
\textbf{Case 2: $\alpha \dout/\varepsilon^2<1$.}
Here take $m=1$.  We do not use the conditional linear estimate on
$\delta_m$, instead, Theorem~\ref{thm:delta} gives the exact identity
$\delta_1=0$.  Hence Eqn.~\eqref{eq:master-necessary} gives
\begin{align}
    \frac13
    &\le
    \delta_1
    +
    \frac{n}{\din(1+1)}
   =
    \frac{n}{2\din}
\end{align}
\begin{align}
    \therefore ~~n
    &\ge
    \frac{2\din}{3}
    >
    \frac{\din}{6}
    >
    \frac{\alpha}{6}
    \frac{\din\dout}
         {\varepsilon^2}
    && \alpha \dout/\varepsilon^2<1
    \nonumber\\
    &=
    \frac{1}{18\sqrt{2 \cdot 96}}
    \frac{\din\dout}
         {\varepsilon^2}.
    \label{eq:case2-final}
\end{align}

Thus in both cases
$
    n
    \ge
    \dfrac{1}{18\sqrt{2 \cdot 96}}
    \dfrac{\din\dout}{\varepsilon^2},
$
which proves Theorem~\ref{thm:main_adapt}.
\end{proof}

For $\dout$ being odd, we use the traceless Hermitian operator \[\Delta_0
    =
    \frac{\sqrt{\dout}}{\dout-1}
    \sum_{j=1}^{(\dout-1)/2}|j\rangle\langle j|
    -
    \frac{\sqrt{\dout}}{\dout+1}
    \sum_{j=(\dout+1)/2}^{\dout}|j\rangle\langle j|\]
instead. Most of the analysis remains the same with the output moment bound in Lemma~\ref{lem:odd-output-mixing} which gives a different universal constant. The query complexity remains the same.

\section{Acknowledgments}
We thank Matthias Caro, Nikhil Bansal and Natalie McHugh for fruitful discussions during the early stages of this work. This work was supported, in part, by the EPSRC New Horizons grant (EP/X018180/1), and the UK Hub for Quantum Computing via Integrated and Interconnected Implementations (QCi3) (EP/Z53318X/1). 

\section{AI Statement}
In Appendix~\ref{app:h(k)_bound}, bounding $h(k)$ using a beta-function representation in Eqn.~\eqref{eq:h-integral} which leads to the subsequent integral bound in Eqn.~\eqref{eq:h-lambda-bound}--\eqref{eq:hk-large-din} was suggested by ChatGPT 5.6 Sol via NebulaOne provided by the University of Warwick. It also suggested the Latex codes for Figures~\ref{fig:tester} and~\ref{fig:adaptive-tester}. Authors take full responsibility of the results presented in this paper.

\newpage

% \appendices
\appendices
\numberwithin{equation}{section}
\renewcommand{\theequation}{\thesection\arabic{equation}}
\renewcommand{\theHequation}{\thesection.\arabic{equation}}

\section*{\huge Appendices}
\bigskip

\section{Low hit contribution}
\label{app:Low-Hit-Contri}
\LowHit*
\begin{proof}
$X_m$ is Hermitian. Since
$
    \sigma_0
    =
    \tauout^{\otimes n}\otimes\Psi_R,
$
and $\tauout$ is full rank, we have
$
    \operatorname{supp}(\sigma_0)
    =
    B^{\otimes n}\otimes\operatorname{supp}(\Psi_R).
$
Moreover,
$
    \sum_s \bm p(s)\bm\omega_R^s
    =
    \Psi_R,
$
and hence
$
    \bm p(s)\bm\omega_R^s
    \preceq
    \Psi_R.
$
Thus, whenever $\bm p(s)>0$,
$
    \operatorname{supp}(\bm\omega_R^s)
    \subseteq
    \operatorname{supp}(\Psi_R).
$
Each nonzero summand in $X_m$ is supported on
$
B^{\otimes n}\otimes\operatorname{supp}(\Psi_R)
$.
Therefore,
$\operatorname{supp}(X_m)\subseteq\operatorname{supp}(\sigma_0).
$
All inverses below are understood on the corresponding supports. By H\"older's
inequality,
\begin{align}
    \|X_m\|_1
    &=
    \left\|
        \sigma_0^{1/4}
        \left(
            \sigma_0^{-1/4}
            X_m
            \sigma_0^{-1/4}
        \right)
        \sigma_0^{1/4}
    \right\|_1
   \le
    \|\sigma_0^{1/4}\|_4^2
    \left\|
        \sigma_0^{-1/4}
        X_m
        \sigma_0^{-1/4}
    \right\|_2
    =
    \left\|
        \sigma_0^{-1/4}
        X_m
        \sigma_0^{-1/4}
    \right\|_2,
\end{align}
where
$
    \|\sigma_0^{1/4}\|_4^4
    =
    \tr(\sigma_0)
    =
    1.
$
Since $X_m$ is Hermitian, squaring the preceding bound gives
\begin{equation}
    \|X_m\|_1^2
    \le \left\|
        \sigma_0^{-1/4}
        X_m
        \sigma_0^{-1/4}
    \right\|_2^2=
    \tr\left[
        \sigma_0^{-1/2}
        X_m
        \sigma_0^{-1/2}
        X_m
    \right].
    \label{eq:weighted-l2}
\end{equation}

To expand the two copies of $X_m$ in Eqn.~\eqref{eq:weighted-l2}, let
$\bm v'$ be an independent copy of $\bm v$, and let
$s,s'\in\{0,1\}^n$ denote the corresponding hit-miss records.
$
    \bar\rho_k=\E_{\bm U}[\bm\rho^{\otimes k}].
$
Let
$
    \bm U,\bm U'\sim\operatorname{Haar}(U(\dout))
$
be independent, and define
\[
    \bm\Delta:=\bm U\Delta_0\bm U^\dagger,
    \quad
    \bm\Delta':=\bm U'\Delta_0(\bm U')^\dagger,
\]
and
\[
    \bm\rho:=\tauout+\frac{\varepsilon}{\sqrt{\dout}}\bm\Delta,
    \quad
    \bm\rho':=\tauout+\frac{\varepsilon}{\sqrt{\dout}}\bm\Delta'.
\]

To expand the right-hand side of Eqn.~\eqref{eq:weighted-l2}, recall that
\[
X_m
=
\E_{\bm v,\bm U}
\sum_{k(s)<m}
\bm p(s)\,
\Pi_s^\dagger
\left[
    \left(
        \bm\rho^{\otimes k(s)}-\tauout^{\otimes k(s)}
    \right)
    \otimes\tauout^{\otimes(n-k(s))}
\right]
\Pi_s
\otimes\bm\omega_R^s.
\]
Let $(\bm v',\bm U')$ be an independent copy of $(\bm v,\bm U)$, with corresponding
$\bm p'(s')$, $\bm\rho'$, and $\bm\omega_R^{\prime s'}$. Since
\[
\sigma_0^{-1/2}
=
\dout^{n/2}I_{B_1\cdots B_n}\otimes\Psi_R^{-1/2},
\]
substituting the two independent copies of $X_m$, using linearity of the trace, and
factorizing the trace over $B_1\cdots B_n$ and $R$ gives
\begin{align}
&\tr\left[
    \sigma_0^{-1/2}
    X_m
    \sigma_0^{-1/2}
    X_m
\right]
\nonumber\\
&=
\E_{\bm v,\bm v',\bm U,\bm U'}
\sum_{\substack{s,s'\in\{0,1\}^n\\
                 k(s)<m,\;k(s')<m}}
\bm p(s)\bm p'(s')
\nonumber\\
&\quad\times
\dout^n
\tr\Bigg[
    \Pi_s^\dagger
    \left[
        \left(
            \bm\rho^{\otimes k(s)}
            -
            \tauout^{\otimes k(s)}
        \right)
        \otimes
        \tauout^{\otimes(n-k(s))}
    \right]
    \Pi_s
\nonumber\\
&\hspace{34mm}\times
    \Pi_{s'}^\dagger
    \left[
        \left(
            (\bm\rho')^{\otimes k(s')}
            -
            \tauout^{\otimes k(s')}
        \right)
        \otimes
        \tauout^{\otimes(n-k(s'))}
    \right]
    \Pi_{s'}
\Bigg]
\nonumber\\
&\quad\times
\tr\left[
    \Psi_R^{-1/2}
    \bm\omega_R^s
    \Psi_R^{-1/2}
    \bm\omega_R^{\prime s'}
\right].
\label{eq:Xm-quadratic-expanded}
\end{align}
Since
$
    \Pi_s^\dagger
    \left[
        \tauout^{\otimes k(s)}
        \otimes
        \tauout^{\otimes(n-k(s))}
    \right]
    \Pi_s
    =
    \tauout^{\otimes n},
$
we have
\[
    \Pi_s^\dagger
    \left[
        \left(
            \bm\rho^{\otimes k(s)}
            -
            \tauout^{\otimes k(s)}
        \right)
        \otimes
        \tauout^{\otimes(n-k(s))}
    \right]
    \Pi_s
    =
    \Pi_s^\dagger
    \left[
        \bm\rho^{\otimes k(s)}
        \otimes
        \tauout^{\otimes(n-k(s))}
    \right]
    \Pi_s
    -
    \tauout^{\otimes n},
\]
which analogously holds for $s'$. We now expand the middle term of Eqn.~\eqref{eq:Xm-quadratic-expanded}. Observe that the last three trace terms here are each equal to $1$, since
\[
    \dout^n
    \tr\Bigg[
        \Pi_s^\dagger
        \left[
            \bm\rho^{\otimes k(s)}
            \otimes
            \tauout^{\otimes(n-k(s))}
        \right]
        \Pi_s
        \tauout^{\otimes n}
    \Bigg]
    =
    \left(\dout\,\tr(\bm\rho\tauout)\right)^{k(s)}
    \left(\dout\,\tr(\tauout^2)\right)^{n-k(s)}
    =
    1,
\]
and similarly for the term containing $\bm\rho'$, while
$
    \dout^n
    \tr\left[(\tauout^{\otimes n})^2\right]
    =
    1.
$ For the remaining term, the permutations place $\bm\rho$ and $\bm\rho'$ precisely at the hit positions of $s$ and $s'$, respectively. Hence the trace factorizes over the $n$ output registers as
\begin{align}
&\dout^n
\tr\Bigg[
    \Pi_s^\dagger
    \left[
        \bm\rho^{\otimes k(s)}
        \otimes
        \tauout^{\otimes(n-k(s))}
    \right]
    \Pi_s
    \Pi_{s'}^\dagger
    \left[
        (\bm\rho')^{\otimes k(s')}
        \otimes
        \tauout^{\otimes(n-k(s'))}
    \right]
    \Pi_{s'}
\Bigg]
\nonumber\\
&\quad=
\prod_{i=1}^n
\begin{cases}
\dout\,\tr(\bm\rho\bm\rho'),
    & s_i=s_i'=1,\\
\dout\,\tr(\bm\rho\tauout),
    & s_i=1,\ s_i'=0,\\
\dout\,\tr(\tauout\bm\rho'),
    & s_i=0,\ s_i'=1,\\
\dout\,\tr(\tauout^2),
    & s_i=s_i'=0,
\end{cases}
\nonumber\\
&\quad=
\prod_{i=1}^n
\left(
    1+\varepsilon^2\bm\beta\,s_is_i'
\right),
\label{eqn:betaprod}
\end{align}
where we have used the definition $\bm\beta:=\tr(\bm\Delta\bm\Delta')$ and the identities:
\begin{align}
    \dout\,\tr(\bm\rho\bm\rho')
    &=
    1+\varepsilon^2\bm\beta
   \quad \text{and}\quad
    \dout\,\tr(\bm\rho\tauout)
    =
    \dout\,\tr(\tauout\bm\rho')
    =
    \dout\,\tr(\tauout^2)
    =
    1.
    \label{eqn-beta_relations}
\end{align}

Therefore,
\begin{align}
&\dout^n
\tr\Bigg[
    \Pi_s^\dagger
    \left[
        \left(
            \bm\rho^{\otimes k(s)}
            -
            \tauout^{\otimes k(s)}
        \right)
        \otimes
        \tauout^{\otimes(n-k(s))}
    \right]
    \Pi_s
    \Pi_{s'}^\dagger
    \left[
        \left(
            (\bm\rho')^{\otimes k(s')}
            -
            \tauout^{\otimes k(s')}
        \right)
        \otimes
        \tauout^{\otimes(n-k(s'))}
    \right]
    \Pi_{s'}
\Bigg]
\nonumber\\
&\quad=
\prod_{i=1}^n
\left(
    1+\varepsilon^2\bm\beta\,s_is_i'
\right)-1.
\label{eq:output-overlap-expanded}
\end{align}

Finally, expanding over subsets $S\subseteq[n]$ of query positions,
\[
    \prod_{i=1}^n(1+x_i)
    =
    \sum_{S\subseteq[n]}
    \prod_{i\in S}x_i,
\]
This follows because $\sum_{S\subseteq[n]}\prod_{i\in S}x_i
=
1+\sum_{|S|=1}\prod_{i\in S}x_i+\sum_{|S|=2}\prod_{i\in S}x_i+\cdots+\prod_{i=1}^n x_i
=
1+\sum_i x_i+\sum_{i<j}x_ix_j+\cdots+x_1\cdots x_n
=
\prod_{i=1}^n(1+x_i)$ and
 $S=\varnothing$ gives the term $1$. We thus obtain
\begin{align}
    \prod_{i=1}^n
    \left(
        1+\varepsilon^2\bm\beta\,s_is_i'
    \right)-1
    &=
    \sum_{\emptyset\neq S\subseteq[n]}
    \prod_{i\in S}
    \left(
        \varepsilon^2\bm\beta\,s_is_i'
    \right)
    =
    \sum_{\emptyset\neq S\subseteq[n]}
    (\varepsilon^2\bm\beta)^{|S|}
    \prod_{i\in S}s_is_i'.
    \label{eq:output-subset-expansion}
\end{align}
For $j\ge1$, we define
\[
    \mu_j
    :=\varepsilon^{2j}
    \E\left[
        \bm\beta^j
    \right].
\]
Substituting Eqn.~\eqref{eq:output-subset-expansion} into
Eqn.~\eqref{eq:Xm-quadratic-expanded} and first averaging over
$\bm U,\bm U'$ gives
\begin{align}
&\tr\left[
    \sigma_0^{-1/2}
    X_m
    \sigma_0^{-1/2}
    X_m
\right]
\nonumber\\
&=
\sum_{\emptyset\neq S\subseteq[n]}
\mu_{|S|}
\E_{\bm v,\bm v'} 
\sum_{\substack{s,s'\in\{0,1\}^n\\
                 k(s)<m,\;k(s')<m}}
\prod_{i\in S}s_i
\prod_{i\in S}s_i'
\bm p(s)\bm p'(s')
\tr\left[
    \Psi_R^{-1/2}
    \bm\omega_R^s
    \Psi_R^{-1/2}
    \bm\omega_R^{\prime s'}
\right]
\nonumber\\
&=
\sum_{\emptyset\neq S\subseteq[n]}
\mu_{|S|}
\tr\Bigg[
    \Psi_R^{-1/2} \times 
    \left(\E_{\bm v}\sum_{k(s)<m}
    \prod_{i\in S}s_i\,
    \bm p(s)\bm\omega_R^s \right) \times 
    \Psi_R^{-1/2}
    \times \left(\E_{\bm v'}\sum_{k(s')<m}
    \prod_{i\in S}s_i'\,
    \bm p'(s')\right) \times \bm \omega_R^{\prime s'}
\Bigg].
\label{eq:Xm-after-output-average}
\end{align}
Let $(\bm v',s')$ be an independent copy of $(\bm v,s)$. For each nonempty $S\subseteq[n]$, define
\begin{align}
    A_S
    :=
    \E_{\bm v}
    \sum_{k(s)<m}
    \left(\prod_{i\in S}s_i\right)
    \bm p(s)\bm\omega_R^s
    =
    \E_{\bm v'}
    \sum_{k(s')<m}
    \left(\prod_{i\in S}s_i'\right)
    \bm p'(s')\bm\omega_R^{\prime s'},
    \label{eqn:As}
\end{align}

Hence,
Eqn.~\eqref{eq:Xm-after-output-average} becomes
\begin{equation}
    \tr\left[
        \sigma_0^{-1/2}
        X_m
        \sigma_0^{-1/2}
        X_m
    \right]
    =
    \sum_{\emptyset\neq S\subseteq[n]}
    \mu_{|S|}
    \tr\left[
        \Psi_R^{-1/2}
        A_S
        \Psi_R^{-1/2}
        A_S
    \right].
    \label{eqn:pairbound}
\end{equation}

Now fix $S\subseteq[n]$ with $|S|=j$. Recall from Eqn.~\eqref{eq:Ps} that
$\bm P_s=\bigotimes_{i=1}^n\bm P_{s_i}$, with
$\bm P_1=\bm P$ and $\bm P_0=I_A-\bm P$, and from
Eqns.~\eqref{eq:ps-def}--\eqref{eq:omega-sv} that
$\bm p(s)\bm\omega_R^s
=
\tr_{A_1\cdots A_n}
[(\bm P_s\otimes I_R)\Psi(\bm P_s\otimes I_R)]$.
Since $\prod_{i\in S}s_i$ is the indicator of $s_i=1$ for all $i\in S$,
\[
    \sum_{k(s)<m}
    \left(\prod_{i\in S}s_i\right)
    \bm p(s)\bm\omega_R^s
    =
    \sum_{\substack{k(s)<m\\ s_i=1\;\forall i\in S}}
    \bm p(s)\bm\omega_R^s
    \preceq
    \sum_{\substack{s_i=1\;\forall i\in S}}
    \bm p(s)\bm\omega_R^s,
\]
because $\bm p(s)\bm\omega_R^s\succeq0$. This follows as
$\sum_{s_i=1\ \forall i\in S}\bm p(s)\bm\omega_R^s
=
\sum_{\substack{k(s)<m\\ s_i=1\ \forall i\in S}}
\bm p(s)\bm\omega_R^s
+
\sum_{\substack{k(s)\ge m\\ s_i=1\ \forall i\in S}}
\bm p(s)\bm\omega_R^s$. Thus, we have
\begin{align}
    A_S
    &=
    \E_{\bm v}
    \sum_{k(s)<m}
    \left(\prod_{i\in S}s_i\right)
    \bm p(s)\bm\omega_R^s
    \nonumber\\
    &\preceq
    \E_{\bm v}
    \sum_{\substack{s\in\{0,1\}^n\\s_i=1\;\forall i\in S}}
    \bm p(s)\bm\omega_R^s
    \nonumber\\
    &=
    \E_{\bm v}
    \sum_{\substack{s\in\{0,1\}^n\\s_i=1\;\forall i\in S}}
    \tr_{A_1\cdots A_n}\left[
        (\bm P_s\otimes I_R)
        \Psi
        (\bm P_s\otimes I_R)
    \right]
    \nonumber\\
    &=
    \E_{\bm v}
    \sum_{\substack{s\in\{0,1\}^n\\s_i=1\;\forall i\in S}}
    \tr_{A_1\cdots A_n}\left[
        (\bm P_s\otimes I_R)\Psi
    \right]
    \nonumber\\
    &=
    \E_{\bm v}
    \tr_{A_1\cdots A_n}\left[
        \left(
            \sum_{\substack{s\in\{0,1\}^n\\s_i=1\;\forall i\in S}}
            \bm P_s
            \otimes I_R
        \right)\Psi
    \right],
\end{align}
where
$\bm P_s^2=\bm P_s$ was used. The remaining sum over $s$ then can be decomposed as 
\[
    \sum_{\substack{s\in\{0,1\}^n\\s_i=1\;\forall i\in S}}\bm P_s
    =
    \bigotimes_{i\in S}\bm P_1
    \otimes
    \bigotimes_{i\notin S}(\bm P_0+\bm P_1)
    =
    \bm P^{\otimes j}\otimes I_{S^c},
\]
where $\bm P_1=\bm P$, $\bm P_0+\bm P_1=I_A$ and $S^c$ denotes the complement of the set $S$. Hence, using the standard Haar moment bound (see Eqn.~\eqref{eq:Haar-pure-moment}) 
\[
    \E_{\bm v}\bm P^{\otimes j}
    =
    c_j\Pi_{\mathrm{sym}}^{(S)},
    \quad
    c_j
    =
    \binom{\din+j-1}{j}^{-1},
\]
we get
\[
    A_S
    \preceq
    c_j\,
    \tr_{A_1\cdots A_n}\left[
        (\Pi_{\mathrm{sym}}^{(S)}\otimes I_R)
        \Psi
        (\Pi_{\mathrm{sym}}^{(S)}\otimes I_R)
    \right]
    \preceq
    c_j\Psi_R. 
\]
which follows since \[\Psi_R =
\tr_{A_1\cdots A_n}\left[
(\Pi_{\mathrm{sym}}^{(S)}\otimes I_R)\Psi
(\Pi_{\mathrm{sym}}^{(S)}\otimes I_R)
\right]
+
\tr_{A_1\cdots A_n}\left[
((I-\Pi_{\mathrm{sym}}^{(S)})\otimes I_R)\Psi
((I-\Pi_{\mathrm{sym}}^{(S)})\otimes I_R)
\right]
\]
Thus
$0\preceq A_S\preceq c_j\Psi_R$, and therefore, on
$\operatorname{supp}(\Psi_R)$,
\[
    \Psi_R^{-1/2}A_S\Psi_R^{-1/2}
    \preceq
    c_j I.
\]
Since $A_S\succeq0$,
\[
    \tr\left[
        \Psi_R^{-1/2}A_S\Psi_R^{-1/2}A_S
    \right]
    \le
    c_j\tr(A_S).
\]
Finally, from the definition of $A_S$ and
$\tr(\bm\omega_R^s)=1$,
\[
    \tr(A_S)
    =
    \E_{\bm v}
    \sum_{k(s)<m}
    \left(\prod_{i\in S}s_i\right)\bm p(s).
\]
Therefore
\begin{align}
    \|X_m\|_1^2
    &\le
    \sum_{\emptyset\neq S\subseteq[n]}
    \mu_{|S|}c_{|S|}\tr(A_S)
    \nonumber\\
    &=
    \sum_{\emptyset\neq S\subseteq[n]}
    \mu_{|S|}c_{|S|}
    \E_{\bm v}
    \sum_{k(s)<m}
    \left(\prod_{i\in S}s_i\right)
    \bm p(s)
    \nonumber\\
    &=
    \E_{\bm v}
    \sum_{k(s)<m}
    \bm p(s)
    \sum_{\emptyset\neq S\subseteq[n]}
    \mu_{|S|}c_{|S|}
    \prod_{i\in S}s_i
    \nonumber\\
    &=
    \E_{\bm v}
    \sum_{k(s)<m}
    \bm p(s)
    \sum_{\substack{\emptyset\neq S\subseteq[n]\\
                     s_i=1\;\forall i\in S}}
    \mu_{|S|}c_{|S|}
    && \because\text{ $\prod_{i\in S}s_i=1$ iff $s_i=1$ for all $i\in S$}
    \nonumber\\
    &=
    \E_{\bm v}
    \sum_{k(s)<m}
    \bm p(s)
    \sum_{j=1}^{k(s)}
    \binom{k(s)}{j}\mu_jc_j
    && \because \text{there are $\binom{k(s)}{j}$ such $S$ with $|S|=j$.}
    \label{eq:Xm-before-h}
\end{align}
\end{proof}

\section{$h(k)$ bound}
\label{app:h(k)_bound}
\HkBound*
\begin{proof}
    Let
$
    \bm\beta
    :=
    \tr(\bm\Delta\bm\Delta'),
$
where $\bm\Delta$ and $\bm\Delta'$ are independent Haar conjugates of
$\Delta_0$. By Lemma~\ref{lem:output-moment},
\[
    \E[\bm\beta^{2r+1}]=0,
    \quad
    \E[\bm\beta^{2r}]
    \le
    K^r r!\frac{1}{\dout^{2r}},
    \quad K=96.
\]
Recall,
$
    \mu_j
    :=
    \varepsilon^{2j}\E[\bm\beta^j].
$
Therefore we have
\[
    \mu_{2r+1}=0,
    \quad
    \mu_{2r}
    \le
    K^r r!\frac{\varepsilon^{4r}}{\dout^{2r}}.
\]

Using the beta-function representation
$
    c_j
    =
    (\din-1)\int_0^1 t^j(1-t)^{\din-2}\,dt
$
and the vanishing of the odd moments, we obtain
\begin{align}
    h(k)
    &=
    \sum_{r\ge1}\binom{k}{2r}\mu_{2r}c_{2r}
    =
    (\din-1)\int_0^1
    \sum_{r\ge1}
    \binom{k}{2r}
    \varepsilon^{4r}\E[\bm\beta^{2r}]
    t^{2r}(1-t)^{\din-2}\,dt.
    \label{eq:h-integral}
\end{align}
For $2r\le k$,
\[
    \binom{k}{2r}
    <
    \frac{k^{2r}}{(2r)!}
    =
    \frac{k^{2r}}{(r!)^2\binom{2r}{r}}
    \le
    \frac{k^{2r}}{2^r(r!)^2}.
\]
Hence, using
$
\E[\bm\beta^{2r}]
\le
r!(K/\dout^2)^r,
$
we obtain
\begin{align}
    \binom{k}{2r}
    \varepsilon^{4r}
    \E[\bm\beta^{2r}]
    t^{2r}
    &\le
    \frac{k^{2r}}{2^r(r!)^2}
    \varepsilon^{4r}
    r!\left(\frac{K}{\dout^2}\right)^r
    t^{2r}
    =
    \frac{1}{r!}
    \left(
        \frac{Kk^2\varepsilon^4}{2\dout^2}
        t^2
    \right)^r.
    \label{eq:h-term-bound}
\end{align}
Therefore,
\begin{align}
    h(k)
    &<
    (\din-1)\int_0^1
    \sum_{r\ge1}
    \frac{(\lambda_k t^2)^r}{r!}
    (1-t)^{\din-2}\,dt
    \nonumber\\
    &=
    (\din-1)\int_0^1
    \left(e^{\lambda_k t^2}-1\right)
    (1-t)^{\din-2}\,dt,
    \quad
    \lambda_k
    :=
    \frac{Kk^2\varepsilon^4}{2\dout^2}.
    \label{eq:h-lambda-bound}
\end{align}

We now bound this integral.

\begin{enumerate}
    \item If $\din\in\{2,3\}$, then for $0\le t\le1$,
$
    e^{\lambda_k t^2}-1
    \le
    \lambda_k t^2 e^{\lambda_k}.
$
Hence Eqn.~\eqref{eq:h-lambda-bound} gives
\begin{align}
    h(k)
    &<
    \lambda_k e^{\lambda_k}
    (\din-1)
    \int_0^1
    t^2(1-t)^{\din-2}\,dt
    =
    \frac{2\lambda_k e^{\lambda_k}}
         {\din(\din+1)}.
    \label{eq:hk-small-din}
\end{align}

    \item If $\din\ge4$, then for $0\le t\le1$,
    \[
        e^{\lambda_k t^2}-1
        \le
        \lambda_k t^2e^{\lambda_k t},
        \quad
        (1-t)^{\din-2}
        \le
        e^{-(\din-2)t}.
    \]
    Hence, provided $\lambda_k<\din-2$,
    \begin{align}
        h(k)
        &<
        \lambda_k(\din-1)
        \int_0^1
        t^2e^{-(\din-2-\lambda_k)t}\,dt
       \le
        \lambda_k(\din-1)
        \int_0^\infty
        t^2e^{-(\din-2-\lambda_k)t}\,dt
        =
        \frac{2\lambda_k(\din-1)}
             {(\din-2-\lambda_k)^3}.
        \label{eq:hk-large-din}
    \end{align}
\end{enumerate}

\end{proof}

\section{Odd $\dout$ proof for Theorem~\ref{thm:main_non-adapt}}
The following proof establishes the lower bound for odd $\dout$ in Theorem~\ref{thm:main_non-adapt}.
\begin{proof}
\label{app:odd-analysis}

For odd $\dout$, we use
\[
    \Delta_0
    =
    \frac{\sqrt{\dout}}{\dout-1}
    \sum_{j=1}^{(\dout-1)/2}|j\rangle\langle j|
    -
    \frac{\sqrt{\dout}}{\dout+1}
    \sum_{j=(\dout+1)/2}^{\dout}|j\rangle\langle j|.
\]
Then
$
    \tr(\Delta_0)=0,
$
$\|\Delta_0\|_1=\sqrt{\dout},$
$\|\Delta_0\|_\infty=\sqrt{\dout}/(\dout-1),$
and
$
    \|\Delta_0\|_2^2=\dout^2/(\dout^2-1).
$

In the analysis of the low-hit contribution, the distribution of
$
    \bm\beta=\tr(\bm\Delta\bm\Delta')
$
is no longer symmetric. Although $\E\bm\beta=0$, the higher odd moments
need not vanish. In Lemma~\ref{lem:odd-output-moment}, for every $j\ge2$ we show,
\[
    |\mu_j|
    =
    \varepsilon^{2j}|\E[\bm\beta^j]|
    \le
    \varepsilon^{2j}\E|\bm\beta|^j
    \le
    \left(
        \frac{C\sqrt j\,\varepsilon^2}{\dout}
    \right)^j.
\]
Since $\mu_1=0$, we bound low-hit contribution by
\[
    \widetilde h(k)
    :=
    \sum_{j=2}^k
    \binom{k}{j}|\mu_j|c_j,
\]
such that
$
    \|X_m\|_1^2\le\widetilde h(m-1).
$
Using the Beta-function representation
\[
    c_j
    =
    (\din-1)\int_0^1 t^j(1-t)^{\din-2}\,dt,
\]
we obtain
\begin{align}
    \widetilde h(k)
    &\le
    (\din-1)\int_0^1
    \sum_{j=2}^k
    \binom{k}{j}
    \varepsilon^{2j}\E|\bm\beta|^j
    t^j(1-t)^{\din-2}\,dt
    \le
    (\din-1)\int_0^1
    \sum_{j=2}^{\infty}
    \frac{(k\varepsilon^2t)^j}{j!}
    \E|\bm\beta|^j
    (1-t)^{\din-2}\,dt.
\end{align}
Using 
$
    \E|\bm\beta|^j
    \le
    \left(\frac{C\sqrt j}{\dout}\right)^j,
    ~\text{for}~j\ge2.
$
(See Lemma~\ref{lem:odd-output-moment}) and writing
$
    x:={Ck\varepsilon^2t}/{\dout},
$
we have
\[
    \sum_{j=2}^{\infty}
    \frac{(k\varepsilon^2t)^j}{j!}
    \E|\bm\beta|^j
    \le
    \sum_{j=2}^{\infty}
    \frac{j^{j/2}x^j}{j!}.
\]
For the even terms, using
$
    {(2r)^r}/{(2r)!}
    \le
   {2^r}/{r!},
$
gives
$
    \sum_{r\ge1}
    {(2r)^r x^{2r}}/{(2r)!}
    \le
    e^{2x^2}-1.
$
For the odd terms, for $r\ge1$,
$
    {(2r+1)^{r+1/2}}/{(2r+1)!}
    \le
    {2^r}/{r!},
$
and therefore
$
    \sum_{r\ge1}
    {(2r+1)^{r+1/2}x^{2r+1}}/{(2r+1)!}
    \le
    x\bigl(e^{2x^2}-1\bigr)
    \le
    e^{\frac52x^2}-1.
$
Combining the two bounds and choosing a large enough universal constant $C' >0$ and substituting back $x$ we get
\[
    \sum_{j=2}^{\infty}
    \frac{j^{j/2}x^j}{j!}
    \le
    \exp\left(
        \frac{C'k^2\varepsilon^4t^2}{\dout^2}
    \right)-1.
\]
Therefore
\[
    \widetilde h(k)
    \le
    (\din-1)\int_0^1
    \left[
        \exp\left(
            \frac{C'k^2\varepsilon^4t^2}{\dout^2}
        \right)-1
    \right]
    (1-t)^{\din-2}\,dt.
\]
This is the same Beta-function integral as in the even-dimensional case in Eqn.~\eqref{eq:h-lambda-bound},
with only the universal constant changed. Consequently,
\[
    k
    \le
    c\,
    \frac{\sqrt{\din}\,\dout}{\varepsilon^2}
\]
for a sufficiently small universal constant $c>0$ implies
$
    \widetilde h(k)<1/9
$.
Thus one may again choose
$
    m
    =
    \Theta\left(
        {\sqrt{\din}\dout}/{\varepsilon^2}
    \right),
$
and the odd-dimensional construction yields the same
$
    \Omega\left(
        {\din^{3/2}\dout}/{\varepsilon^2}
    \right)
$
query lower bound.
\end{proof}

\section{Proof of Theorem~\ref{thm:din2-lower-bound}}
\label{app:din2bound}
\dinbound*
\begin{proof}

We use the hard family from Eqn.~\eqref{eqn:hard-case-non-adapt-v1} with
$U=I$, thereby removing the Haar randomness on the output. Define
\[
    \rho_*:=\tauout+\frac{\varepsilon}{\sqrt{\dout}}\Delta_0
\]
and, for fixed $v$, the hard channel instance
\[
    \M_v(X)
    :=
    \langle v|X|v\rangle\rho_*
    +
    \tr[(I-|v\rangle\langle v|)X]\tauout.
\]
The CPTP and diamond-distance checks follow as usual. For a fixed
tester input $\Psi$, let $\sigma_0 = (\N^{\otimes n}\otimes\Id_R)(\Psi)$ and $\sigma_{\bm v}
    :=
    (\M_{\bm v}^{\otimes n}\otimes\Id_R)(\Psi)$ denote the pre-measurement states under the null and the alternate hypothesis respectively. Define: $\bar\sigma_*:=\E_{\bm v}\sigma_{\bm v}$. Applying Lemma~\ref{lem:randomized-to-deterministic-tester} we get $1/3\le 1/2\|\bar\sigma_*-\sigma_0\|_1$,
    
Following the hit--miss notation we have already established in Section~\ref{sec:hit,miss}, for a fixed $m\ge1$,
we define the new low-hit contribution as
\[
    X_m^{(*)}
    :=
    \E_{\bm v}
    \sum_{k(s)<m}
    p_{\bm v}(s)
    \Bigg(
        \Pi_s^\dagger
        \left[
            \rho_*^{\otimes k(s)}
            \otimes\tauout^{\otimes(n-k(s))}
        \right]
        \Pi_s
        -
        \tauout^{\otimes n}
    \Bigg)
    \otimes\omega_R^{s,\bm v}.
\]
The high-hit contribution is bounded above by a Markov's inequality exactly as in
Eqn.~\eqref{eq:high-hit-bound}, giving
\begin{equation}
    \frac12\|\bar\sigma_*-\sigma_0\|_1
    \le
    \frac12\|X_m^{(*)}\|_1+\frac{n}{m\din}.
    \label{eq:din2-high-hit}
\end{equation}

$X_m^{*}$ is Hermitian and its support lies in the support of $\sigma_0$. 
Hence 
\begin{equation}
    \|X_m^{(*)}\|_1^2
    \le
    \tr\left[
        \sigma_0^{-1/2}X_m^{(*)}
        \sigma_0^{-1/2}X_m^{(*)}
    \right].
    \label{eq:din2-weighted-l2}
\end{equation}

We proceed to bound the RHS of Eqn.~\eqref{eq:din2-weighted-l2}, the same way we did in Eqns.~\eqref{eq:Xm-quadratic-expanded},~\eqref{eqn:betaprod} and~\eqref{eq:output-overlap-expanded}. Since
$\tr(\Delta_0^2)=1$, the analogue of
Eqn.~\eqref{eq:output-overlap-expanded} is
\[
    \prod_{i=1}^n(1+\varepsilon^2s_is_i')-1
    =
    \sum_{\emptyset\neq S\subseteq[n]}
    \varepsilon^{2|S|}
    \prod_{i\in S}s_is_i'.
\]
Thus Eqns.~\eqref{eq:Xm-after-output-average}--\eqref{eq:Xm-before-h}
apply with $\mu_j$ replaced by $\varepsilon^{2j}$. In particular,
with $c_j:=\binom{\din+j-1}{j}^{-1}$, the same Haar average over
$\bm v$ gives $A_S\preceq c_{|S|}\Psi_R$, and therefore
\[
    \|X_m^{(*)}\|_1^2
    \le
    \E_{\bm v}
    \sum_{k(s)<m}p_{\bm v}(s)
    \sum_{j=1}^{k(s)}
    \binom{k(s)}{j}\varepsilon^{2j}c_j.
\]
For $k:=k(s)$,
\[
    \binom{k}{j}c_j
    =
    \prod_{r=0}^{j-1}\frac{k-r}{\din+r}
    \le
    \left(\frac{k}{\din}\right)^j,
\]
and hence, whenever $k\varepsilon^2<\din$,
\[
    \sum_{j=1}^{k}
    \binom{k}{j}\varepsilon^{2j}c_j
    \le
    \sum_{j=1}^{\infty}
    \left(\frac{k\varepsilon^2}{\din}\right)^j
    =
    \frac{k\varepsilon^2/\din}
         {1-k\varepsilon^2/\din}.
\]
Since $k(s)\le m-1$ on every low-hit branch,
\begin{equation}
    \|X_m^{(*)}\|_1^2
    \le
    \frac{(m-1)\varepsilon^2/\din}
         {1-(m-1)\varepsilon^2/\din}.
    \label{eq:din2-low-hit}
\end{equation}
Choosing $m-1\le\din/(10\varepsilon^2)$ forces
$\|X_m^{(*)}\|_1<1/3$, and combining this with
Eqn.~\eqref{eq:din2-high-hit} yields our desired complexity.

For odd $\dout$, take
\[
    \Delta_0
    :=
    \frac{\sqrt{\dout}}{\dout-1}
    \sum_{i=1}^{(\dout-1)/2}|i\rangle\langle i|
    -
    \frac{\sqrt{\dout}}{\dout+1}
    \sum_{i=(\dout+1)/2}^{\dout}|i\rangle\langle i|.
\]
Here
\[
    \tr(\Delta_0)=0,
    \quad
    \|\Delta_0\|_1=\sqrt{\dout},
    \quad
    \tr(\Delta_0^2)
    =
    \frac{\dout^2}{\dout^2-1}
    \le\frac98.
\]
The preceding steps  therefore apply with $\varepsilon^2$ replaced
by $\tr(\Delta_0^2)\varepsilon^2$, changing only the constants.
Choosing $m=\Theta(\din/\varepsilon^2)$ again yields the same
complexity, completing the proof.
\end{proof}

\section{The output-state bound}\label{sec:output}

Assume $\dout$ is even, and define
\[
    \Delta_0
    :=
    \frac{1}{\sqrt{\dout}}
    \left(
        \sum_{j=1}^{\dout/2}|j\rangle\langle j|
        -
        \sum_{j=\dout/2+1}^{\dout}|j\rangle\langle j|
    \right).
\]
Let $B\cong\C^{\dout}$ be the output space,
$\tauout:=I_B/\dout$, and let
$\bm U\sim\operatorname{Haar}(U(\dout))$. Recall that
\[
    \bm\rho
    :=
    \tauout+\frac{\varepsilon}{\sqrt{\dout}}\,
    \bm U\Delta_0\bm U^\dagger
    \in\mathsf D(B).
\]
For a hit-miss record $s\in\{0,1\}^n$, let
$k:=k(s):=\sum_{i=1}^n s_i$. Define
\[
    \bar\rho_k
    :=
    \E_{\bm U}\left[\bm\rho^{\otimes k}\right]
    \in\mathsf D(B^{\otimes k}),
    \quad
    \delta_k
    :=
    \frac12
    \left\|
        \bar\rho_k-\tauout^{\otimes k}
    \right\|_1.
\]

\begin{lemma}[Output-state]\label{lem:output-mixing}
For every $k\ge0$,
$
    \delta_k
    \le
    \sqrt{K}\,\dfrac{k\varepsilon^2}{\dout},
$
where $K=96$.
\end{lemma}

\begin{proof}
$\delta_0 = 0$ so assume in the rest of the proof $k\ge1$.

\paragraph{Step 1: Using Chi-squared divergence and reducing to the purity of $\bar\rho_k$ - }

Since $\tauout^{\otimes k}=I_{\dout^k}/\dout^k$ is full rank, using Lemma~\ref{lem:chi} we get
\begin{equation}
    4\delta_k^2
    \le
    \chi^2(\bar\rho_k\|\tauout^{\otimes k})=
    \dout^k
    \tr \left[
        (\bar\rho_k-\tauout^{\otimes k})^2
    \right] =
    \dout^k\tr(\bar\rho_k^2)-1,
    \label{eq:chi-purity}.
\end{equation}

where we used,
$
    (\tauout^{\otimes k})^{-1}
    =
    \dout^k I_{\dout^k}
$ and 
$
    \tr(\bar\rho_k\tauout^{\otimes k})
    =
    \tr((\tauout^{\otimes k})^2)
    =
    \frac1{\dout^k}.
$

The reason for using the $\chi^2$-divergence is that, with the maximally mixed reference state over $k$ registers, it reduces to a quadratic Hilbert-Schmidt expression and hence to the purity $\tr(\bar\rho_k^2)$, which is convenient for the Haar-moment calculation below. 

\paragraph{Step 2: Expressing the purity as an overlap moment - }

Let $\bm U'$ be an independent Haar unitary and define the independent copy
\[
    \bm\rho'
    :=
    \tauout+\frac{\varepsilon}{\sqrt{\dout}}\bm\Delta'.
\]
where $\bm\Delta'
    :=
    \bm U'\Delta_0(\bm U')^\dagger.$
By independence,
\begin{align}
    \tr(\bar\rho_k^2)
    =
    \E_{\bm U,\bm U'}
    \tr \left[
        \bm\rho^{\otimes k}(\bm\rho')^{\otimes k}
    \right]=
    \E_{\bm U,\bm U'}
    \left[\left[\tr(\bm\rho\bm\rho')\right]^k\right].
\end{align}
Therefore
\begin{equation}\label{eq:purity-overlap}
    \dout^k\tr(\bar\rho_k^2)
    =
    \E_{\bm U,\bm U'}
    \left[ \left[
        \dout\,\tr(\bm\rho\bm\rho')
    \right]^k\right].
\end{equation}

\paragraph{Step 3: Computing the one-copy overlap -}

Using
\[
    \bm\rho
    =
    \tauout+\frac{\varepsilon}{\sqrt{\dout}}\bm\Delta,
    \quad
    \bm\rho'
    =
    \tauout+\frac{\varepsilon}{\sqrt{\dout}}\bm\Delta',
\]
and $\tr\bm\Delta=\tr\bm\Delta'=0$, we obtain
\begin{align}
    \tr(\bm\rho\bm\rho')
    =
    \tr(\tauout^2)
    +
    \frac{\varepsilon^2}{\dout}
    \tr(\bm\Delta\bm\Delta') =
    \frac1{\dout}
    +
    \frac{\varepsilon^2}{\dout}
    \tr(\bm\Delta\bm\Delta').
\end{align}
Thus
\begin{equation}\label{eq:one-overlap}
    \dout\,\tr(\bm\rho\bm\rho')
    =
    1+\varepsilon^2\bm \beta,
    \quad
    \bm \beta:=\tr(\bm\Delta\bm\Delta').
\end{equation}

By Lemma~\ref{lem:output-moment}, the distribution of $\bm \beta$ is symmetric and
\begin{equation}\label{eq:Xmoment-use}
    \E[\bm \beta^{2r}]
    \le
    K^r r!\,\dout^{-2r},
    \quad r\ge1,
\end{equation}
with $K=96$.

Thus, using the $\chi^2$-divergence we have now turned a quantum-state distinguishability problem into a scalar moment problem for $\bm \beta$ defined above.

\paragraph{Step 4: Bounding the chi-squared divergence - }

Combining Eqns.~\eqref{eq:chi-purity}, \eqref{eq:purity-overlap}, and \eqref{eq:one-overlap} we get
\begin{align}
    \chi^2(\bar\rho_k\|\tauout^{\otimes k})
    &=
    \E(1+\varepsilon^2\bm \beta)^k-1 =
    \sum_{r\ge1}
    \binom{k}{2r}
    \varepsilon^{4r}\E[\bm \beta^{2r}] \le
    \sum_{r\ge1}
    \binom{k}{2r}
    \varepsilon^{4r}
    \frac{K^r r!}{\dout^{2r}}.
    \label{eq:chi-series}
\end{align}
Using
$
    \binom{k}{2r}\le {k^{2r}/}{(2r)!}$ and $
    {r!}/{(2r)!}\le1/{r!},
$
we obtain
\[
    \binom{k}{2r}
    \varepsilon^{4r}
    \frac{K^r r!}{\dout^{2r}}
    \le
    \frac1{r!}
    \left(
        \frac{Kk^2\varepsilon^4}{\dout^2}
    \right)^r.
\]
Therefore,
\begin{align}
    \chi^2(\bar\rho_k\|\tauout^{\otimes k})
    \le
    \sum_{r=1}^{\infty}
    \frac1{r!}
    \left(
        \frac{Kk^2\varepsilon^4}{\dout^2}
    \right)^r =
    \exp \left(
        \frac{Kk^2\varepsilon^4}{\dout^2}
    \right)-1.
    \label{eq:chi-exp}
\end{align}

\paragraph{Step 5: Concluding the trace-distance bound -}

Let
$
    t:=\sqrt K\,\frac{k\varepsilon^2}{\dout}.
$
If $t\le1$, then
$
    e^{t^2}-1
    \le
    (e-1)t^2
    <
    4t^2.
$
Together with Eqn.~\eqref{eq:chi-purity} and Eqn.~\eqref{eq:chi-exp}, this gives
$
    4\delta_k^2<4t^2,$
     hence $
    \delta_k\le t.
$
If $t\ge1$, the trivial bound $\delta_k\le1\le t$ gives the same
conclusion, therefore proving the Lemma.
\end{proof}

\begin{lemma}[Output-state bound for odd output dimension]
\label{lem:odd-output-mixing}
Assume that $\dout\ge3$ is odd, and let
\[
    \Delta_0
    =
    \frac{\sqrt{\dout}}{\dout-1}
    \sum_{j=1}^{(\dout-1)/2}|j\rangle\langle j|
    -
    \frac{\sqrt{\dout}}{\dout+1}
    \sum_{j=(\dout+1)/2}^{\dout}|j\rangle\langle j|.
\]
Let
\[
    \bm\rho
    :=
    \tauout+\frac{\varepsilon}{\sqrt{\dout}}\,
    \bm U\Delta_0\bm U^\dagger,
    \quad
    \bar\rho_k
    :=
    \E_{\bm U}[\bm\rho^{\otimes k}],
    \quad
    \delta_k
    :=
    \frac12\|\bar\rho_k-\tauout^{\otimes k}\|_1.
\]
Then, for every $k\ge0$,
\[
    \delta_k
    \le
    \sqrt{K_{\mathrm{odd}}}\,
    \frac{k\varepsilon^2}{\dout},
\]
where $K_{\mathrm{odd}}>0$ is a universal constant.
\end{lemma}

\begin{proof}
The case $k=0$ is immediate, so assume $k\ge1$. As in the proof of
Lemma~\ref{lem:output-mixing},
\[
    4\delta_k^2
    \le
    \chi^2(\bar\rho_k\|\tauout^{\otimes k})
    =
    \dout^k\tr(\bar\rho_k^2)-1.
\]
Let $\bm U'$ be an independent Haar unitary and set
$\bm\Delta:=\bm U\Delta_0\bm U^\dagger$,
$\bm\Delta':=\bm U'\Delta_0(\bm U')^\dagger$, and
$\bm\beta:=\tr(\bm\Delta\bm\Delta')$. Since $\tr(\Delta_0)=0$, the same
overlap calculation gives
$
    \dout\,\tr(\bm\rho\bm\rho')
    =
    1+\varepsilon^2\bm\beta,
$
and hence
$
    \chi^2(\bar\rho_k\|\tauout^{\otimes k})
    =
    \E(1+\varepsilon^2\bm\beta)^k-1.
$

Unlike the even-dimensional case, the distribution of $\bm\beta$ need not
be symmetric. By Lemma~\ref{lem:odd-output-moment},
\[
    \E\bm\beta=0,
    \quad
    \E|\bm\beta|^j
    \le
    \left(C\sqrt j/\dout\right)^j,
    \quad j\ge2,
\]
for a universal constant $C>0$. Therefore,
\begin{align}
    \chi^2(\bar\rho_k\|\tauout^{\otimes k})
    &=
    \sum_{j=2}^k
    \binom{k}{j}
    \varepsilon^{2j}\E[\bm\beta^j]
    \le
    \sum_{j=2}^{\infty}
    \frac{(k\varepsilon^2)^j}{j!}\E|\bm\beta|^j \le
    \sum_{j=2}^{\infty}
    \frac{j^{j/2}}{j!}
    \left(Ck\varepsilon^2/\dout\right)^j.
\end{align}
Set $x:=Ck\varepsilon^2/\dout$. Splitting the series according to the
parity of the moment order $j$, for $j=2r$,
$
    (2r)^r/(2r)!
    \le
    2^r/r!,
$
therefore
$
    \sum_{r\ge1}(2r)^r x^{2r}/(2r)!
    \le
    e^{2x^2}-1.
$
For $j=2r+1$, $r\ge1$,
$
    (2r+1)^{r+1/2}/(2r+1)!
    \le
    3^r/r!,
$
and hence
$
    \sum_{r\ge1}
    (2r+1)^{r+1/2}x^{2r+1}/(2r+1)!
    \le
    x(e^{3x^2}-1).
$
Since $x\le e^{x^2/2}$ for $x\ge0$, we get
$
    x(e^{3x^2}-1)
    \le
    e^{(7/2)x^2}-1.
$
Combining the two contributions,
\[
    \chi^2(\bar\rho_k\|\tauout^{\otimes k})
    \le
    2\bigl(e^{(7/2)x^2}-1\bigr)
    \le
    e^{7x^2}-1.
\]
Thus, with $K_{\mathrm{odd}}:=7C^2$,
\[
    \chi^2(\bar\rho_k\|\tauout^{\otimes k})
    \le
    \exp\left(
        K_{\mathrm{odd}}k^2\varepsilon^4/\dout^2
    \right)-1.
\]

Finally, let
$t:=\sqrt{K_{\mathrm{odd}}}\,k\varepsilon^2/\dout$. If $t\le1$, then
$e^{t^2}-1<(4t^2)$, and hence $4\delta_k^2<4t^2$. Therefore
$\delta_k\le t$. If $t\ge1$, the trivial bound
$\delta_k\le1\le t$ gives the same conclusion.
\end{proof}
\bigskip
\section{Haar output moment bound}\label{app:output-moment}

We prove the bound used in Lemma~\ref{lem:output-mixing}. The argument
uses concentration of Haar measure on $U(\dout)$ with respect to the
Hilbert-Schmidt metric.

\begin{lemma}[Haar concentration, \cite{fawzi2023quantum, 10.1214/ECP.v18-2551}]\label{lem:Haar-concentration} 
Let $\bm V$ be Haar-random on $U(\dout)$, and let
$F:U(\dout)\to\mathbb R$ be $L$-Lipschitz with respect to the
Hilbert-Schmidt norm. Then, for every $t>0$,
\begin{equation}\label{eq:Haar-concentration}
    \Pr \left(
        |F(\bm V)-\E F(\bm V)|\ge t
    \right)
    \le
    2\exp \left(
        -\frac{\dout t^2}{12L^2}
    \right).
\end{equation}
\end{lemma}

\begin{lemma}[Haar output moment for even dimension]
\label{lem:output-moment}
Assume that $\dout$ is even, and let
$\bm U,\bm U'$ be independent Haar-random unitaries on $U(\dout)$. Set
\[
    \bm\Delta:=\bm U\Delta_0\bm U^\dagger,
    \quad
    \bm\Delta':=\bm U'\Delta_0(\bm U')^\dagger,
    \quad
    \bm\beta:=\tr(\bm\Delta\bm\Delta').
\]
Then
\[
    \E\bm\beta=0,
    \quad
    \E[\bm\beta^{2r+1}]=0
    \quad (r\ge0),
\]
and, for every $r\ge1$,
\begin{equation}
    \E[\bm\beta^{2r}]
    \le
    96^r r!\,\dout^{-2r}.
    \label{eq:output-moment-final}
\end{equation}
\end{lemma}

\begin{proof}

\medskip
\noindent\textbf{Step 1: Reduction to one Haar unitary.}

Let
$
    \bm V:=\bm U^\dagger\bm U'.
$
Since $\bm U$ and $\bm U'$ are independent Haar-random unitaries,
$\bm V$ is Haar-random on $U(\dout)$. By cyclicity of the trace,
\begin{align}
    \bm \beta
    =
    \tr \left(
        \bm U\Delta_0\bm U^\dagger
        \bm U'\Delta_0(\bm U')^\dagger
    \right)=
    \tr \left(
        \Delta_0\bm V\Delta_0\bm V^\dagger
    \right).
\end{align}
Thus, with
$
    F(V):=\tr(\Delta_0V\Delta_0V^\dagger),
$
we have $\bm \beta=F(\bm V)$.

\medskip
\noindent\textbf{Step 2: Mean and symmetry.}

We first compute the mean of $\bm \beta=F(\bm V)$. 
Therefore, by linearity of expectation and trace and using Eqn.~\eqref{eqn:Haarfirstmoment} we get
\begin{align}
    \E_{\bm V}[\bm \beta]
    =
    \E_{\bm V}
    \tr(\Delta_0\bm V\Delta_0\bm V^\dagger) =
    \tr \left(
        \Delta_0
        \E_{\bm V}[\bm V\Delta_0\bm V^\dagger]
    \right)
    =
    0.
    \label{eq:X-mean-zero}
\end{align}

We next show that all odd moments vanish. Since $\dout$ is even, define
\[
    W
    :=
    \sum_{j=1}^{\dout/2}
    \left(
        |2j-1\rangle\langle 2j|
        +
        |2j\rangle\langle 2j-1|
    \right).
\]
This unitary swaps the $+1/\sqrt{\dout}$ and
$-1/\sqrt{\dout}$ eigenspaces of $\Delta_0$, yielding
\[
    W\Delta_0W^\dagger=-\Delta_0.
\]
For every fixed $V\in U(\dout)$,
\begin{align}
    F(VW)
    =
    \tr\left(
        \Delta_0VW\Delta_0W^\dagger V^\dagger
    \right)
    =
    -\tr(\Delta_0V\Delta_0V^\dagger)
    =
    -F(V).
\end{align}

Since right multiplication by the fixed unitary $W$ preserves Haar measure,
$\bm V W$ has the same distribution as $\bm V$. Thus, for every
$r\ge0$,
\begin{align}
    \E_{\bm V}[\bm \beta^{2r+1}]
    &=
    \E_{\bm V}[F(\bm V)^{2r+1}] =
    \E_{\bm V}[F(\bm V W)^{2r+1}] =
    \E_{\bm V}[(-F(\bm V))^{2r+1}] =
    -\E_{\bm V}[F(\bm V)^{2r+1}]\nonumber\\
    &=
    -\E_{\bm V}[\bm \beta^{2r+1}].
\end{align}
Hence $
    \E_{\bm V}[\bm \beta^{2r+1}]=0,
    \quad r\ge0.$

\medskip
\noindent\textbf{Step 3: Lipschitz bound.}
For arbitrary fixed $V,V'\in U(\dout)$,
\begin{align}
    &|F(V)-F(V')|
    \\&=
    \left|
        \tr(\Delta_0V\Delta_0V^\dagger)
        -
        \tr(\Delta_0V'\Delta_0(V')^\dagger)
    \right|
    && \text{definition of $F$}
    \nonumber\\[2pt]
    &=
    \left|
        \tr\left(\Delta_0(V-V')\Delta_0V^\dagger\right)
        +
        \tr\left(\Delta_0V'\Delta_0
        (V^\dagger-(V')^\dagger)\right)
    \right|
    && \pm\,\tr(\Delta_0V'\Delta_0V^\dagger)
    \nonumber\\[2pt]
    &\le
    \left|
        \tr\left(\Delta_0(V-V')\Delta_0V^\dagger\right)
    \right|
    +
    \left|
        \tr\left(\Delta_0V'\Delta_0
        (V^\dagger-(V')^\dagger)\right)
    \right|
    && \text{triangle inequality}
    \nonumber\\[2pt]
    &=
    \left|
        \tr\left((V-V')\Delta_0V^\dagger\Delta_0\right)
    \right|
    +
    \left|
        \tr\left((V^\dagger-(V')^\dagger)\Delta_0V'\Delta_0\right)
    \right|
    && \text{cyclicity of trace}
    \nonumber\\[2pt]
    &\le
    \|V-V'\|_2\,
    \|\Delta_0V^\dagger\Delta_0\|_2
    +
    \|V^\dagger-(V')^\dagger\|_2\,
    \|\Delta_0V'\Delta_0\|_2
    && \text{refer Eqn.~\eqref{HS-norm}}
    \nonumber\\[2pt]
    &\le
    \|\Delta_0\|_\infty\|\Delta_0\|_2
    \left(
        \|V-V'\|_2+
        \|V^\dagger-(V')^\dagger\|_2
    \right)
    && \text{refer Eqn.~\eqref{eq:2 to operator norm}}
    \nonumber\\[2pt]
    &=
    2\|\Delta_0\|_\infty\|\Delta_0\|_2
    \|V-V'\|_2
    && \|V^\dagger-(V')^\dagger\|_2=\|V-V'\|_2
    \nonumber\\[2pt]
    &=
    \frac{2}{\sqrt{\dout}}\,
    \|V-V'\|_2
    && \|\Delta_0\|_2=1,\ 
       \|\Delta_0\|_\infty=\dout^{-1/2}.
    \label{eq:X-Lipschitz}
\end{align}
Since the preceding estimate holds for all fixed $V,V'\in U(\dout)$,
the deterministic function $F$ is $L=2/\sqrt{\dout}$-Lipschitz.
Evaluating $F$ at the Haar-random unitary $\bm V$, we have
$\bm \beta=F(\bm V)$. Therefore, Lemma~\ref{lem:Haar-concentration}
applies to the random variable $\bm \beta$.

\medskip
\noindent\textbf{Step 4: Tail bound.}

Applying Lemma~\ref{lem:Haar-concentration} with
$L^2=4/\dout$ and $\E F(\bm V)=0$ gives
\begin{align}
    \Pr(|\bm \beta|\ge t)
    &\le
    2\exp \left(
        -\frac{\dout t^2}{12(4/\dout)}
    \right) =
    2\exp \left(
        -\frac{\dout^2t^2}{48}
    \right).
    \label{eq:X-tail}
\end{align}

\medskip
\noindent\textbf{Step 5: Moment bound.}

For any nonnegative random variable $Y$ and $p>0$,
\[
    \E[Y^p]
    =
    p\int_0^\infty t^{p-1}\Pr(Y\ge t)\,dt.
\]
Applying this with $Y=|\bm \beta|$ and $p=2r$, then using
Eqn.~\eqref{eq:X-tail},
\begin{align}
    \E|\bm \beta|^{2r}
    =
    2r\int_0^\infty
    t^{2r-1}\Pr(|\bm \beta|\ge t)\,dt \le
    4r\int_0^\infty
    t^{2r-1}
    \exp \left(
        -\frac{\dout^2t^2}{48}
    \right)dt.
    \label{eq:X-moment-integral}
\end{align}

Let
$
    u:=\frac{\dout^2t^2}{48}.
$
Such that
$
    t^{2r-1}dt
    =
    \dfrac12
    \left(\dfrac{48}{\dout^2}\right)^r
    u^{r-1}du.
$
Substituting into Eqn.~\eqref{eq:X-moment-integral} yields
\begin{align}
    \E|\bm \beta|^{2r}
    &\le
    2r
    \left(\frac{48}{\dout^2}\right)^r
    \int_0^\infty u^{r-1}e^{-u}\,du
    =
    2r
    \left(\frac{48}{\dout^2}\right)^r
    \Gamma(r)
    =
    2\,r!
    \left(\frac{48}{\dout^2}\right)^r,
\end{align}
Since $r\ge1$ implies $2\le2^r$ and
$
    2\cdot48^r\le96^r.
$
Therefore
\[
    \E[\bm \beta^{2r}]
    =
    \E|\bm \beta|^{2r}
    \le
    96^r r!\,\dout^{-2r},
\]
which proves Eqn.~\eqref{eq:output-moment-final}.
\end{proof}

\begin{lemma}[Odd-dimensional Haar absolute moments]
\label{lem:odd-output-moment}
Let $\dout\ge3$ be odd and let
\[
    \Delta_0
    =
    \frac{\sqrt \dout}{\dout-1}
    \sum_{j=1}^{(\dout-1)/2}|j\rangle\langle j|
    -
    \frac{\sqrt \dout}{\dout+1}
    \sum_{j=(\dout+1)/2}^{\dout}|j\rangle\langle j|.
\]
Let $\bm U,\bm U'\sim\operatorname{Haar}(U(\dout))$ be independent and define
\[
    \bm\Delta:=\bm U\Delta_0\bm U^\dagger,
    \quad
    \bm\Delta':=\bm U'\Delta_0(\bm U')^\dagger,
    \quad
    \bm\beta:=\tr(\bm\Delta\bm\Delta').
\]
Then
$
    \E\bm\beta=0,
$
and there exists a universal constant $C>0$ such that, for every $j\ge2$,
\begin{equation}
    \E|\bm\beta|^j
    \le
    \left(
        \frac{C\sqrt j}{\dout}
    \right)^j.
    \label{eq:odd-absolute-moment}
\end{equation}
\end{lemma}

\begin{proof}
By Haar invariance,
$
    \bm\beta
    \overset{\dout}{=}
    f(\bm W), $ and $
f(W):=\tr(\Delta_0W\Delta_0W^\dagger),
$
where $\bm W\sim\operatorname{Haar}(U(\dout))$. Since
$\tr(\Delta_0)=0$,
\[
    \E f(\bm W)
    =
    \tr\left[
        \Delta_0\,
        \E_{\bm W}(\bm W\Delta_0\bm W^\dagger)
    \right]
    =0.
\]
Moreover,
\begin{align}
    |f(U)-f(V)|
    &\le
    \|\Delta_0\|_2
    \|U\Delta_0U^\dagger-V\Delta_0V^\dagger\|_2
    \le
    2\|\Delta_0\|_2\|\Delta_0\|_\infty\|U-V\|_2.
\end{align}
Since
\[
    \|\Delta_0\|_2^2=\frac{\dout^2}{\dout^2-1},
    \quad
    \|\Delta_0\|_\infty=\frac{\sqrt \dout}{\dout-1},
\]
the Lipschitz constant of $f$ is $O(\dout^{-1/2})$. Standard Haar
concentration on $U(\dout)$ therefore gives universal constants
$c,C_0>0$ such that
\[
    \Pr(|\bm\beta|\ge t)
    \le
    C_0 e^{-c\dout^2t^2}.
\]
Hence, for $j\ge2$,
\begin{align}
    \E|\bm\beta|^j
    &=
    j\int_0^\infty
    t^{j-1}\Pr(|\bm\beta|\ge t)\,\dout t
    \le
    C_0j\int_0^\infty
    t^{j-1}e^{-c\dout^2t^2}\,\dout t
    =
    \frac{C_0j}{2(c\dout^2)^{j/2}}
    \Gamma\left(\frac j2\right)
    \le
    \left(
        \frac{C\sqrt j}{\dout}
    \right)^j,
\end{align}
for a universal constant $C>0$.
\end{proof}
\bigskip
\section{Construction of the common truncation map}
\label{app:gamma}

This appendix proves Lemma~\ref{lem:truncation}. We give the formal construction and verification of the common truncation map. 

Fix a unit vector $v\in\C^{\din}$ and a truncation threshold $m\ge1$. We recall:
\begin{itemize}
    \item \textbf{Hard instance and hit-miss decomposition.}
    For fixed $v,U$, recall the hard channel instance
    \[
        \M_{v,U}(X)
        =
        \langle v|X|v\rangle\,\rho_U
        +
        \tr\left[(I_A-|v\rangle\langle v|)X\right]\tauD.
    \]
    The channel under null hypothesis can also be similarly decomposed as 
    \[
        \N(X)
        =
        \langle v|X|v\rangle\,\tauD
        +
        \tr\left[(I_A-|v\rangle\langle v|)X\right]\tauD.
    \]
    Thus each query admits a hit-miss decomposition indexed by $s_i\in\{0,1\}$, where $s_i=1$ denotes a hit associated with $|v\rangle\langle v|$, and $s_i=0$ denotes a miss associated with $I_A-|v\rangle\langle v|$. This decomposition is used only in the analysis and is not revealed to the tester.

    \item \textbf{Outcome records and hit count.}
    A fixed outcome record is
    $s_{1:n}=(s_1,\ldots,s_n)\in\{0,1\}^n$, while
    $\bm S_{1:n}=(\bm S_1,\ldots,\bm S_n)$ denotes the corresponding
    random variable for the outcome record. For a fixed record, define
    $
        \mathsf K(s_{1:i})
        :=
        \sum_{t=1}^i s_t,
    $
    and, for fixed $v$,
    $
        \bm{\mathsf K}_v
        :=
        \mathsf K(\bm S_{1:n})
        =
        \sum_{i=1}^n \bm S_i.
    $
    denotes its random variable.
    \item \textbf{Subnormalized contributions.}
    Let $\psi_i^{v,U}(s_{1:i})$ and
    $\psi_i^{v,0}(s_{1:i})$ denote the subnormalized tester states
    associated with the record $s_{1:i}$ under $\M_{v,U}$ and $\N$,
    respectively. We split the pre-measurement states according to the
    hit count:
    \[
        \sigma_{\le m}^{v,U}
        :=
        \sum_{\mathsf K(s_{1:n})\le m}
        \psi_n^{v,U}(s_{1:n}),
        \quad
        \sigma_{>m}^{v,U}
        :=
        \sum_{\mathsf K(s_{1:n})>m}
        \psi_n^{v,U}(s_{1:n}),
    \]
    with $\sigma_{\le m}^{v,0}$ and $\sigma_{>m}^{v,0}$ defined
    similarly under the null.
\end{itemize}

If an outcome record contains at most $m$ hits, then $\M_{v,U}$ has
supplied the tester with at most $m$ copies of the $U$-dependent state
$\rho_U$; at the corresponding hit locations, $\N$ supplies $\tauD$
instead. Thus the only hypothesis-dependent output states appearing along
such a record are copies of $\rho_U$ versus copies of $\tauD$, suggesting
a reduction to the $m$-copy comparison
\[
    \rho_U^{\otimes m}\quad\text{versus}\quad
    \tauD^{\otimes m}.
\] The difficulty is that the tester is
adaptive and earlier copies of $\rho_U$ can influence later channel inputs, the locations and probabilities of subsequent hits. It is therefore
not immediate that the entire low-hit evolution can be written as a single
$U$-independent map applied to these $m$ source states. We give existence of such a map below.

\subsection{Truncation Map: An Intuition}

The truncation map (formally defined in the next section) may be viewed as a simulation of the original adaptive
experiment as a non-adaptive CP trace-nonincreasing map, restricted to outcome records containing at most $m$ hits.
It has access to additional source registers
$C_1,\ldots,C_m$, initialized either in
$\rho_U^{\otimes m}$ or in $\tauD^{\otimes m}$.

Recall that the original experiment (see Figure~\ref{fig:adaptive-tester}) starts from $\Psi$ and alternates channel uses
$\Phi\otimes\Id$ with the adaptive maps $\mathsf V_i$, where
$\Phi=\N$ under the null hypothesis and $\Phi=\M_{v,U}$ under the
alternate hypothesis. The truncation map follows the same adaptive
evolution, but implements each channel use according to the following rule:
A miss of the projector onto the hidden input direction outputs $\tauD$, while on the $r$-th hit the state stored in the source register $C_r$ is supplied as the corresponding channel output to the tester. Thus, if the source registers $C_1,\ldots,C_m$ are initialized in the product state
$\rho_U^{\otimes m}=\rho_U^{C_1}\otimes\cdots\otimes\rho_U^{C_m}$, the construction reproduces the alternative evolution; initializing them instead in
$\tauD^{\otimes m}=\tauD^{C_1}\otimes\cdots\otimes\tauD^{C_m}$ reproduces the null evolution, for every outcome record containing at most $m$ hits.

This idea is extremely close in spirit to teleportation stretching, similar to what Das-Wilde proposes in their paper \cite[Prop.~2]{DasWilde2019}, where one replaces an adaptive channel protocol by a fixed (non-adaptive) CPTP map acting on copies of a parameter-dependent state. Our truncation construction follows the same principle, that is, we replace the adaptive interaction with the channel queries by a simulation in which the unknown $U$-dependence is carried by copies of $\rho_U$ loaded into the source registers $C_{1:m}$, while the tester, the $(v)-$dependent routing of these hits, and the remaining channel action are then absorbed into a $(U)$-independent CP trace-nonincreasing map $\Gamma_{v,m}$.

\subsection{Truncation Map: Formal Construction}
\label{formal_const}
Recall our standing abuse of notation that the register labels
$A_i,B_i,R_i$ also denote their corresponding Hilbert spaces, with
$A_i\cong\C^{\din}$, $B_i\cong\C^{\dout}$, and $R_i$ an arbitrary finite-dimensional Hilbert space.
Introduce source
registers $C_1,\ldots,C_m\cong\C^{\dout}$, with the same
convention, and write
\[
    C_{r:l}:=C_r\otimes\cdots\otimes C_l,
    \quad\text{if }r\le l\quad~~~~\text{and }~~~
    C_{r:l}:=\C\quad\text{if }r>l.
\]

For a positive, possibly sub-normalized operator $X$ on
$A_i\otimes R_{i-1}$, recall the residual states on the memory
$R_{i-1}$ associated with the hit-miss outcomes are respectively:
\begin{align*}
    \mathsf{Res}_i^{\,1,v}[X]
    &:=
    (\langle v|\otimes I_{R_{i-1}})
    X
    (|v\rangle\otimes I_{R_{i-1}}),\\
    \mathsf{Res}_i^{\,0,v}[X]
    &:=
    \tr_{A_i}\left[
        \bigl((I_{A_i}-|v\rangle\langle v|)\otimes I_{R_{i-1}}\bigr)X
    \right].
\end{align*}
Where we call
\[
    \mathsf{Res}_i^{\,s_i,v}:
    \Lop(A_i\otimes R_{i-1})
    \longrightarrow
    \Lop(R_{i-1}),
    \quad s_i\in\{0,1\},
\]
residual maps. When auxiliary source registers are present, these maps act trivially on them, that is,
for any register $C$,
\begin{equation}
    (\mathsf{Res}_i^{\,s_i,v}\otimes\Id_C)[X],
    \quad
    s_i\in\{0,1\},
    \label{eq:Res-extended}
\end{equation}
with the residual map acting only on $A_i$ and leaving the memory and all
unused source registers unchanged. For every surviving outcome record sequence $s_{1:i}$, with
$k:=\mathsf K(s_{1:i})$, define
\begin{equation}
    \mathsf T_{s_{1:i}}:
    \Lop(C_{1:m})
    \longrightarrow
    \begin{cases}
        \Lop(A_{i+1}\otimes R_i\otimes C_{k+1:m}),
        & i<n,\\
        \Lop(R_n\otimes C_{k+1:m}),
        & i=n.
    \end{cases}
    \label{eq:T-prefix-type}
\end{equation}
For $Y\in\mathcal L(C_{1:m})$, initialize $\mathsf T_{\varnothing}(Y):=\Psi\otimes Y.$

\paragraph{Miss recursion: }
Define
\begin{equation}
    \mathsf T_{s_{1:i}}(Y)
    :=
    (\mathsf V_i\otimes\Id_{C_{k+1:m}})
    \left(
        \tauD\otimes
        (\mathsf{Res}_i^{\,0,v}\otimes\Id_{C_{k+1:m}})
        [\mathsf T_{s_{1:i-1}}(Y)]
    \right).
    \label{eq:T-miss}
\end{equation}
The miss prepares $\tauD$ on $B_i$, does not consume a source register, and
leaves the hit count equal to $k$.

\paragraph{Hit recursion: }

If $s_i=1$ and $k<m$, query $i$ is the $(k+1)$-st hit. Using the fixed
identification $C_{k+1}\cong B_i \cong \mathbb C^{\dout}$, let
\begin{equation}
    \mathsf S_{i,k+1}:
    R_{i-1}\otimes C_{k+1}
    \longrightarrow
    B_i\otimes R_{i-1},
    \quad
    \mathsf S_{i,k+1}(|x\rangle\otimes|y\rangle)
    :=|y\rangle\otimes|x\rangle,
    \label{eq:S-swap}
\end{equation}
and let
$\widehat{\mathsf S}_{i,k+1}:=\mathsf S_{i,k+1}\otimes I_{C_{k+2:m}}$.
Define
\begin{equation}
    \mathsf T_{s_{1:i}}(Y)
    :=
    (\mathsf V_i\otimes\Id_{C_{k+2:m}})
    \left(
        \widehat{\mathsf S}_{i,k+1}
        (\mathsf{Res}_i^{\,1,v}\otimes\Id_{C_{k+1:m}})
        [\mathsf T_{s_{1:i-1}}(Y)]
        \widehat{\mathsf S}_{i,k+1}^{\dagger}
    \right).
    \label{eq:T-hit}
\end{equation}

If $k=m$ and $s_i=1$, the new prefix contains $m+1$ hits and is omitted from
the recursion.

\paragraph{Complete Map: }

For the complete surviving record $s_{1:n}$, with
$k_n:=\mathsf K_v(s_{1:n})\le m$ and define
\begin{equation}
    \mathsf G_{s_{1:n}}(Y)
    :=
    \tr_{C_{k_n+1:m}}\mathsf T_{s_{1:n}}(Y)
    \in\mathcal L(\Hfin).
    \label{eq:G-record}
\end{equation}
The partial trace removes precisely the source registers not consumed by hits. Finally, define
\begin{equation}
    \Gamma_{v,m}(Y)
    :=
    \sum_{\substack{s_{1:n}\in\{0,1\}^n\\
                    \mathsf K_v(s_{1:n})\le m}}
    \mathsf G_{s_{1:n}}(Y).
    \label{eq:Gamma-formal-definition}
\end{equation}
The map $\Gamma_{v,m}$ depends on $v$, $m$, and the tester, but not on $U$.
\subsection[Truncation map action]{Truncation Map operating on $\rho_U^{\otimes m}$ and
$\tauD^{\otimes m}$}

For every surviving prefix $s_{1:i}$, with
$k:=\mathsf K_v(s_{1:i})\le m$. We prove by induction on $i$ that
\begin{align}
    \mathsf T_{s_{1:i}}(\rho_U^{\otimes m})
    &=
    \psi_i^{v,U}(s_{1:i})
    \otimes
    \rho_U^{C_{k+1}}\otimes\cdots\otimes\rho_U^{C_m},
    \label{eq:T-rho-prefix}\\
    \mathsf T_{s_{1:i}}(\tauD^{\otimes m})
    &=
    \psi_i^{v,0}(s_{1:i})
    \otimes
    \tauD^{C_{k+1}}\otimes\cdots\otimes\tauD^{C_m}.
    \label{eq:T-tau-prefix}
\end{align}
For $i=0$, both identities follow immediately from
$\mathsf T_{\varnothing}(Y)=\Psi\otimes Y$ and
$\psi_0^{v,U}(\varnothing)=\psi_0^{v,0}(\varnothing)=\Psi$.

Suppose now that the identities hold for $s_{1:i-1}$, and write
$
    k:=\mathsf K_v(s_{1:i-1}).
$
\paragraph{Miss: $s_i=0$: }
The hit count remains $k$. For the alternative source,
\begin{align}
    \mathsf T_{s_{1:i}}(\rho_U^{\otimes m})
    &=
    (\mathsf V_i\otimes\Id_{C_{k+1:m}})
    \left(
        \tauD\otimes
        (\mathsf{Res}_i^{\,0,v}\otimes\Id_{C_{k+1:m}})
        [\mathsf T_{s_{1:i-1}}(\rho_U^{\otimes m})]
    \right)
    \nonumber\\
    &=
    (\mathsf V_i\otimes\Id_{C_{k+1:m}})
    \left(
        \tauD\otimes
        \mathsf{Res}_i^{\,0,v}
        [\psi_{i-1}^{v,U}(s_{1:i-1})]
        \otimes
        \rho_U^{C_{k+1}}\otimes\cdots\otimes\rho_U^{C_m}
    \right)
    \nonumber\\
    &=
    \psi_i^{v,U}(s_{1:i})
    \otimes
    \rho_U^{C_{k+1}}\otimes\cdots\otimes\rho_U^{C_m}.
    \label{eq:T-rho-miss-check}
\end{align}
For the null source,
\begin{align}
    \mathsf T_{s_{1:i}}(\tauD^{\otimes m})
    &=
    (\mathsf V_i\otimes\Id_{C_{k+1:m}})
    \left(
        \tauD\otimes
        \mathsf{Res}_i^{\,0,v}
        [\psi_{i-1}^{v,0}(s_{1:i-1})]
        \otimes
        \tauD^{C_{k+1}}\otimes\cdots\otimes\tauD^{C_m}
    \right)
    \nonumber\\
    &=
    \psi_i^{v,0}(s_{1:i})
    \otimes
    \tauD^{C_{k+1}}\otimes\cdots\otimes\tauD^{C_m}.
    \label{eq:T-tau-miss-check}
\end{align}

\paragraph{Hit: $s_i=1$: }
Since the new prefix survives, $k<m$, so query $i$ is the
$(k+1)$-st hit. For the alternative source,
\begin{align}
    \mathsf T_{s_{1:i}}(\rho_U^{\otimes m})
    &=
    (\mathsf V_i\otimes\Id_{C_{k+2:m}})
    \left(
        \widehat{\mathsf S}_{i,k+1}
        \Big(
            \mathsf{Res}_i^{\,1,v}
            [\psi_{i-1}^{v,U}(s_{1:i-1})]
            \otimes
            \rho_U^{C_{k+1}}\otimes\cdots\otimes\rho_U^{C_m}
        \Big)
        \widehat{\mathsf S}_{i,k+1}^{\dagger}
    \right)
    \nonumber\\
    &=
    \psi_i^{v,U}(s_{1:i})
    \otimes
    \rho_U^{C_{k+2}}\otimes\cdots\otimes\rho_U^{C_m}.
    \label{eq:T-rho-check}
\end{align}
For the null source,
\begin{align}
    \mathsf T_{s_{1:i}}(\tauD^{\otimes m})
    &=
    (\mathsf V_i\otimes\Id_{C_{k+2:m}})
    \left(
        \widehat{\mathsf S}_{i,k+1}
        \Big(
            \mathsf{Res}_i^{\,1,v}
            [\psi_{i-1}^{v,0}(s_{1:i-1})]
            \otimes
            \tauD^{C_{k+1}}\otimes\cdots\otimes\tauD^{C_m}
        \Big)
        \widehat{\mathsf S}_{i,k+1}^{\dagger}
    \right)
    \nonumber\\
    &=
    \psi_i^{v,0}(s_{1:i})
    \otimes
    \tauD^{C_{k+2}}\otimes\cdots\otimes\tauD^{C_m}.
    \label{eq:T-tau-check}
\end{align}
Thus a miss consumes no source register, whereas a hit consumes $C_{k+1}$.
The induction proves
Eqns.~\eqref{eq:T-rho-prefix} and~\eqref{eq:T-tau-prefix} for every surviving prefix.
\paragraph{Endpoint identities: }
Let $s_{1:n}$ be a complete surviving record with
$\mathsf K(s_{1:n})\le m$. Since the unused source states have trace one,
the prefix identities and Eqn.~\eqref{eq:G-record} give
\begin{align}
    \mathsf G_{s_{1:n}}(\rho_U^{\otimes m})
    &=
    \psi_n^{v,U}(s_{1:n}),
    &
    \mathsf G_{s_{1:n}}(\tauD^{\otimes m})
    &=
    \psi_n^{v,0}(s_{1:n}).
\end{align}
Summing over all surviving records therefore yields
\begin{align}
    \Gamma_{v,m}(\rho_U^{\otimes m})
    &=
    \sum_{\mathsf K(s_{1:n})\le m}
    \psi_n^{v,U}(s_{1:n})
    =
    \sigma_{\le m}^{v,U},
    \label{eq:Gamma-rho-endpoint}\\
    \Gamma_{v,m}(\tauD^{\otimes m})
    &=
    \sum_{\mathsf K(s_{1:n})\le m}
    \psi_n^{v,0}(s_{1:n})
    =
    \sigma_{\le m}^{v,0}.
    \label{eq:Gamma-tau-endpoint}
\end{align}
\paragraph{Does it enlarge the model? :}
Importantly, $\Gamma_{v,m}$ is not an operation available to the tester and
does not enlarge the tester model. It is introduced only as an external tool to
handle the low-hit contribution. On every outcome record with at most $m$
hits, the tester receives the same output registers $B_i$ and applies the
same adaptive maps $\mathsf V_i$ as in the original evolution; the auxiliary
source registers are never accessible to the tester.

The role of $\Gamma_{v,m}$ is only to connect the original low-hit quantity
to the $m$-copy output-state problem. Recall, the proof has the form
\[
    \underbrace{
    \tr[\mathsf P\sigma_{\le m}^{v,0}]
    -
    \E_{\bm U}\tr[\mathsf P\sigma_{\le m}^{v,\bm U}]
    }_{\text{quantity from the original tester}}
    =
    \underbrace{
    \tr\left[
        \mathsf P\,\Gamma_{v,m}
        (\tauD^{\otimes m}-\bar\rho_m)
    \right]
    }_{\text{auxiliary factorization}}
    \le
    \underbrace{
     \delta_m:=\frac12\|\bar\rho_m-\tauD^{\otimes m}\|_1
    }_{\text{$m$-copy output-state bound}}.
\]
Therefore $\Gamma_{v,m}$ appears only in the intermediate representation as both
the quantity being bounded and the resulting upper bound are independent of
the auxiliary map itself.

For outcome records with more than $m$ hits, $\Gamma_{v,m}$ does not
reproduce the original adaptive evolution. This is irrelevant for the proof,
since the high-hit contribution is controlled separately. Hence
$\Gamma_{v,m}$ serves only to reproduce the adaptive evolution on the
low-hit records satisfying $\mathsf K_v(s_{1:n})\le m$.

\subsection{Complete positivity and trace non-increase}

Each $\mathsf T_{s_{1:i}}$ is completely positive: the map $Y\mapsto\Psi\otimes Y$, the hit-miss maps $\mathsf{Res}_i^{1,v}$ and $\mathsf{Res}_i^{0,v}$, tensoring with $\tauD$, conjugation by $\widehat{\mathsf S}_{i,k+1}$, and application of $\mathsf V_i$ are all completely positive. Since partial trace and summation preserve complete positivity, $\Gamma_{v,m}$ is completely positive.

To prove trace non-increase, let $Y\succeq0$, fix a surviving prefix
$s_{1:i-1}$, and write $k=\mathsf K(s_{1:i-1})$.

If $s_i=0$, then by Eqn.~\eqref{eq:T-miss},
\begin{align}
\tr\mathsf T_{s_{1:i}}(Y)
&=
\tr\left[
(\mathsf V_i\otimes\Id_{C_{k+1:m}})
\left(
\tauD\otimes
(\mathsf{Res}_i^{\,0,v}\otimes\Id_{C_{k+1:m}})
[\mathsf T_{s_{1:i-1}}(Y)]
\right)
\right]
\nonumber\\
&=
\tr\left[
\tauD\otimes
(\mathsf{Res}_i^{\,0,v}\otimes\Id_{C_{k+1:m}})
[\mathsf T_{s_{1:i-1}}(Y)]
\right]
\nonumber\\
&=
\tr\left[
(\mathsf{Res}_i^{\,0,v}\otimes\Id_{C_{k+1:m}})
[\mathsf T_{s_{1:i-1}}(Y)]
\right],
\end{align}
where we used that $\mathsf V_i$ is trace preserving and
$\tr\tauD=1$.

Similarly for $s_i=1$, then by Eqn.~\eqref{eq:T-hit} it follows that
\begin{align}
\tr\mathsf T_{s_{1:i}}(Y)
=
\tr\left[
(\mathsf{Res}_i^{\,1,v}\otimes\Id_{C_{k+1:m}})
[\mathsf T_{s_{1:i-1}}(Y)]
\right],
\end{align}
where we used that $\mathsf V_i$ is trace preserving and
$\widehat{\mathsf S}_{i,k+1}$ is unitary.

Hence
\begin{align}
\sum_{\substack{s_i\in\{0,1\}:\\
\mathsf K(s_{1:i})\le m}}
\tr\mathsf T_{s_{1:i}}(Y)
&\le
\sum_{s_i\in\{0,1\}}
\tr\left[
(\mathsf{Res}_i^{\,s_i,v}\otimes\Id_{C_{k+1:m}})
[\mathsf T_{s_{1:i-1}}(Y)]
\right]
\nonumber\\
&=
\tr\mathsf T_{s_{1:i-1}}(Y),
\label{eq:Gamma-trace-step}
\end{align}
where the last equality follows from
\[
\tr\left[
(\mathsf{Res}_i^{\,1,v}\otimes\Id_{C_{k+1:m}})[X]
\right]
+
\tr\left[
(\mathsf{Res}_i^{\,0,v}\otimes\Id_{C_{k+1:m}})[X]
\right]
=
\tr X.
\]
By definition of $\Gamma_{v,m}$,
\begin{align}
    \tr\Gamma_{v,m}(Y)
    &=
    \sum_{\substack{s_{1:n}\in\{0,1\}^n:\\
          \mathsf K(s_{1:n})\le m}}
    \tr\mathsf T_{s_{1:n}}(Y)
    && \text{after tracing out unused source registers}
    \nonumber\\
    &\le
    \tr\mathsf T_{\varnothing}(Y)
    && \text{by iterating Eqn.~\eqref{eq:Gamma-trace-step}}
    \nonumber\\
    &=
    \tr Y,
    && \text{since $\mathsf T_{\varnothing}(Y)=\Psi\otimes Y$ and
    $\tr\Psi=1$.}
    \label{eq:TNI}
\end{align}
Thus $\Gamma_{v,m}$ is completely positive and trace non-increasing.

\bigskip
\section{Corollary Proof}
\label{corollary_proof}

\srproblem*

Denote the query complexity of Problem~\ref{prob:SR} by
$Q_{\rm SR}(\varepsilon)$.

\srcorollary*

\begin{proof}[Proof of Corollary~\ref{cor:state-replacement-lower-bound}]

Let
$
    \bm v=\bm V|0\rangle,
    \text{ with }
    \bm V\sim\operatorname{Haar}(U(\din)) 
    \text{ and } 
    \bm U\sim\operatorname{Haar}(U(\dout)),
$
with $\bm V$ and $\bm U$ independent. For each realization $v,U$,
define
\[
    \tilde\rho_U
    :=
    \tauD+\frac{2\varepsilon}{\sqrt{\dout}}\,U\Delta_0U^\dagger,
    \quad
    \tauD:=\frac{I_{\dout}}{\dout}.
\]
We now propose the hard channel instance for Problem~\ref{prob:SR}:
\begin{equation}
\label{eq:SR-hard-channel}
    \mathcal O_{v,U}(X)
    :=
    \langle v|X|v\rangle\,\tilde \rho_U
    +
    \tr\left[
        (I_A-|v\rangle\langle v|)X
    \right]\tauD.
\end{equation}

Since $\{|v\rangle\langle v|,\,I_A-|v\rangle\langle v|\}$ is a POVM and
$\tilde \rho_U,\tauD\in\mathsf D(\mathbb C^{\dout})$, $\mathcal O_{v,U}$ is a CPTP for all $0\le 2\varepsilon\le 1$. It remains to prove that each $\mathcal{O}_{v,U}$ is indeed a hard channel instance for Problem~\ref{prob:SR}.
 Fix $v,U$ and
$\sigma\in\mathsf D(\mathbb C^{\dout})$. Operating $\mathcal O_{v,U}$ on the inputs
$
    |v\rangle\langle v|
    \text{ and } 
    \frac{I_A-|v\rangle\langle v|}{\din-1}
$
respectively yields
\begin{align}
    \|\mathcal O_{v,U}-\mathsf R_\sigma\|_\diamond
    &\ge
    \bigl\|
        \mathcal O_{v,U}(|v\rangle\langle v|)
        -
        \mathsf R_\sigma(|v\rangle\langle v|)
    \bigr\|_1
    =
    \|\tilde \rho_U-\sigma\|_1
    \nonumber\\
    \|\mathcal O_{v,U}-\mathsf R_\sigma\|_\diamond
    &\ge
    \left\|
        \mathcal O_{v,U}\left(
            \frac{I-|v\rangle\langle v|}{\din-1}
        \right)
        -
        \mathsf R_\sigma\left(
            \frac{I-|v\rangle\langle v|}{\din-1}
        \right)
    \right\|_1
    =
    \|\tauD-\sigma\|_1
    \nonumber\\
    \therefore~~\quad
    \|\mathcal O_{v,U}-\mathsf R_\sigma\|_\diamond
    &\ge
    \max\left\{
        \|\tilde \rho_U-\sigma\|_1,\,
        \|\tauD-\sigma\|_1
    \right\}
    \nonumber\\
    &\ge
    \frac12
    \left(
        \|\tilde \rho_U-\sigma\|_1
        +
        \|\tauD-\sigma\|_1
    \right) \ge
    \frac12\|\tilde \rho_U-\tauD\|_1
    =
    \varepsilon.
\end{align}
\begin{align}
\therefore~~~\inf_{\mathsf R_\sigma\in
    \mathsf{SR}(\C^{\din},\C^{\dout})}
    \|\mathcal O_{v,U}-\mathsf R_\sigma\|_\diamond
    &
    \ge
    \inf_{\sigma\in\mathsf D(\C^{\dout})}
    \varepsilon
    =
    \varepsilon.
    \label{eq:SR-hard-far}
\end{align}
Hence every $\mathcal O_{v,U}$ belongs to the alternate hypothesis $H_1^{\rm SR}$. Therefore, a valid restriction of Problem~\ref{prob:SR} is the certification problem
\begin{equation}
\label{eq:SR-hard-problem}
H_0^{\rm hard}:
\mathcal M=\mathcal N, ~~~\text{or}~
\quad
H_1^{\rm hard}:
\mathcal M=\mathcal O_{v,U}
\text{ for some }v,U.
\end{equation}
Where $\mathcal N := \mathsf R_{\dout}$ is the completely depolarizing channel. Indeed these hypotheses sets are subsets of the original hypotheses:
\[
    \{\mathcal N\}
    \subseteq
    H_0^{\rm SR},
    \text{ and }
    \{\mathcal O_{v,U}:v,U\}
    \subseteq
    H_1^{\rm SR}.
\]

While the first inclusion is trivial since $\mathcal{N} = \mathsf{R}_{\tauD} \in H_0^{\rm SR}$, the second inclusion follows from Eqn.~\eqref{eq:SR-hard-far}. The set $H_1^{\rm SR}$ comprises all channels whose diamond distance from the complete set of state-replacement channels is at least $\varepsilon$. In contrast, $H_1^{\rm hard}$ constraints the unknown channel to the specific family $\{\mathcal{O}_{v,U}:v,U\}$, wherein each member satisfies this distance criterion by construction.

Let $Q_{\rm hard}(2\varepsilon)$ denote the query complexity of
distinguishing the two hypotheses in Eqn.~\eqref{eq:SR-hard-problem} with
success probability at least $2/3$ in the fully adaptive ancilla-assisted
testing model. Since the hard problem is a restriction of
Problem~\ref{prob:SR}, any tester for the latter also solves the former.
Therefore,
\begin{equation}
\label{eq:SR-hard-restriction}
    Q_{\rm hard}(2\varepsilon)
    \le
    Q_{\rm SR}(\varepsilon).
\end{equation}

Observe that the restricted problem in Eqn.~\eqref{eq:SR-hard-problem} is
exactly the hard-family problem we analyzed in the adaptive depolarizing-channel
lower bound of Theorem~\ref{thm:main_adapt} but with the diamond distance separation
$2\varepsilon$.
Thus the null hypothesis, the alternate hypothesis, tester model, and success
requirements coincide with those used in that proof. Hence, for
$0<\varepsilon\le 1/2$,
\[
    Q_{\rm hard}(2\varepsilon)
    =
    \Omega\left(
        \frac{\din\dout}{(2\varepsilon)^2}
    \right)
    =
    \Omega\left(
        \frac{\din\dout}{\varepsilon^2}
    \right).
\]
Combining this with Eqn.~\eqref{eq:SR-hard-restriction} yields 
$    Q_{\rm SR}(\varepsilon)
    =
    \Omega\left(
        \dfrac{\din\dout}{\varepsilon^2}
    \right).
$

For coherent non-adaptive testers, the same restriction to hard hypotheses argument transfers both non-adaptive lower bounds. First, allowing random $v$ and $U$ gives exactly the hard-family problem used to prove the bound
\[
    \Omega\left(
        \frac{\din^{3/2}\dout}{(2\varepsilon)^2}
    \right)
    =
    \Omega\left(
        \frac{\din^{3/2}\dout}{\varepsilon^2}
    \right)
\]
in Theorem~\ref{thm:main_non-adapt}. Second, fixing $U=I$ gives the restricted family
\[
    \mathcal O_{v,I}(X)
    :=
    \langle v|X|v\rangle\,\tilde\rho_I
    +
    \tr\left[
        (I_A-|v\rangle\langle v|)X
    \right]\tauD,
\]
where
$
    \tilde\rho_I:=\tauD+\frac{2\varepsilon}{\sqrt{\dout}}\Delta_0.
$
Since
$
    \frac12\|\tilde\rho_I-\tauD\|_1 =\varepsilon,
$
it follows that
\[
    \inf_{\mathsf R_\sigma\in\mathsf{SR}(\C^{\din},\C^{\dout})}
    \|\mathcal O_{v,I}-\mathsf R_\sigma\|_\diamond
    \ge\varepsilon.
\]
Thus this family is also contained in $H_1^{\rm SR}$, and the
corresponding non-adaptive depolarizing-channel lower bound in Theorem~\ref{thm:din2-lower-bound} yields
$
    \Omega\left(
        {\din^2}/{(2\varepsilon)^2}
    \right)
    =
    \Omega\left(
        {\din^2}/{\varepsilon^2}
    \right).
$
Consequently, every coherent non-adaptive tester for
Problem~\ref{prob:SR} requires
\[
    \Omega\left(
        \max\left\{
            \frac{\din^{3/2}\dout}{\varepsilon^2},
            \frac{\din^2}{\varepsilon^2}
        \right\}
    \right)
\]
queries.
\end{proof}

\bibliographystyle{alpha}

\bibliography{references}

\end{document}